\documentclass[conference]{IEEEtran}

\usepackage{hyperref}
\usepackage{listings}
\usepackage{stmaryrd}
\usepackage{environ}
\usepackage{pifont}

\date{}

 \newcommand{\exclude}[1]{
 }

\newcommand{\nop}[1]{} % pretty much the same as exclude, to disable blocks

\NewEnviron{partold}{
 }
\NewEnviron{partnew}{
         \BODY
 }

\usepackage{amsfonts,amssymb,amsmath}

\lstdefinelanguage{promela}
  {morekeywords={do,od,init,proctype,if,fi,byte,bool,atomic},
  morecomment=[s]{/*}{*/},escapechar=\@,
  basicstyle=\small\ttfamily,
  commentstyle=\itshape\rmfamily\small,
  keywordstyle=\ttfamily\small\underbar
}

\usepackage{tikz}
\usetikzlibrary{arrows}
\usetikzlibrary{shapes}
\usetikzlibrary{calc}
\usetikzlibrary{positioning}

\usepackage{array}
\newcounter{rowno}
\usepackage{amssymb,amsmath}
\newtheorem{proposition}{Proposition}
\newtheorem{theorem}[proposition]{Theorem}

\newtheorem{corollary}[proposition]{Corollary}
\newtheorem{definition}[proposition]{Definition}

\newtheorem{remark}[proposition]{Remark}
\newtheorem{example}[proposition]{Example}
\newcommand{\qed}{} % IEEE adds QED

\title{
Counter Attack on Byzantine Generals: Parameterized Model
Checking of Fault-tolerant Distributed Algorithms
}

\author{\IEEEauthorblockN{Annu John,
Igor Konnov, Ulrich Schmid,
    Helmut Veith, Josef Widder
\thanks{Supported in part by the Austrian National Research
    Network S11403-N23 (RiSE) of the Austrian Science Fund (FWF),
    and by the Vienna Science and Technology Fund (WWTF)
    grant PROSEED.}
\IEEEauthorblockA{Vienna University of Technology (TU Wien)}
}%
}%

\IEEEoverridecommandlockouts % to show thanks when in conference mode

\newcommand{\mypara}[1]{\smallskip\noindent\emph{#1.}}

\newcommand{\ident}[1]{\textit{#1\/}\rule{0cm}{1ex}}
\gdef\dash---{\thinspace---\hskip.16667em\relax}
\gdef\ndash---{\thinspace--\hskip.16667em\relax}

\newcommand{\quotes}[1]{\textnormal{[}#1\textnormal{]}}

\newcommand{\Natural}{{\mathbb N}}
\newcommand{\NatZero}{{\mathbb N}_0}
\newcommand{\Int}{{\mathbb Z}}

\newcommand{\trmove}{{\sc (move)}}
\newcommand{\trmaintain}{{\sc (frame)}}

\newcommand{\paraset}{\Pi}
\newcommand{\globset}{\Gamma}
\newcommand{\locset}{\Lambda}
\newcommand{\pcval}{Z}

\newcommand{\AdmP}{\mathbf{P}_{RC}}
\newcommand{\param}{\mathbf{p}}

\newcommand{\pc}{\ident{sv}}
\newcommand{\PC}{\ident{SV}}
\newcommand{\rcvd}{\ident{rcvd}}
\newcommand{\sent}{\ident{nsnt}}
\newcommand{\sentf}{\ident{nsntf}}
\newcommand{\numparam}{{|\paraset|}}
\newcommand{\Next}[1]{{#1}^{\prime}}

\newcommand{\restrict}[2]{#1 |_{#2}}

\newcommand{\syssize}{N}

\newcommand{\fromstate}{{\text{\sc from}}}
\newcommand{\tostate}{{\text{\sc to}}}

\newcommand{\Prop}{\text{AP}}
\newcommand{\PropPC}{\Prop_{\PC}}
\newcommand{\PropVAR}{\Prop_\ident{D}}
\newcommand{\aprop}{\ident{p}}

\newcommand{\absMap}{h^{dat}_{\param}}
\newcommand{\absMapSys}{\bar{h}^{dat}_{\param}}
\newcommand{\vassMap}{h^{\textsf{VASS}}_{\param}}
\newcommand{\cacMap}{\bar{h}^{cnt}_{\param}}
\newcommand{\wrapMap}{\bar{h}^{dc}}
\newcommand{\absMax}{\mu}

\newcommand{\IntAbs}[1]{\hat{#1}}

\newcommand{\threshSet}{{\mathcal T}}
\newcommand{\thresh}{\vartheta}

\newcommand{\gleich}[1]{=_{#1}}
\newcommand{\nichtgleich}[1]{\ne_{#1}}

\newcommand{\ResCond}{{\ident{RC}}} %% resiliency condition
\newcommand{\abst}{\alpha}
\newcommand{\afunc}[2]{\alpha_{#1}(#2)}
\newcommand{\conc}{\gamma}
\newcommand{\cfunc}[2]{\gamma_{#1}(#2)}

\newcommand{\absEx}[0]{\alpha_E}

\newcommand{\somedomain}{\tilde{D}}

\newcommand{\absdomain}{\widehat{D}}
\newcommand{\absSym}{I}
\newcommand{\absZero}{\absSym_{0}}

\newcommand{\CFA}{{\cal A}}

\newcommand{\Sk}{\textsf{Sk}}
\newcommand{\SkAut}{\textsf{Sk}(\CFA)}

\newcommand{\SkAAut}{\textsf{Sk}_{\ident{abs}}(\CFA)}
\newcommand{\SkSAAut}{\textsf{Sk}_{\locset}(\CFA)}

\newcommand{\instBLA}{\textsf{Inst}}

\newcommand{\sConcSys}{{\instBLA(\param,\Sk)}}
\newcommand{\ConcSys}{{\instBLA(\param,\SkAut)}}
\newcommand{\AbstSys}{{\instBLA(\param,\SkAAut)}}
\newcommand{\SemiAbstSys}{{\instBLA(\param,\SkSAAut)}}
\newcommand{\ConcSysP}[1]{{\instBLA(#1,\SkAut)}}

\newcommand{\LetterConcSys}[0]{I}
\newcommand{\LetterWrapSys}[0]{\omega}
\newcommand{\WrapSys}[0]{\ensuremath{{\mbox{\textsf{Sys}}}}_\LetterWrapSys}
\newcommand{\LVassSys}[0]{\ensuremath{{\mbox{\textsf{VS}}_{\textsf{$\locset$}}}}}

\newcommand{\LetterAbstSys}{{\hat{I}}}

\newcommand{\CntASys}{{\textsf{C}}}
\newcommand{\CntSkAut}{{\CntASys(\SkAAut)}}

\newcommand{\newreftheorem}[2]{
 \newenvironment{#1}[1]{\par\vspace{3mm}\noindent\textbf{#2~\ref{##1}.}
\em}{\rm}
}
\newreftheorem{reflemma}{Lemma}
\newreftheorem{refproposition}{Proposition}
\newreftheorem{reftheorem}{Theorem}
\newreftheorem{refcorollary}{Corollary}

\newcommand{\CFAtemplate}[1]{{\mathtt{#1}}}
\newcommand{\CFAreserved}[1]{{\underline{\mathtt{#1}}}}

\newcommand{\CFAguard }{\CFAtemplate{guard}}

\newcommand{\CFApc}{{\pc}}
\newcommand{\CFAinc}{\CFAreserved{inc}}

\newcommand{\CFAdummy}{\varepsilon}
\newcommand{\CFApick}{\; \CFAreserved{where} \;}
\newcommand{\CFApickOp}[3]{#1 #2 #3}

\newcommand\labelfun{\lambda}

\newcommand{\absCntSymb}[0]{\kappa}
\newcommand{\absCnt}[1]{\absCntSymb [#1]}

\newcommand\localstates{{L}}

\newcommand\gst{\sigma}
\newcommand{\absgst}{\hat{\gst}}

\newcommand{\IT}{\mathrm{V0}}
\newcommand{\RI}{\mathrm{V1}}
\newcommand{\SE}{\mathrm{SE}}
\newcommand{\AC}{\mathrm{AC}}
\newcommand{\DD}{\mathrm{CR}}

\newcommand\ltlF{\textsf{\textbf{F}}\,}
\newcommand\ltlG{\textsf{\textbf{G}}\,}
\newcommand\ltlU{\,\textsf{\textbf{U}}\,}

\newcommand\LTLX{$\mbox{\textsf{LTL}}_{\textsf{{-X}}}$}

\newcommand\ACTL{$\mbox{\textsf{ACTL}}^\ast$}

\newcommand\Wedge[2]{\ensuremath{\bigwedge\limits_{#1}^{#2}}}
\newcommand\Vee[2]{\ensuremath{\bigvee\limits_{#1}^{#2}}}
\newcommand\Union[2]{\ensuremath{\bigcup\limits_{#1}^{#2}}}

\renewcommand\vec[1]{\mathbf{#1}}

\newcommand\offu[0]{\ensuremath{\quotes{\mathit{off\ } U}}}

\usepackage{color}
\usepackage{ifthen}

\usepackage{algorithm}
\usepackage[noend]{algorithmic}

\makeatletter

\newcommand{\newlinetag}[3]{\newcommand{#1}[#2]{\item[#3]}}
\newcommand{\newconstruct}[5]{%%
  \newenvironment{ALC@\string#1}{\begin{ALC@g}}{\end{ALC@g}}
   \newcommand{#1}[2][default]{\ALC@it#2\ ##2\ #3%%
     \ALC@com{##1}\begin{ALC@\string#1}}
   \ifthenelse{\boolean{ALC@noend}}{
     \newcommand{#4}{\end{ALC@\string#1}}
   }{
     \newcommand{#4}{\end{ALC@\string#1}\ALC@it#5}
   }
}

\newcommand{\ALCEXT@linenosize}{\small}
\newcommand{\ALCEXT@linenofont}{\rm}
\newcommand{\renew@ALC@linenosize}{\renewcommand{\ALC@linenosize}{\ALCEXT@linenosize\ALCEXT@linenofont}}
\newcommand{\setlinenosize}[1]{\renewcommand{\ALCEXT@linenosize}{#1}\renew@ALC@linenosize}
\newcommand{\setlinenofont}[1]{\renewcommand{\ALCEXT@linenofont}{#1}\renew@ALC@linenosize}

\renew@ALC@linenosize
\renewcommand{\ALC@linenodelimiter}{:}

\newcounter{ALCEXT@lineno}

\let\ALCEXT@endalgorithmic=\endalgorithmic
\def\endalgorithmic{\setcounter{ALCEXT@lineno}{\value{ALC@line}}\ALCEXT@endalgorithmic}

\newlinetag{\CODE}{1}{\textrm{Code for processes #1:}}

\newlinetag{\PARAM}{0}{\textbf{Parameters}}
\newlinetag{\VAR}{0}{\textbf{Variables}}
\newlinetag{\TRANS}{0}{\textbf{Rules}}
\makeatother

\makeatletter
\newcommand{\MESGitem}[1]{\ensuremath{#1}}
\newcommand{\MESGsep}{, }

\def\MESG#1{{(\/\let\theMessage={}\parsefirst#1,\@end,\theMessage\/)}}
\def\parsefirst#1,{%%
  \ifx#1\@end \let\next=\relax\else%%
  \let\next=\parsemessage%%
  \let\theMessage={\theMessage\MESGitem{#1}}\fi \next}
\def\parsemessage#1,{%%
  \ifx#1\@end \let\next=\relax \else%%
  \let\next=\parsemessage%%
  \let\theMessage={\theMessage\MESGsep\MESGitem{#1}}\fi \next}

\def\new@MESG#1[#2]{\expandafter\gdef\csname #1\endcsname##1{%%
    \MESG{\textsc{#2},##1}}}

\def\newMESG#1{\@ifnextchar [{\new@MESG{#1}}{\new@MESG{#1}[#1]}}
\makeatother

\newcommand{\recallthm}[2]{%
  {\bfseries Theorem~\ref{#1}.}{\itshape #2}
}
\newcommand{\recallproposition}[2]{%
  {\bfseries Proposition~\ref{#1}.}{\itshape #2}
}

\begin{document}

\pagestyle{plain}

\maketitle

\begin{abstract}
We introduce an automated parameterized verification method for
fault-tolerant distributed algorithms (FTDA). FTDAs are parameterized
by both the number of processes and the assumed maximum number of
Byzantine faulty processes.  At the center of our technique is a
parametric interval abstraction (PIA) where the interval boundaries
are arithmetic expressions over parameters.  Using PIA for both data
abstraction and a new form of counter abstraction, we reduce the
parameterized problem to finite-state model checking.  We demonstrate
the practical feasibility of our method by verifying several variants
of the well-known distributed algorithm by Srikanth and Toueg. Our
semi-decision procedures are complemented and motivated by an
undecidability proof for FTDA verification which holds even in the
absence of interprocess communication.  To the best of our knowledge,
this is the first paper to achieve parameterized automated
verification of Byzantine FTDA.
\end{abstract}

\IEEEpeerreviewmaketitle

%% Local Variables: %%
%% tex-main-file : paper %%

\section{Introduction}\label{sec:Intro}

\mypara{Parameterized Model Checking} %.
In its original formulation~\cite{ClarkeE81}, Model Checking was
     concerned with efficient procedures for the evaluation of a
     temporal logic specification~$\varphi$ over a finite Kripke
     structure $K$, i.e., decision procedures for~$K \models \varphi$.
Since $K$ can be extremely large, a multitude of logic-based
     algorithmic methods including symbolic verification
     \cite{McM93,BiereCCFZ99} and predicate abstraction \cite{GrafS97}
     were developed to make this decidable problem tractable for
     practical applications.
Finite-state models are, however, not always an adequate modeling
     formalism for software and hardware:

(i) Infinite-state models.
Many programs and algorithms are most naturally modeled by unbounded
     variables such as integers, lists, stacks etc.
Modern model checkers are using predicate abstraction \cite{GrafS97}
     in combination with SMT solvers to reduce an infinite-state model
     $I$ to a finite state model~$h(I)$ that is amenable to finite
     state model checking.
The construction of $h$ assures soundness, i.e., for a given
     specification logic such as \ACTL, we can assure by construction
     that $h(I) \models \varphi$ implies $I \models \varphi$.
The major drawback of abstraction is incompleteness: if $h(I)
     \not\models \varphi$ then it does in general not follow that $I
     \not\models \varphi$.
(Note that \ACTL\ is not closed under negation.) Counterexample-guided
     abstraction refinement (CEGAR) \cite{Clarke2003,BallMMR01}
     addresses this problem by an adaptive procedure, which analyzes
     the abstract counterexample for $h(I) \not\models \varphi$ on
     $h(I)$ to find a concrete counterexample or obtain a better
     abstraction $h'(I)$.
For abstraction to work in practice, it is crucial that the abstract
     domain from which $h$ and $h'$ are chosen is tailored to the
     problem class and possibly the specification.
Abstraction thus is a semi-decision procedure whose usefulness has to
     be demonstrated by practical examples.

(ii) An orthogonal modeling and verification problem is
     parameterization: Many software and hardware artifacts are
     naturally represented by an infinite class of structures ${\bf K}
     = \{K_1, K_2, \dots \}$ rather than a single structure.
Thus, the verification question is $\forall i K_i \models \varphi$,
     where~$i$ is called the parameter.
In the most important examples of this class, the parameter $i$ is
     standing for the number of replications of a concurrent
     component, e.g., the number of processes in a distributed
     algorithm, or the number of caches in a cache coherence protocol.
It is easy to see that even in the absence of concurrency,
     parameterized model checking is undecidable \cite{AK86}; more
     interestingly, undecidability even holds for networks of constant
     size processes arranged in a ring with a single token for
     communication \cite{S88,EN95}.
Although several approaches have been made to identify decidable
     classes for parameterized verification
     \cite{EN95,EmersonK00,WT07}, no decidable formalism has been
     found which covers a reasonably large class of interesting
     problems.
The diversity of problem domains for parameterized verification and
     the difficulty of the problem gave rise to many approaches
     including regular model checking \cite{A12} and abstraction
     \cite{PXZ02,CTV2008}\dash---the method discussed in this paper.
Again, the challenge in abstraction is to find an abstraction $h({\bf
     K})$ such that $h({\bf K}) \models \varphi$ implies~$K_i \models
     \varphi$~for~\emph{all}~$i$.

Most previous research on parameterized model checking focused on
     concurrent systems with $n+c$ processes where~$n$ is the
     parameter and $c$ is a {\em constant}: $n$ of the processes are
     {\em identical} copies; $c$ processes represent the
     non-replicated part of the system, e.g., cache directories,
     shared memory, dispatcher processes
     etc.~\cite{GS1992,Ip1996,McMillan01,CTV2008}.
%% Many approaches abstract the $n$ processes to a finite system, e.g.,
%%      counter abstraction \cite{PXZ02}, and environment
%%      abstraction~\cite{CTV2008}.
Most of the work on parameterized model checking considers only
     safety.
Notable exceptions are~\cite{KP2000,PXZ02} where several notions of
     fairness are considered in the context of abstraction to
     verify~liveness.

\mypara{Fault-tolerant Distributed Algorithms} In this paper, we are
     addressing the problem of parameterized verification of
     fault-tolerant distributed algorithms (FTDA).
This work is part of an interdisciplinary effort by the authors to develop a tool
     basis for the automated verification, and, in the long run,
     deployment of FTDAs~\cite{JKSVW12b,KVW12}.
FTDAs constitute a core topic
of the distributed algorithms community with a rich body of
     results~\cite{Lyn96,AW04}.
%% As discussed in~\cite{JKSVW12b,KVW12}, the verification of FTDA has to
%%      address two challenges, (i) the formalization problem, i.e., the
%%      question how to move from a mathematically intricate, but usually
%%      quite informal description in pseudocode to an {\em adequate}
%%      formal model, and (ii) the verification problem, i.e., {\em how
%%      to verify FTDA by an automated model checking based method}.
%% Although step (i) is very important, it involves many lengthy and
%%      technical arguments from distributed computing theory, which would
%%      exceed the format of this paper (cf.\ \cite{JKSVW12b} for
%%      extensive discussions of these issues).
%% This paper is exclusively concerned with the verification problem.
FTDAs are more difficult than the standard setting of parameterized model checking
     because {\em a certain number $t$ of the $n$
     processes can be faulty}.
In the case of e.g.~Byzantine faults, this means that the faulty
     processes can send messages in an unrestricted manner.
Importantly, the upper bound $t$ for the faulty processes is also a
     parameter, and is essentially a fraction of $n$.
The relationship between $t$ and~$n$ is given by a \emph{resilience
     condition}, e.g., $n > 3t$.
Thus, one has to reason about all systems with $n-f$ non-faulty and
     $f$ faulty processes, where $f \leq t$ and~$n > 3t$.

From a more operational viewpoint, FTDAs typically consist of multiple
     processes that communicate by message passing over a completely
     connected communication graph.
Since a sender can be faulty, a receiver cannot wait for a message
     from a specific sender process.
Therefore, most FTDAs use counters to reason about their environment.
If, for instance, a process receives a certain message $m$ from more
     than~$t$ distinct processes, it can conclude that one of the
     senders is non-faulty.
A large class of FTDAs expresses these counting arguments using
     \emph{threshold guards:}

\begin{minipage}{.8\linewidth}
  \begin{lstlisting}[language=pascal,label=Lst:ThresGC]
if received <m> from t+1 distinct processes
then action(m);
  \end{lstlisting}
\end{minipage}

Note that threshold guards generalize existential and universal guards
     \cite{EmersonK00}, i.e., rules that wait for messages from at
     least one or all processes, respectively.
As can be seen from the above example, and as discussed in
     \cite{JKSVW12b}, existential and universal guards are not
     sufficient to capture advanced~FTDAs.

%% After formalizing FTDAs that contain such rules using control flow
%%      automata \cite{HenzingerJMS02}, we will see that model checking
%%      of FTDAs falls in the class of undecidable problems
%%      from~\cite{AK86}, that is, parallel composition of parameterized
%%      processes.

\mypara{Contribution} %.
We consider parameterized verification of FTDAs with  threshold guards
     and resilience conditions.
We start by introducing a framework based on a new form of control
     flow automata that captures the semantics of threshold-guarded
     fault-tolerant distributed algorithms.
Based on this framework, we show that the parameterized model
     checking problem under consideration is undecidable, even
     for FTDAs {\em without} interprocess communication and without
     arithmetic operations.
Thus, we are led to propose a novel two-step abstraction
     technique.
Both steps are based on \emph{parametric interval abstraction} (PIA),
     a generalization of interval abstraction where the interval
     borders are parameters rather than constants.
Using the PIA domain, we obtain a finite-state model checking problem
     in two steps:

\noindent \textbf{Step 1: PIA data abstraction.} We evaluate the
     threshold guards over the parametric intervals.
Thus, we abstract away unbounded variables and parameters from the
     process code.
We obtain a parameterized system where the replicated processes are
     finite-state and independent of the parameters.

%%Our undecidability results applies to the latter system as well.

%% Following ideas by German and Sistla \cite{GS1992}, systems that
%%      results from Step~1 can be expressed as  vector addition state
%%      systems (Petri net), and similar arguments as introduced by
%%      Esparza~\cite{E97} can be used to show that liveness model
%%      checking is undecidable.\footnote{To verify liveness we require
%%      the modalities $\ltlF$, $\ltlG$, as well as atomic proposition
%%      expressing a test for zero, which is enough to follow the
%%      proof~in~\cite{E97}.} We are thus led to conduct a second
%%      abstraction step.

\noindent
\textbf{Step 2: PIA counter abstraction.}  We use a new form of
     counter abstraction where the process counters are abstracted to
     PIA.
As Step~1 guarantees that we need only finitely many counters, PIA
     counter abstraction yields a finite-state system.

It is interesting to note that the intermediate model obtained by Step 1
still falls in the undecidable class obtained above.

To evaluate the precision of our abstractions, we implemented our
     abstraction technique in a tool chain, and conducted experiments
     on several FTDAs.
Our experiments showed the need for abstraction refinement to deal
     with spurious counterexamples~\cite{Clarke2003}.
We encountered spurious behaviors that are due to parameterized
     abstraction and fairness; this required novel refinement
     techniques, which we also discuss in this paper.
In addition to refinement of PIA counter abstraction, which is
     automated in a loop using a model checker and an SMT solver, we
     are also exploiting simple user-provided invariant candidates to
     refine the abstraction similar to the CMP
     method~\cite{McMillan01,TalupurT08}.

Thus, we are able to verify several variants of the well-known
     distributed broadcast algorithm by Srikanth and
     Toueg~\cite{ST87,ST87:abc} in the Byzantine setting as well as
     the (simpler) algorithm  verified by Fisman et
     al.~\cite{FismanKL08}.
To the best of our knowledge, this is the first paper to achieve
     parameterized automated verification of Byzantine FTDA.

\mypara{Related work} %.
Traditionally, correctness of FTDAs was shown by handwritten
     proofs~\cite{Lyn96,AW04}, and, in some cases, by proof
     assistants~\cite{LR93,SWR02:hom,Charron-BostM09,Lamport11a}.
Completely automated model checking or synthesis methods are usually
     not parameterized~\cite{TS11,SteinerRSP04,BKA12}.
Our work stands in the tradition of parameterized model checking for
     protocols \cite{BCG1989,GS1992,EmersonK03,PXZ02,CTV2008}, i.e.,
     for mutual exclusion, cache coherence etc.
In particular, the techniques by Pnueli et al.~\cite{PXZ02}, namely,
     counter abstraction and justice preservation are keystones of our
     work.

%% Regular model checking is a well-established parameterized model checking
%%     technique that addresses both safety and liveness~\cite{AJNOS12};
%%     cf.~\cite{A12} for a recent survey.
%% The crucial idea is to represent a global state of a system of $n$
%%     processes as a word of length~$n$; one letter per local state of a
%%     finite state process.
%% By varying the length $n$, one can instantiate concrete systems consisting
%%     of $n$ processes.

%% Thus a set of global states is captured by a finite automaton, and the
%%     transition relation is a length preserving transducer.
%% By performing a backward exploration from a set of bad states, one can
%%     check the backward reachability of an initial state.
%% To render the technique applicable to practical examples, the techniques
%%     is extended with acceleration and approximation methods that enforce
%%     termination.

The first work that addresses parameterized model checking of FTDA
     uses regular model checking \cite{AJNOS12,A12}, and was conducted
     by Fisman et al.~\cite{FismanKL08}.
They model a parameterized system consisting of $n$ processes as a
     transducer, which translates a global state\dash---modeled as a
     word of length~$n$\dash---into the next global state of the same
     length.
Consequently, their models are limited to processes whose local state space and
     transition relation are {\em finite, fixed, and independent of
     parameters}, in particular, of~$n$ in their case.
Such models were sufficient to verify a folklore reliable broadcast
     algorithm RBC (cf.\ e.g., \cite{CT96}) that tolerates crash
     faults, and where every process stores whether it has received at
     least one message.
However, these models are not sufficient to capture more involved
     FTDAs that contain threshold guards as in our case.

Moreover, as \cite{FismanKL08} explain, the presence of a resilience
condition such as $n>3t$ would require them to intersect the regular
languages which describe sets of states with context-free languages
which enforce the resilience condition.

%     algorithms requiring a resilience condition like~$n>3t$, Fisman
%     et al.\ explained that one has to extend regular model checking,
%     with a preliminary step: %.
%one has to intersect regular languages expressing sets of states with
%     the context-free language expressing the resilience condition.
%% However, it is not clear whether there are FTDAs that require a
%%      specific resilience condition, and where the processes' local
%%      state space is fixed, finite, and independent of parameters.

Our framework captures the RBC algorithm, and more advanced algorithms
     that use threshold guards over parameters and resilience conditions.

To the best of our knowledge, the current paper is the first in which
     a distributed algorithm that tolerates Byzantine faults has been
     automatically verified for \emph{all} system sizes and \emph{all}
     admissible numbers of faulty processes.

Our technique applies to FTDAs, which are an important aspect
of distributed systems, but by far not the only one. While there is
other work in the verification and synthesis of distributed systems,
they do not focus on algorithmic verification of safety and liveness properties
for fault tolerant distributed algorithms. In this broader class of distributed systems literature, the work  by
     Abdulla and Johnson~\cite{AJ93} and Mayr~\cite{Mayr03} appears
     most closely related to ours because they address faults.
However, their ``lossy systems'' contain very different fault
     assumptions, which are not part of the problem class we consider
     here.

%%  who consider extreme
%%      cases where channels have no guarantees such that any message can
%%      be lost, or processes can reset their states arbitrarily, etc.
%% In this setting reachability analysis becomes decidable, while
%%      liveness is still undecidable.
%% Our problem is quite different in that we do not consider the both
%%      extrema of no faults at all or totally unrestricted faults.
%% For instance, while not requiring that all messages are delivered, we
%%      have to express that a certain fraction of processes may
%%      communicate.
%% Working with such intermediate assumptions\dash---as required to
%%      capture the subtle details of distributed computing theory
%%      adequately\dash---requires us to develop techniques to express
%%      specific fault behaviors, and restriction on fault
%%      occurrences (e.g., the resilience condition).

%% In the case of a symmetric system with finite-state processes, one can
%%     express a parameterized system as a Vector Addition System with States
%%     (or, equivalently, a Petri net)~\cite{GS92}.
%% We are using this idea after constructing the data abstraction, but our
%%     setting is very different from~\cite{GS92}.

Regarding our abstraction technique, an abstract domain
     similar to PIA was developed in~\cite{SIG07}.
It was used in the framework of abstract interpretation~\cite{CC1977},
     and was developed as a generalization of the polyhedra domain.
Starting from a similar domain,~\cite{SIG07} is thus taking a
     direction that is substantially different from parameterized
     model~checking.

Let us conclude the introduction with a note on terminology: Fault-tolerant
broadcast protocols are distributed algorithms which {\em achieve} reliable
all-to-all communication on top of (partially) unreliable communication, or 
in the presence of processes that may send conflicting information
     to different processes.
This notion of ``broadcast'' should not be confused with broadcast
     systems where the computational model contains broadcast
     \emph{primitives} that ensure that processes can send information
     to all others, as e.g., in~\cite{Esparza99}.

%%\paragraph{Organization of the paper.}

  \nop{
In Section~\ref{sec:prelim}, we formalize distributed systems and the
     parameterized model checking problem with multiple parameters.
In Section~\ref{sec:DA}, we specialize this problem to distributed
     algorithms, by introducing  a variant of control flow automata
     (CFA) which is suitable to describe distributed algorithms.
In Section~\ref{sec:Scheme}, we present our abstraction techniques
     that allow to reduce the parameterized model checking problem of
     threshold guarded algorithms to finite state model checking.
We also prove soundness of the abstractions.
Section~\ref{sec:casestudy} discusses, from a distributed algorithms
     perspective, how to model message passing
	 distributed algorithms in our framework by referring to our case
	 study, a variant of the broadcasting algorithm from~\cite{ST87}.
Finally, we discuss an example implementation in Section~\ref{sec:exp}.
  }

\section{System Model with Multiple Parameters}
\label{sec:prelim}
\label{sec:DA}

We  define the parameters, local variables of the processes, and
     shared variables referring to a single \emph{domain} $D$ that is
     totally ordered and has the operations addition and subtraction.
In this paper we will assume that $D = \Natural_0$.

% NOTE: we should take care when speaking about variables and their values.
% Values always range over D (except for pc), while variables serve only 
% information purposes, i.e., they help to understand which element of 
% a tuple refers to what variable.

We start with some notation.
Let $Y$ be a finite set of variables ranging over $D$.
We will denote by $D^{|Y|}$, the set of all $|Y|$-tuples of variable
     values.
In order to simplify notation, given $\vec{s}\in D^{|Y|}$, we use the
     expression $\vec{s}.y$, to refer to the value of a variable $y
     \in Y$ in vector $\vec{s}$.
For two vectors of variable values $\vec{s}$ and $\vec{s}'$, by
     $\vec{s} \gleich{X} \vec{s}'$ we denote the fact that for all
     $x\in X$, $\vec{s}.x = \vec{s}'.x$ holds.

\newcommand{\vars}{V}

\mypara{Process}
The  set of variables $\vars$ is $ \{\pc\} \cup\locset\cup
      \globset \cup \paraset$:
The variable $\pc$ is the  \emph{status variable} that ranges over a
     finite set $\PC$ of \emph{status values}.
%% (For simplicity, we assume that only one status variable is used;
%%      however, multiple finite domain status variables can be encoded into
%%      $\pc$.) %.
The finite set $\locset$ contains variables that range over the
     domain~$D$.
The variable $\pc$ and the variables from $\locset$ are \emph{local
     variables}.
The finite set $\globset$ contains the \emph{shared variables} that
     range over $D$.
The finite set $\paraset$ is a set of \emph{parameter variables} that
     range over~$D$, and the \emph{resilience condition} $\ResCond$ is
     a predicate over $D^{\numparam}$.
In our example, $\paraset = \{n,t,f\}$, and the resilience condition
     $\ResCond(n,t,f)$ is $n>3t \;\wedge\; f \le t \;\wedge\; t>0$.
Then, we denote the set of \emph{admissible parameters} by  $\AdmP =
     \{ \param\in D^\numparam \mid RC(\param)\}$.

A process operates on states from the set $S=\PC \times D^{|\locset|}
     \times D^{|\globset|} \times D^{\numparam}$.
Each process starts its computation in an initial state from a set
     $S^0\subseteq S$.
A relation $R \subseteq S \times S$ defines \emph{transitions} from
     one state to another, with the restriction that the  values of
     parameters remain unchanged, i.e., for all  $(s,
     t) \in R$, $s \gleich{\paraset} t$.
Then, a \emph{parameterized process skeleton} is a tuple~$\Sk = (S, S^0,
     R)$.

We get a process instance by fixing the parameter values $\param
     \in D^{\numparam}$: %.
one can restrict the set of process states to $\restrict{S}{\param} =
     \{ s \in S \mid s \gleich{\paraset} \param \}$ as well as the set
     of transitions to  
%% $\restrict{R}{\param} = \{ (s, t) \in R \mid s \in
%%      \restrict{S}{\param}, t \in \restrict{S}{\param}  \}$, that is,
     $\restrict{R}{\param} = R \cap (\restrict{S}{\param} \times
     \restrict{S}{\param})$.
Then, a \emph{process instance}  is a process skeleton $\restrict{\Sk}{\param} =
     (\restrict{S}{\param}, \restrict{S^0}{\param},
     \restrict{R}{\param})$ where $\param$ is constant.

\mypara{System Instances} For fixed admissible parameters $\param$, a
     distributed system is modeled as an asynchronous parallel
     composition of identical processes $\restrict{\Sk}{\param}$.
The number of processes depends on the parameters.
To formalize this, we define the size of a system (the number of
     processes) using a  function $\syssize\colon \AdmP \rightarrow
     \NatZero$, for instance, when modeling only correct processes
     explicitly, $n - f$ for $\syssize(n,t,f)$.

\nop{
Finally, given $\param \in \AdmP$, and a parameterized process
     skeleton $\Sk = (S, S^0, R)$, a  \emph{system instance}
     $\sConcSys = (S_\LetterConcSys, S^0_\LetterConcSys,
     R_\LetterConcSys, \Prop, \labelfun_\LetterConcSys)$ is a Kripke
     structure defined as an asynchronous parallel composition
%%     $\restrict{Sk}{\param} \parallel \cdots \parallel
%%     \restrict{Sk}{\param}$ 
of $\syssize(\param)$ process instances, indexed by $i \in \{1, \dots,
     \syssize(\param) \}$, following standard interleaving semantics.
Given a state $\gst$ of $\sConcSys$, we denote the state of process
     $i$ by $\sigma[i]$.
(The formal definition is given in Appendix~\ref{sec:proofs}.)
}

Given $\param \in \AdmP$, and a process skeleton
    $\Sk = (S, S^0, R)$, a system instance is defined as an asynchronous
    parallel composition of $\syssize(\param)$ process instances, indexed
    by $i \in \{1, \dots, \syssize(\param) \}$, with standard
    interleaving semantics\nop{ as a Kripke structure}. 
Let $\Prop$ be a set of atomic propositions. 
\nop{(The specific atomic propositions and labeling function that we will
    consider in this paper will be introduced in Section~\ref{sec:APs}.)}
    A \emph{system instance} $\sConcSys$ is a Kripke structure
    $(S_\LetterConcSys, S^0_\LetterConcSys, R_\LetterConcSys, \Prop,
    \labelfun_\LetterConcSys)$ where:     
\begin{itemize}
%\addtolength{\itemsep}{-0.45\baselineskip}
\item The set of \emph{(global) states} is $S_\LetterConcSys = \{ (\gst[1], \dots, \gst[\syssize(\param)]) \in (\restrict{S}{\param})^{\syssize(\param)}
\mid \forall i, j \in \{1,
\dots, \syssize(\param) \},  \gst[i] \gleich{\globset \cup  \paraset}
     \gst[j] \}$.
Informally, a global state $\gst$ is a Cartesian product of the
     state $\gst[i]$ of each process~$i$, with identical values of
     parameters and shared variables at each process.

\item $S^0_\LetterConcSys = (S^0)^{\syssize(\param)} \cap S_\LetterConcSys$ is the set of
     \emph{initial (global) states}, where $(S^0)^{\syssize(\param)}$
     is the Cartesian product of initial states of individual
     processes.

\item A transition $(\gst, \gst')$ from a global state $\gst \in S_\LetterConcSys$
     to  a global state $\gst' \in S_\LetterConcSys$ belongs to $R_\LetterConcSys$ iff there is
     an index $i$, $1\le i \le \syssize(\param)$, such that: 
    \begin{itemize}
      \item \trmove\ 
          The $i$-th process \emph{moves}: 
$(\gst[i], \gst'[i]) \in \restrict{R}{\param}$.
      \item \trmaintain\ 
    The values of the local variables of the other processes are preserved:
        for  every
        process index $j \ne i$, $1 \le j \le \syssize(\param)$, it
        holds that $\gst[j]
        \gleich{\{ \pc \} \cup \locset}  \gst'[j]$. 
    \end{itemize}

    \item $\labelfun_\LetterConcSys: S_\LetterConcSys \rightarrow 2^{\Prop}$ is a state labeling function.
\end{itemize}

%% Given a state $\gst$ of $\sConcSys$, we denote the state of process
%%      $i$ by $\sigma[i]$.

\begin{remark}\label{rem:symmetry}
The set of global states $S_\LetterConcSys$ and the transition relation $R_\LetterConcSys$ are
     preserved under every transposition $i \leftrightarrow j$ of
     process indices $i$ and $j$ in $\{1, \dots, \syssize(\param)\}$.
That is, every system $\sConcSys$ is \emph{fully symmetric} by
     construction.
\end{remark}

\begin{remark}
We call a pair of resilience condition and system size function
     $(\ResCond,\syssize)$ \emph{natural} if $\{\syssize(\param)\mid
     \ResCond(\param) \}$ is infinite.
From now on we consider only families of system instances with natural
     $(\ResCond,\syssize)$, as this implies that there is no bound on
     the number of processes.
\end{remark}

\mypara{Atomic Propositions}  
We  define the set of atomic proposition $\Prop$ to be the disjoint
     union of $\PropPC$ and $\PropVAR$.
The set $\PropPC$ contains propositions that capture comparison
     against a given status value $\pcval \in \PC$, i.e.,  $
     \left[\forall i.\; \pc_i = \pcval \right]$  and $\left[\exists
     i.\; \pc_i = \pcval \right]$.
\nop{ This allows us to express specifications of distributed
     algorithms.
To express the mentioned relay property, we may identify the status
     values where a process has accepted the message.
(We may quantify over all processes as we will only model correct
     process explicitly, while faults are modelled via
     non-determinism.
For a more detailed discussion see Section~\ref{sec:casestudy}) } %.
Further, the set of atomic propositions $\PropVAR$ captures
     comparison of variables  $x$, $y$, and a linear combination $c$
     of parameters from $\paraset$; $\PropVAR$ consists of
     propositions of the form $\left[\exists i.\; x_i + c < y_i
     \right]$  and $\left[\forall i.\; x_i + c \ge y_i \right]$.

The labeling function $\labelfun_{\LetterConcSys}$ of a  system
     instance $\sConcSys$ maps a state~$\gst$ to expressions $\aprop$
     from~$\Prop$ as follows (the existential case is defined
     accordingly using disjunctions):   

\begin{equation*}
\left[\forall i.\; \pc_i = \pcval \right] \in \labelfun_{\LetterConcSys}({\gst})
\text{ iff }
\Wedge{1 \le i \le \syssize(\param)}{}
\left(\gst[i].\pc = \pcval \right) 
\end{equation*}
\begin{multline*}
%% &\left[\exists i.\; \pc_i = \pcval \right] \in \labelfun_{\LetterConcSys}({\gst})
%%  \text{ iff }
%% \Vee{1 \le i \le \syssize(\param)}{}
%% \left(\gst[i].\pc = \pcval  \right)
%% \nonumber
%\end{align*}
%\begin{align*}
\left[\forall i.\; x_i + c \ge y_i \right] \in \labelfun_{\LetterConcSys}({\gst})
 \text{ iff } \\
\Wedge{1 \le i \le \syssize(\param)}{}
    \left(\gst[i].x + c(\param) \ge \gst[i].y \right)
%% &\left[\exists i.\; x_i + c < y_i \right] \in \labelfun_{\LetterConcSys}({\gst})
%%  \text{ iff }
%% \Vee{1 \le i \le \syssize(\param)}{}
%% \left(\gst[i].x + c < \gst[i].y \right)
\end{multline*}

\mypara{Temporal Logic}  
We specify properties using formulas of temporal logic \LTLX\ over
     $\PropPC$.
We use the standard definitions of paths and \LTLX\
     semantics~\cite{CGP1999}.
A formula of \LTLX{} is defined inductively as:     
(i) a literal $\aprop$ or $\neg\aprop$, where $\aprop \in \PropPC$, or
(ii) $\ltlF \varphi$, $\ltlG \varphi$, $\varphi
     \ltlU \psi$, $\varphi \vee \psi$, and $\varphi \wedge \psi$,
     where $\varphi$ and $\psi$ are \LTLX\ formulas.

\nop{An \LTLX\ formula is said to be negation-free, if it does not contain
literals of the form $\neg\aprop$.}

\mypara{Fairness} We are interested in verifying safety and liveness
     properties.
The latter can be usually proven only in the presence of fairness
     constraints.
As in~\cite{KP2000,PXZ02}, we consider verification of safety and
     liveness in systems with \emph{justice} fairness constraints.
We define fair paths of a system instance
     $\sConcSys$ using a set of justice constraints $J \subseteq
     \PropVAR$.
A path $\pi$ of a system $\sConcSys$ is $J$-\emph{fair} iff for every
     $\aprop \in J$ there are infinitely many states $\sigma$ in $\pi$
     with $\aprop \in \labelfun_I(\sigma)$.
By $\sConcSys \models_J \varphi$ we denote that the formula $\varphi$
     holds on all $J$-fair paths of $\sConcSys$.

\begin{definition}[PMCP]
\label{def:PMCP}
Given a parameterized system description containing
\begin{itemize}
%\addtolength{\itemsep}{-0.45\baselineskip}
\item a domain $D$,
\item a parameterized process skeleton $\Sk = (S, S_0, R)$,
\item a resilience condition $\ResCond$
  (generating a set of admissible parameters $\AdmP$),
\item a system size function $\syssize$,
\item justice requirements $J$, 
\end{itemize}
and an \LTLX\ formula $\varphi$,  the \emph{parameterized model
     checking problem} (PMCP) is to verify 
$$ \forall \param \in \AdmP.\ \sConcSys \models_J \varphi.$$
\end{definition}

\section{Threshold-guarded Fault-tolerant Distributed Algorithms}
\label{sec:TGFTDA}

\tikzstyle{node}=[circle,draw=black,thick,minimum size=4mm,font=\normalsize]
\tikzstyle{init}=[circle,draw=black!90,fill=white!20,thick,minimum size=4mm,font=\normalsize]
\tikzstyle{dest}=[draw=black!90,circle,double,fill=white!20,thick,minimum size=4mm,font=\normalsize]
\tikzstyle{post}=[->,thick]
\tikzstyle{pre}=[<-,thick]
\tikzstyle{cond}=[rounded
  corners,rectangle,minimum
  width=1cm,draw=black,fill=white,font=\normalsize]
\tikzstyle{asign}=[rectangle,minimum
  width=1cm,draw=black,fill=gray!5,font=\normalsize]

\begin{figure}[t]

\begin{minipage}{.45\columnwidth}
%    \centering
~
\hskip -.7cm
    \scalebox{0.6}{
      \tikzstyle{node}=[circle,draw=black,thick,minimum size=4mm,font=\normalsize]
\tikzstyle{init}=[circle,draw=black!90,fill=white!20,thick,minimum size=4mm,font=\normalsize]
\tikzstyle{dest}=[draw=black!90,circle,double,fill=white!20,thick,minimum size=4mm,font=\normalsize]
\tikzstyle{post}=[->,thick]
\tikzstyle{pre}=[<-,thick]
\tikzstyle{cond}=[rounded
  corners,rectangle,minimum
  width=1cm,draw=black,fill=white,font=\normalsize]
\tikzstyle{asign}=[rectangle,minimum
  width=1cm,draw=black,fill=gray!5,font=\normalsize]

\begin{tikzpicture}[>=latex,rotate=0]

  \node at ( 0,12) [init] (I) {$q_I$};
  \draw [thick,->] (-.75,12) --(I);

  \node at ( 0,10) [node] (D) {$q_1$};
  \node at ( 3.5,9.5) [node] (1) {$q_2$} ;
  \node at ( 3.5,7.5) [node] (2) {$q_3$};
  \node at ( 0,7) [node] (3) {$q_4$};  
\draw [post] (D) -- node[cond] {$\CFApc = \RI$} (1);
\draw [post] (D) -- node[cond,text width=2.2cm]
    {$\CFApc \ne \RI\, \wedge$\\
     $ \sent^0=\sent\, \wedge$\\
     $ \pc^0=\pc$} (3);
\draw [post] (1) -- node[asign] {$\sent^0 = \sent + 1$} (2);
\draw [post] (2) -- node[asign] {$\CFApc^0 = \SE$} (3);

  \node at ( 3.5,6) [node] (4)  {$q_{5}$};
  \node at ( 1,4.5) [node] (5) {$q_6$};
  \node at ( 3.5,3) [node] (6) {$q_7$} ;

  \node at ( 1,1.5) [node] (7) {$q_8$};

  \node at ( 3.5,0.25) [node] (8) {$q_9$};
  \node at ( 0,0) [dest] (9) {$q_F$};

\draw [post] (I) --  (D);
\node at ( 1.5,11) [asign] {$\rcvd \le \rcvd'  \; \wedge \;$ $\rcvd' \le\sent+f$};

\draw [post] (3) -- node[cond,xshift=-.6cm,yshift=-1cm,text width=2.5cm]
    {$(t+1 > \rcvd')\, \wedge$ \\
     $\CFApc' = \CFApc^0\, \wedge$ \\
     $ \sent' = \sent^0$}  (9);  
%%\node[below=.75 of 3,cond] {$\neg(t+1 \le \rcvd)$};
\draw [post] (3) -- node[cond]
    {$t+1 \le \rcvd'$} (4);

\draw [post] (4) -- node[cond,yshift=.2cm] {$\CFApc^0 = \IT$} (5);
\draw [post] (4) -- node[cond,yshift=.1cm,text width=2cm]
    {$\CFApc^0 \ne \IT\, \wedge$\\
     $ \sent' = \sent^0$} (6);
\draw [post] (5) -- node[asign,xshift=-.3cm] {$\sent' = \sent^0 + 1$} (6);

\draw [post] (6) -- node[cond] {$n-t > \rcvd'$} (8); 
\draw [post] (6) -- node[cond,xshift=-.25cm] {$n-t \le \rcvd'$} (7); 

\draw [post] (8) -- node[asign] {$\CFApc' = \SE$} (9);
\draw [post] (7) -- node[asign,right=-0.15cm] {$\CFApc' = \AC$} (9);

\end{tikzpicture}
    }
    \caption{CFA of our case study for Byzantine faults.}
    \label{Fig:CFA_SSA}
\end{minipage}
\begin{minipage}{.05\columnwidth}
\end{minipage}
\begin{minipage}{.45\columnwidth}
~
\hskip -.7cm
\scalebox{0.6}{
    \tikzstyle{node}=[circle,draw=black,thick,minimum size=4mm,font=\normalsize]
\tikzstyle{init}=[circle,draw=black!90,fill=white!20,thick,minimum size=4mm,font=\normalsize]
\tikzstyle{dest}=[draw=black!90,circle,double,fill=white!20,thick,minimum size=4mm,font=\normalsize]
\tikzstyle{post}=[->,thick]
\tikzstyle{pre}=[<-,thick]
\tikzstyle{cond}=[rounded
  corners,rectangle,minimum
  width=1cm,draw=black,fill=white,font=\normalsize]
\tikzstyle{asign}=[rectangle,minimum
  width=1cm,draw=black,fill=gray!5,font=\normalsize]

\begin{tikzpicture}[>=latex,rotate=0]

  \node at ( 1.5,12.5) [init] (I) {$q_I$};

  \node at ( 1.5,10) [node] (D) {$q_1$};
  \node at ( 3,7.5) [node] (1) {$q_2$} ;
 \node at ( 0,7.5) [node] (2) {$q_3$};

\node at ( 1.5,5.5) [node] (3) {$q_{4}$};

\node at ( 3,4) [node] (4) {$q_{5}$};

\node at ( 1.5,0.5) [dest] (9) {$q_F$};

\draw [post] (I) --  (D);
\node at ( 1.5,11.3) [asign,text width=3cm]
    {$\rcvd \le \rcvd'\; \wedge$\\ $\rcvd' \le \sent + \sentf$ };

\draw [post] (D) -- node[cond,xshift=.1cm]
    {$\CFApc = \RI$} (1);
\draw [post] (D) -- node[cond,xshift=-.1cm]
    {$\CFApc = \IT$} (2);

\draw [post] (D) .. controls (-2.5,9) and (-2.5,2) ..
        node[cond,xshift=-.2cm,yshift=1cm]
    {$\CFApc = \AC$} (9);

\draw [post] (D) .. controls (-2,9) and (-1.2,2) ..
        node[cond,xshift=.1cm,yshift=0cm]
    {$\CFApc = \DD$} (9);

\draw [post] (2) .. controls (-0,6) and (0,2) .. node[cond,anchor=south,yshift=.5cm]
    {$1 > \rcvd'$} (9);

\draw [post] (2) -- node[cond] {$1 \le \rcvd'$} (1);

\draw [post] (1) -- node[asign] {$\CFApc' = \DD$} (4);
\draw [post] (4) .. controls (3,3) and (3,1.5) .. node[asign,text width=1.5cm]
    {$\sentf'=$\\$\sentf+1$} (9);

\draw [post] (1) -- node[asign,xshift=-.2cm] {$\CFApc' = \AC$} (3);
\draw [post] (3) -- node[asign,yshift=.25cm,text width=1.5cm]
    {$\sent' =$\\$ \sent+1$} (9);

%%   \node at ( 3,7.5) [node] (2) {$q_3$};
%%   \node at ( 0,7) [node] (3) {$q_4$};  

%% \draw [post] (D) -- node[cond] {$\neg(\CFApc = \RI$)} (3);
%% \draw [post] (1) -- node[asign] {$\CFAinc \; \sent$} (2);
%% \draw [post] (2) -- node[asign] {$\CFApc := \SE$} (3);

%%   \node at ( 3,3.5) [node] (4) {$q_{5}$};

%%   \node at ( 0,0) [dest] (9) {$q_F$};

%% \draw [post] (I) --  (D);
%% \node at ( 1.5,10.9) [asign] {$\rcvd :=
%%   \CFApickOp{\CFAdummy}{\CFApick}{\rcvd\le\CFAdummy  \; \wedge \;$
%%     $\CFAdummy \le\sent + \sent{f}}$};

%% \draw [post] (3) -- node[cond] {$\neg(n-t \le \rcvd)$}  (9);  
%% %%\node[below=.75 of 3,cond] {$\neg(t+1 \le \rcvd)$};
%% \draw [post] (3) -- node[cond] {$n-t\le\rcvd$} (4);

%% \draw [post] (4) -- node[cond] {$\CFApc := \AC$} (9);

\end{tikzpicture}
}
\caption{CFA of FTDA from~\cite{FismanKL08}
    (if $x'$ is not assigned, then $x'=x$).}
    \label{Fig:CFA_RBC}
\end{minipage}
\end{figure}

\subsection{Framework for FTDAs} %.
\label{sec:CFA}

We give a formalization that is adequate for threshold-guarded FTDAs, and
     suitable for verification.
It captures threshold guards as discussed in Section~\ref{sec:Intro} as
     a core primitive.
FTDAs are usually described in steps that consist of a
     loop-free sequence of small steps.

Further, we model faults (e.g., Byzantine) that have the effect that
     correct processes receive more or less messages than actually
     should have been sent.
We model the send operation by an increase of a global
     variable, and the receive by a non-deterministic choice  
%% between
%%      the previous value of a local variable, and a value depending on
%%      a global variable and the number of faulty processes.
that captures faults and asynchrony in communication.
By this, we model the effect of faulty processes rather than modeling
     them explicitly.
The soundness of the modeling approach requires involved arguments
     discussed in \cite{JKSVW12b}.
These arguments are in the area of distributed computing theory and
     out of scope of the current~paper.

To address all these issues,  we propose a variant of control flow
     automata.
Henzinger et al.~\cite{HenzingerJMS02} introduced CFA as a framework
     to describe the control flow of a program, using a graph where
     the edges are labeled with instructions.

%% \mypara{Control flow automata}   In analogy to software model checking
%%      where programs are translated into Kripke structures, below we
%%      define control flow automata (CFA) to represent distributed
%%      algorithms that contain threshold guards, and then show how they
%%      induce process skeletons.
%% Different from recognizing finite automata, Henzinger et
%%      al.~\cite{HenzingerJMS02} introduced CFA as a framework to
%%      describe the control flow of a program, that intuitively have
%%      the edges labeled with the instructions.
%% To capture the step semantics of FTDA, a step of the distributed
%%      algorithm is defined via a path from the initial location of the
%%      CFA to the final location.
%% Moreover, we are required to capture these steps as a transition
%%      relation.
%% In particular, we need to compute a formula that represents this
%%      relation (in some sense, a transducer for an individual process).
%% To this end, we use the standard  technique introduced by Cytron et
%%      al.~\cite{Cytron1991} to construct CFAs that are in the form of
%%      single static assignment (SSA).
%% Figure~\ref{Fig:CFA_SSA} is an example CFA in SSA  of the Byzantine
%%      fault-tolerant broadcasting algorithm by Srikanth and
%%      Toueg~\cite{ST87:abc} that is obtained from the the formalization
%%      in~\cite{JKSVW12b} (which is provided in Figure~\ref{Fig:STCFA}),
%%      using the SSA transformation algorithm
%%      from~\cite{Cytron1991}.

Formally, a \emph{guarded control flow automaton}  (CFA) is an
     edge-labeled directed acyclic graph $\CFA = (Q, q_I, q_F, E)$
     with a finite set $Q$ of nodes called locations, an initial
     location $q_I\in Q$, and a final location $q_F\in Q$.
A path from $q_I$ to $q_F$ is used to describe one step of a
     distributed algorithm.
The edges have the form $E \subseteq Q \times \CFAguard \times Q$,
     where $\CFAguard$ is defined as an expression of one of the
     following~forms where $a_0, \dots, a_\numparam \in \Int$, and
     $\paraset = \{p_1, \dots, p_\numparam \}$:       

\begin{itemize}
  \item if $Z \in \PC$, then $\pc = Z$ and $\pc \neq Z$ are
    \emph{status guards};

  \item if $x$ is a variable in $D$ and $\lhd \in \{\le, > \}$, then
$$a_0 +
        \sum_{1 \le i \le \numparam} a_i \cdot p_i\ \lhd\ x $$ 
    is a \emph{threshold guard};

  \item if $y, z_1, \dots, z_k$ are variables in $D$ for $k \ge 1$,
    and $\lhd \in \{ =, \ne, <,$ $\le, >, \ge\}$,
    and $a_0, \dots, a_\numparam \in \Int$, then
    $$y\ \lhd\ z_1 + \dots + z_k + \big( a_0 +
        \sum_{1 \le i \le \numparam} a_i \cdot p_i \big)$$
    is a \emph{comparison guard};

  \item a conjunction $g_1 \wedge g_2$ of guards $g_1$ and $g_2$ is a
      guard.
\end{itemize}

Status guards are used to capture the basic control flow.
Threshold guards capture the core primitive of the FTDAs we consider.
Finally, comparison guards are used to model send and receive
     operations.

\mypara{Obtaining a Skeleton from a CFA} %.
One step of a process skeleton is defined by a path from  $q_I$ to
     $q_F$ in a CFA.
Given $\PC$, $\locset$, $\globset$, $\paraset$, $\ResCond$, and a CFA
     $\CFA$, we define the process skeleton $\SkAut = (S, S^0, R)$
     induced by~$\CFA$ as follows: %.
The set of variables used by the CFA is $W\supseteq
     \paraset \cup \locset \cup \globset \cup\{\pc\} \cup \{ x' \mid x \in \locset
     \cup \globset \cup \{sv\} \}$.

Informally, a variable $x$ corresponds to the value before a step and
     the variable $x'$ to the value after the step.
A path $p$ from $q_I$ to $q_F$ of CFA induces a conjuction of all the
     guards along it.
We call a mapping~$v$ from $W$ to the values from the respective
     domains a \emph{valuation}.
We may write $v \models p$ to denote that the valuation $v$ satisfies
     the guards of the path~$p$.
We are now in the position to define the mapping between a CFA~$\CFA$
     and the transition relation of a process skeleton $\SkAut$: If
     there is a path $p$ and a valuation $v$ with $v \models p$, then
     $v$ defines a single transition $(s, t)$ of a process skeleton
     $\SkAut$, if for each variable $x \in \locset \cup \globset \cup
     \{\pc\}$ it holds that $s.x = v(x)$ and $t.x = v(x')$ and for
     each parameter variable $z \in \paraset$, $s.z=t.z=v(z)$.

Finally, to specify $S^0$, all variables of the skeleton that range
     over~$D$ are initialized to~$0$, and $\pc$ ranging over $\PC$
     takes an initial value from a fixed subset of~$\PC$.    

\begin{remark} 
It might seem restrictive that our guards do not contain, e.g.,
     increment, assignments, non-deterministic choice from a range of
     values.
However, all these statements can be translated in our form using the
     SSA transformation algorithm from~\cite{Cytron1991}.
For instance, Figure~\ref{Fig:CFA_SSA} has been obtained from a CFA
     (given in the appendix in Figure~\ref{Fig:STCFA}) that contains
     the mentioned statements.
\end{remark}

\begin{definition}[PMCP for CFA]
\label{def:PMCP_CFA}
We define the Parameterized Model Checking Problem for CFA $\CFA$ by
     specializing Definition~\ref{def:PMCP} to the parameterized
     process skeleton $\SkAut$.
\end{definition}

\subsection{Undecidability of PMCP for CFA}

We call a CFA where all guards are status guards a
     \emph{non-communicating} CFA.
In this section we argue that the problem from
     Definition~\ref{def:PMCP_CFA} is undecidable even if the CFA is
     non-communicating.\footnote{The detailed construction and proof
     are given in Appendix~\ref{sec:undec}.}    

The outline of the proof is similar to the undecidability proofs
     in~\cite{GS1992,EN95}: one of the processes plays the role of a
     \emph{control process} that simulates the program of a two
     counter machine (2CM), and the other processes are \emph{data
     processes} that each store at most one digit of one of the two
     counters encoded in unary representation.
The control processes increments or decrements a counter by a
     handshake with a data process.
In system instances that contain $n$ data processes, this is
     sufficient to simulate $n$ steps of a 2CM.
If the parameterized model checking problem under consideration is
     defined for a natural $(\ResCond,\syssize)$, then for arbitrarily
     many steps there is some system instance that simulates at least
     that many steps of the 2CM.
Undecidability of the parameterized model checking problem then
     follows from the undecidability of the non-halting problem for
     2CMs.

Esparza \cite{E97} has shown that test for zero statements of a 2CM
     can be simulated with a temporal logic specification using an
     atomic proposition ``test for zero''.
Given the proof strategies from~\cite{GS1992,EN95}, the only technical
     difficulty that remains is to ensure a handshake between
     non-communicating CFAs.
We do so by enforcing a handshake using sequences of status values the
     CFAs go through, and an \LTLX{} specification which acts as a
     scheduler that ensures a specific interleaving between the
     updates of the status variables of the two~CFAs.

\newcommand\thmUndecBLA{
Let ${\cal M}$ be a two counter machine, and $(\ResCond,\syssize)$ be
     a natural pair of resilience condition and system size function.
One can efficiently construct a non-communicating CFA $\CFA({\cal M})$
     and an \LTLX{} property $\varphi_{\mathrm{nonhalt}}({\cal M})$
     such that the following two statements are equivalent:    
\begin{itemize}
\item ${\cal M}$ does not halt. 
%when the two counters are initially zero
\item $\forall \param \in \AdmP$, $\ConcSys \models_\emptyset
  \varphi_{\mathrm{nonhalt}}({\cal M})$.
\end{itemize}
}

\begin{theorem} \label{thm:thmUndecBLA}
\thmUndecBLA
\end{theorem}

\begin{corollary}\label{cor:Undec}
PMCP for CFAs is undecidable even if CFAs contain status guards only.
\end{corollary}

As discussed in Section~\ref{sec:CFA}, we model faults by the
     influence they have on values of variables in the domain $D$.
As we do not restrict the set of local and global variables, the
     result also applies if these sets are non-empty.
Moreover, in this kind of modeling, the atomic propositions
     $\left[\exists i.\; \pc_i = \pcval \right]$ range over correct
     processes only.
Hence, the undecidability result also holds for FTDAs.
If one chooses to model faults differently, i.e., by changing the
     transition relation of a process, then the decidability depends
     on the way the transition relation is modified.
For instance, certain problems are decidable in lossy
     systems~\cite{AJ93,Mayr03}.
However, lossy systems are not suitable for modeling the FTDAs we
     consider.

Notwithstanding this undecidability results, the rest of the paper is
     concerned with abstraction techniques for threshold-based FTDA.
In this context, Corollary~\ref{cor:Undec} shows that the model
     checking problem we obtain after the first abstraction step
     (mentioned in the introduction) is still undecidable.
As discussed in the introduction, abstraction always has to be
     accompanied by a case study along with practical experiments.

\subsection{Case Study} 
\label{sec:CaseStudy}

The distributed algorithm by Srikanth and Toueg~\cite{ST87,ST87:abc}
     is one of the most basic distributed algorithms that has
     applications in a wide area of distributed computing.
It is a basic building block that has been used in various
     environments (degrees of synchrony, fault assumptions, etc.) and
     many other more complicated distributed algorithms, such as
     consensus \cite{DLS88,ADFT06:DSN}, software and hardware clock
     synchronization \cite{ST87,FSFK06:EDCC}, approximate agreement
     \cite{DolevLPSW86}, and $k$-set agreement \cite{PriscoMR01}.
Figure~\ref{Fig:CFA_SSA} shows the guarded control flow automaton of
     the core of  Srikanth and Toueg's algorithm.

In our experiments we consider additional three algorithms that are
     based on similar algorithmic ideas.\footnote{Their CFA is given
     in Figure~\ref{Fig:manyCFA} in the appendix.} %.
They deal with different fault models and resilience conditions; the
     algorithms are: \textsc{\sc (Byz)}, which is the algorithm from
     Figure~\ref{Fig:CFA_SSA}, for $t$ Byzantine faults if $n>3t$,
     {\sc (symm)} for $t$ symmetric (identical Byzantine~\cite{AW04})
     faults if $n>2t$, \textsc{\sc (omit)} for $t$ send omission
     faults if $n>2t$, and \textsc{\sc (clean)} for $t$ clean crash
     faults if $n>t$.
For comparison with the results by Fisman et al.~\cite{FismanKL08}, we
     also verified the RBC algorithm whose CFA is given in
     Figure~\ref{Fig:CFA_RBC}.
%.
In this paper we verify the following safety and liveness
     specifications for the algorithms: \footnote{In addition, for the
     RBC algorithm, we also verified the specification, which was
     verified in~\cite{FismanKL08} (called A in Table~\ref{tab:exp})}
\begin{align}
  \tag{U}
 \left[ \forall i.\; \pc_i \ne \RI \right]
     \rightarrow & \ltlG
   \left[\forall j.\;  \pc_j \ne \AC
     \right] \label{ST:u}
\\
  \tag{C}
\left[\forall i.\; \pc_i = \RI \right]
      \rightarrow &
   \ltlF  \left[ \exists j.\; \pc_j = \AC\right]
   \label{ST:corr}
 \\
\tag{R}
  \ltlG (\neg \left[\exists i.\; \pc_i = \AC \right])
      \vee\ &\ltlF \left[
   \forall j.\; \pc_j = \AC \right]
\label{ST:rel}
\end{align}

However, from the literature we know that we cannot expect to verify
     these FTDAs without putting additional constraints on the
     environment, e.g., communication fairness, i.e., every message
     sent is eventually received.
To capture this, we use justice requirements, e.g., $J = \{
     \quotes{\forall i.\; \rcvd_i \ge \sent } \}$ in the Byzantine
     case.

%% After presenting our abstraction techniques in the following section,
%%      Section~\ref{sec:exp} discusses the experimental evaluation.

%% \input{undecidability}

%%%%%%%%%\input{casestudy.tex}

\section{Abstraction Scheme}\label{sec:Scheme}

%%%%% GRAPH %%%%%
% Define block styles

The input to our abstraction method is the infinite  parameterized
     family ${\cal F} = \{\ConcSys \mid \param\in\AdmP \}$ of Kripke
     structures specified via a CFA $\CFA$.
The  family ${\cal F}$ has two principal sources of
     unboundedness: unbounded variables in the process skeleton
     $\SkAut$, and the unbounded number of processes
     $\syssize(\param)$.
We deal with these two aspects separately, using two abstraction
     steps, namely the \emph{PIA data abstraction} and the \emph{PIA
     counter abstraction}.
In both abstraction steps we use the parametric interval abstraction
     PIA that we introduce in Section~\ref{sec:AbsDomain}.

\subsection{Abstract Domain of Parametric Intervals (PIA)}
\label{sec:AbsDomain}

\newcommand\gset{{\cal G}_\CFA} 

Given a CFA $\CFA$, let $\gset$ be the set of all linear combinations
     $a_0 + \sum_{1 \le i \le \numparam} a_i \cdot p_i$ that are met
     in the left-hand sides of $\CFA$'s threshold guards.
Every expression $\varepsilon$ of $\gset$ defines a function
     $f_\varepsilon\colon \AdmP \rightarrow D$.
Let $\threshSet = \{0, 1\} \cup \{ f_\varepsilon \mid \varepsilon \in
     \gset \}$ be a finite \emph{threshold set} of cardinality
     $\absMax + 1$.
For convenience, we name elements of $\threshSet$ as $\thresh_0,
     \thresh_1, \dots, \thresh_\absMax$ with $\thresh_0$ corresponding
     to the constant function~$0$, and~$\thresh_1$ corresponding to
     the constant function~$1$.
For instance, the CFA in Figure~\ref{Fig:CFA_SSA} has the threshold
     set $\{\thresh_0, \thresh_1, \thresh_2, \thresh_3\}$, where
     $\thresh_2(n,t,f) = t+1$ and $\thresh_3(n,t,f) = n-t$.

Then, we define the domain of parametric intervals: $$ \absdomain =
    \{\absSym_j \mid 0\le j \le \absMax \} $$

Our abstraction rests on an implicit property of many fault-tolerant
     distributed algorithms, namely, that the resilience condition
     $\ResCond$ induces an order on the thresholds used in the
     algorithm (e.g., $t+1<n-t$).
Assuming such an order does not limit the application of our approach:
     In cases where only a partial order is induced by $\ResCond$, one
     can simply enumerate all finitely many total orders.
As parameters, and thus thresholds, are kept unchanged in a run, one
     can verify an algorithm for each threshold order separately, and
     then combine the results.
We may thus restrict the threshold sets we consider by:

\begin{definition}\label{def:totalOrder}
The finite set $\threshSet$ is uniformly ordered if for all $\param
     \in \AdmP$, and all $\thresh_j(\param)$ and $\thresh_k(\param)$
     in~$\threshSet$ with $0 \le j<k \le \absMax$,  it holds that
     $\thresh_j(\param) < \thresh_k(\param)$.
\end{definition}

Definition~\ref{def:totalOrder} allows us to properly define the
     \emph{parameterized abstraction function} $\abst_{\param} \colon
     D \rightarrow \absdomain$ and the \emph{parameterized
     concretization function}  $\conc_{\param} \colon \absdomain
     \rightarrow 2^{D}$.
\begin{align}
\afunc{\param}{x} &=
\begin{cases}
\absSym_j  &\text{if }
 x \in {[ \thresh_j(\param) , \thresh_{j+1}(\param) [} \text{ for some }  0\le j < \absMax\\
\absSym_{\absMax} & \text{otherwise}.
\end{cases}\nonumber\\
\cfunc{\param}{\absSym_j} &=
\begin{cases}
{[\thresh_j(\param),  \thresh_{j+1}(\param) [}
  & \text{if } j < \absMax\\
{[ \thresh_{\absMax}(\param), \infty [} & \text{otherwise}.
\end{cases}\nonumber
\end{align}

From $\thresh_0(\param) = 0$  and~$\thresh_1(\param) = 1$, it
     immediately follows that for all $\param \in \AdmP$, we have
     $\afunc{\param}{0} = \absSym_0$, $\afunc{\param}{1} = \absSym_1$,
     and $\cfunc{\param}{\absSym_0} = \{ 0 \}$.
Moreover, from the definitions of $\alpha$, $\gamma$, and
     Definition~\ref{def:totalOrder} one immediately obtains:

\begin{proposition}\label{prop:RussianDolls}
For all $\param$ in $\AdmP$, and for all $a$ in $D$, it holds
that $a \in \cfunc{\param}{\afunc{\param}{a}}$.
\end{proposition}
\begin{definition} \label{def:torder}
$\absSym_k \le \absSym_\ell \mbox{ iff } k \le \ell$.
\end{definition}

The PIA domain has similarities to predicate abstraction since the
     interval borders are naturally expressed as predicates, and
     computations over PIA are directly reduced to SMT solvers.
However, notions such as the order of
     Definition~\ref{def:torder} are not naturally expressed in terms
     of predicate abstraction.

\subsection{PIA data abstraction}\label{sec:FirstAbs}

% *******************************************************************

Our parameterized data abstraction is based on two abstraction ideas.
First, the variables used in a process skeleton are unbounded and we
     have to map those unbounded variables to a fixed-size domain.
If we fix parameters $\param \in \AdmP$, then an interval
     abstraction~\cite{CC1977} is a natural solution to the problem of
     unboundedness.
Second, we want to produce a single process skeleton that does not
     depend on parameters $\param \in \AdmP$ and captures the behavior
     of \emph{all} process instances.
This can be done by using ideas from existential
     abstraction~\cite{Clarke1994,DGG1997,KP2000} and sound
     abstraction of fairness constraints~\cite{KP2000}.
Our contribution consists of combining these two ideas to arrive at
     parametric interval data abstraction. 

Our abstraction maps values of unbounded variables to parametric
     intervals~$\absSym_j$, whose boundaries are symbolic expressions
     over parameters.
This abstraction differs from interval abstraction \cite{CC1977} in that
     the interval bounds are not numeric.
However, for every instance, the boundaries are \emph{constant}
     because the parameters are fixed.
We hence do not have to deal with symbolic ranges over
     \emph{variables} in the sense of~\cite{SIG07}.

We now discuss an existential abstraction of a formula~$\Phi$
    that is either a threshold or a comparison guard (we consider
    other guards later).
To this end we introduce notation for sets of vectors
     satisfying~$\Phi$.
According to Section~\ref{sec:CFA}, formula $\Phi$ has two kinds
     of free variables: parameter variables from $\paraset$ and data
     variables from $\locset \cup \globset$.
Let $\vec{x}^p$ be a vector of parameter variables $(x^p_1, \dots,
     x^p_{|\paraset|})$ and $\vec{x}^v$ be a vector of variables
     $(x^v_1, \dots, x^v_k)$ over $D^k$.
Given a $k$-dimensional vector~$\vec{d}$ of values from $D$, by
     $$\vec{x}^p=\param,\vec{x}^v = \vec{d} \models \Phi$$ 
we denote
     that $\Phi$ is satisfied on concrete values $x^v_1=d_1, \dots,
     x^v_k=d_k$ and parameter values~$\param$.
%Let $||\Phi|| \subseteq \AdmP \times D^k$ be the maximum set of vectors
%satisfying $\Phi$. Further, using $||\Phi||$
We define:  % $||\Phi||_E \subseteq \absdomain^k$:
\begin{multline*}
||\Phi||_E =
    \{  \IntAbs{\vec{d}} \in \absdomain^k
        \mid 
        \exists \param \in \AdmP \exists \vec{d} = (d_1, \dots, d_k) \in D^k .\ \\
        \IntAbs{\vec{d}} = (\afunc{\param}{d_1}, \dots, \afunc{\param}{d_k})
        \wedge \vec{x}^p=\param, \vec{x}^v=\vec{d} \models \Phi
    \} %\\
%||\Phi||_A =
%    \{ \IntAbs{\vec{d}} \in \absdomain^k
%        \mid &
%        \forall \param \in \AdmP \forall \vec{d} = (d_1, \dots, d_k) \in D^k .\ \\
%        &\IntAbs{\vec{d}} = (\afunc{\param}{d_1}, \dots, \afunc{\param}{d_k})
%        \wedge \paraset=\param, \vec{x}=\vec{d} \models f
%    \}
\end{multline*}

Hence, $||\Phi||_E$ contains all vectors of abstract values that correspond to
some concrete values satisfying $\Phi$. Note carefully, that parameters do not
appear anymore due to existential quantification.
A PIA \emph{existential abstraction} of $\Phi$ is defined to be a formula
     $\IntAbs{\Phi}$  over a vector of variables $\IntAbs{\vec{x}} =
     \IntAbs{x}_1, \dots, \IntAbs{x}_k$ over $\absdomain^k$ such that
     $\{ \IntAbs{\vec{d}} \in \absdomain^k \mid
     \IntAbs{\vec{x}}=\IntAbs{\vec{d}} \models \IntAbs{\Phi} \}
     \supseteq ||\Phi||_E$.

\mypara{Computing PIA abstractions} The central property of our
     abstract domain is that it allows to abstract comparisons against
     thresholds (i.e., threshold guards) in a precise way.
That is, we can abstract formulas of the form $\thresh_j(\param) \le
     x_1$ by $\absSym_j \le \IntAbs{x}_1$ and $\thresh_j(\param) >
     x_1$ by $\absSym_j > \IntAbs{x}_1$.
In fact, this abstraction is precise in the following~sense.

\newcommand\propPreciseAbs{
For all $\param \in \AdmP$ and all $a \in D$:\\
$    \thresh_j(\param) \le a  \mbox{ iff } \absSym_j \le \afunc{\param}{a}, \mbox{ and }
    \thresh_j(\param) > a \mbox{ iff } \absSym_j > \afunc{\param}{a}$.
}

\begin{proposition}\label{prop:precise-abs}
    \propPreciseAbs
\end{proposition}

\newcommand\propPreciseAbsProof{
    \begin{proof}
        Fix an arbitrary $\param \in \AdmP$.

        \mypara{Case $a \ge \thresh_j(\param)$}
                $(\Rightarrow)$ Fix an arbitrary $a \in D$ satisfying $a \ge
        \thresh_j(\param)$. Let $k$ be a maximum number such that $a \ge
        \thresh_k(\param)$. Then $\afunc{\param}{a} = \absSym_k$.  By
        Definition of~$\abst_\param$ we have $k \ge j$ and thus, by
        Definition~\ref{def:torder}, $\absSym_k \ge \absSym_j$. It immediately
        gives $\afunc{\param}{a} \ge \absSym_j$.

        $(\Leftarrow)$ Let $a \in D$ be a value satisfying $\afunc{\param}{a}
        \ge \absSym_j$. There is $k$ such that $\afunc{\param}{a} = \absSym_k$
        and $a \ge \thresh_k(\param)$.  From $\afunc{\param}{a} \ge \absSym_j$
        it follows that $\absSym_k \ge \absSym_j$ and, by
        Definition~\ref{def:torder}, $k \ge j$. Then by
        Definition~\ref{def:totalOrder} we have $\thresh_k(\param) \ge
        \thresh_j(\param)$ and by transivity $a \ge \thresh_j(\param)$.

        \mypara{Case $a < \thresh_j(\param)$}
        $(\Rightarrow)$ Fix an arbitrary $a \in D$ satisfying $a <
        \thresh_j(\param)$. Let $k$ be a maximum number such that $a \ge
        \thresh_k(\param)$. Then $\afunc{\param}{a} = \absSym_k$.

        Consider the case when $k \ge j$.
        By Definition~\ref{def:torder} it implies $\absSym_k \ge \absSym_j$. It
        immediately gives $\afunc{\param}{a} \ge \absSym_j$, which contradicts
        the assumption $a < \thresh_j(\param)$. Thus, the only case is
        $k < j$.
        
        By Definition~\ref{def:torder}, $k < j$ implies
        $\absSym_k \le \absSym_j$. As we excluded the case $k = j$ we have
        $\absSym_k \le \absSym_j$, $\absSym_k \ne \absSym_j$ or, equivalently,
        $\afunc{\param}{a} = \absSym_k < \absSym_j$.

        $(\Leftarrow)$ Let $a \in D$ be a value satisfying $\afunc{\param}{a} <
        \absSym_j$ or, equivalently, $\afunc{\param}{a} \le \absSym_j$ and
        $\afunc{\param}{a} \ne \absSym_j$.  There exists $k$ such that
        $\afunc{\param}{a} = \absSym_k$ and either \emph{(a)} $a <
        \thresh_{k+1}(\param)$ or \emph{(b)} $k=\absMax$. From the assumption
        we have $\absSym_k \le \absSym_j$ and $\absSym_k \ne \absSym_j$. From
        this we conclude: \emph{(c)} $k \ne \absMax$ excluding \emph{(b)};
        \emph{(d)} $\absSym_{k+1} \le \absSym_j$. From \emph{(d)} by
        Definition~\ref{def:torder}, $k+1 \le j$. This implies by
        Definition~\ref{def:totalOrder}, $\thresh_{k+1}(\param) \le
        \thresh_j(\param)$. From this and \emph{(a)} we conclude that $a <
        \thresh_j(\param)$.
        \qed
    \end{proof}
}

For all
comparison guards we are going to use a general form
(well-known from the literature), namely:
%Similarly, a formula $\IntAbs{\Phi}$ over $\IntAbs{\vec{x}}$ is a universal
%abstraction of $\Phi$ if and only if $\{ \IntAbs{\vec{x}} \in \absdomain^k
%\mid \IntAbs{\vec{x}} \models \IntAbs{\Phi} \} \subseteq ||\Phi||_A$.
\begin{align*}
    \IntAbs{\Phi}_E = \Vee{(\IntAbs{d}_1, \dots, \IntAbs{d}_k)
\in ||\Phi||_E}{} & \IntAbs{x}_1 = \IntAbs{d}_1 \wedge
    \dots \wedge \IntAbs{x}_k = \IntAbs{d}_k %\\
%    \IntAbs{f}_A = \Vee{(\IntAbs{d}_1, \dots, \IntAbs{d}_k)
%\in ||f||_A}{} & \IntAbs{x}_1 = \IntAbs{d}_1 \wedge
%    \dots \wedge \IntAbs{x}_k = \IntAbs{d}_k \mbox{ is a universal abstraction.}
\end{align*}

\newcommand\propGenAbs{
    If $\Phi$ is a formula over variables $x_1, \dots, x_k$ over $D$, then
    $\IntAbs{\Phi}_E$ is a PIA existential abstraction.
}

\begin{proposition}\label{prop:genabs}
    \propGenAbs
\end{proposition}

\newcommand\propGenAbsProof{
    \begin{proof}
        Consider an arbitrary $\vec{d} \in
        ||\Phi||_E$. As $\vec{d} \in ||\Phi||_E$, it satisfies the conjunct
        $\IntAbs{x}_1 = \IntAbs{d}_1 \wedge \dots \wedge \IntAbs{x}_k = \IntAbs{d}_k$
        and thus satisfies the disjunction $\IntAbs{\Phi}$,
        i.e.
        $\vec{x} = \vec{d} \models \IntAbs{\Phi}_E$. As $\vec{d}$ is chosen
        arbitrarily, we conclude that $||\Phi||_E \subseteq \{ \IntAbs{\vec{x}}
        \in \absdomain^k \mid \IntAbs{\vec{x}} \models \IntAbs{\Phi}_E \}$.
%
%        \emph{Universal abstraction}. Consider an arbitrary
%        $\vec{d}$ such that $\vec{x}=\vec{d} \models \IntAbs{\Phi}_A$.  As a
%        disjunction $\IntAbs{\Phi}_A$ evaluates to true only on $\vec{d} \in
%        ||f||_A$, we immediately arrive at $\vec{d} \in ||f||_A$.  Hence, $\{
%            \IntAbs{\vec{x}} \in \absdomain^k \mid \IntAbs{\vec{x}} \models
%        \IntAbs{\Phi}_A \} \subseteq ||f||_A$.
        \qed
    \end{proof}
}

If the domain $\absdomain$ is small (as it is in our case), then one
     can enumerate all vectors of abstract values in $\absdomain^k$
     and check which belong to our abstraction $||\Phi||_E$,
     using an SMT solver.

%\begin{example}\label{ex:plus}
%Consider
%     the threshold set $\threshSet = \{0,1,n-t \}$ which leads to
%     $\absdomain=\{\absSym_0, \absSym_1, \absSym_2 \}$, and
% consider $RC$ to be $n>2t \wedge t>0$.
%Consider the example for addition of ``$1$'' to $\absSym_1$
%     in Table~\ref{tab:plusone}.
%As $\absPlus{1}$ is defined via existential quantification over
%     parameters~$\param$, $\absPlus{1}(\absSym_1)$ contains
%     $\absSym_1$ (due to the bottom line of the table).
%This result is spurious for the specific abstraction of the system
%     with $n=3$ and $t=1$ (cf.~the corresponding
%     line in the table).
%However, this allows us to have one abstract system that describes all
%     possible behaviors of all concrete ones satisfying $RC$.
%\end{example}

% *******************************************************************

% *******************************************************************

\mypara{Transforming CFA}
We now describe a general method to abstract $\CFAguard$ formulas,
     and thus construct an abstract process skeleton.
To this end, we denote by $\absEx$ a mapping from a concrete formula
     $\Phi$ to some existential abstraction of $\Phi$ (not necessarily
     constructed as above).
By fixing $\absEx$, we can define an abstraction of a $\CFAguard$ of a
     CFA:   
$$
abst(g) =
\begin{cases}
    {\absEx(g)} & \text{if } g \mbox{ is a threshold guard} \\
    {\absEx(g)} & \text{if } g \mbox{ is a comparison guard} \\
    g & \text{if } g \mbox{ is a status guard}\\
    {abst(g_1) \wedge abst(g_2)} & \text{otherwise, i.e., }
            g \mbox{ is } g_1 \wedge g_2 \\
\end{cases}
$$

By slightly abusing the notation, for a CFA $\CFA$ by $abst(\CFA)$ we
     denote the CFA that is obtained from $\CFA$ by replacing every
     guard $g$ with $abst(g)$.
Note that $abst(\CFA)$ contains only guards over $sv$ and over
     abstract variables over $\absdomain$.

\begin{definition}
We define a mapping $\absMap$ from valuations~$v$ of a CFA $\CFA$ 
    to
    valuations~$\IntAbs{v}$ of CFA
    $abst(\CFA)$ as follows: for each
    variable $x$ over $D$, $\IntAbs{v}.x = \afunc{\param}{v.x}$, and for each
    variable~$y$ over $\PC$, $\IntAbs{v}.y = v.y$.
\end{definition}

The following theorem follows immediately from the definition of existential
abstraction and $abst(\CFA)$:

\begin{theorem}\label{thm:sem-abs-new}
    For all guards $g$, all $\param$ in $\AdmP$, and for all valuations $v$ with
    $v \gleich{\paraset} \param$: if $v \models g$,
    then $\absMap(v) \models abst(g)$.
\end{theorem}

% *******************************************************************

For model checking purposes we have to reason about the Kripke
     structures that are built using the skeletons obtained from CFAs.
We denote by $\SkAAut$, the process skeleton that is induced by CFA
     $abst(\CFA)$.
Analogously to $\absMap$, we define the parameterized abstraction
     mapping $\absMapSys$ that maps global states from $\ConcSys$ to
     global states from $\AbstSys$.
After that, we obtain Theorem~\ref{thm:comm} from
     Theorem~\ref{thm:sem-abs-new} and the construction of system
     instances.

\begin{definition}
Let $\gst$ be a state of $\ConcSys$, and $\IntAbs{\gst}$ be a state
     of the abstract instance $\AbstSys$.
Then, $\IntAbs{\gst} = \absMapSys(\gst)$ if for each variable $y\in
     \Lambda \cup \Gamma \cup \paraset$, $\IntAbs{\gst}.y =
     \afunc{\param}{\gst.y}$, and $\IntAbs{\gst}.\pc = \gst.\pc$.
\end{definition}

\newcommand\thmComm{
    For all $\param\in\AdmP$, and for all
    CFA~$\CFA$, if system instance $\ConcSys =
         (S_{\LetterConcSys}, S^0_{\LetterConcSys}, R_{\LetterConcSys},
         \Prop, \labelfun_{\LetterConcSys})$ and system instance $\AbstSys=
         (S_{\LetterAbstSys}, S^0_{\LetterAbstSys}, R_{\LetterAbstSys},
         \Prop, \labelfun_{\LetterAbstSys})$, then:
    $
    \text{if } (\gst,\gst')\in R_{\LetterConcSys} \text{, then }
    (\absMapSys({\gst}),\absMapSys({\gst'})) \in R_{\LetterAbstSys}.
    $
}
\begin{theorem}
%%[data: transition preservation]
\label{thm:comm}
    \thmComm
\end{theorem}

\newcommand\thmCommProof {
    \begin{proof}
        Let $R$ and $R_{abs}$ be the transition relations of $\SkAut$ and
        $\SkAAut$ respectively. From $(\gst, \gst') \in R_{\LetterConcSys}$ and
        the definition of $\ConcSys$ it follows that there is a process index
        $i: 1 \le i \le \syssize(\param)$ such that $(\gst[i], \gst'[i]) \in R$
        and other processes do not change their local states.

        Let $v$ be a valuation of $\CFA$.  By the definition of $\SkAut$ from
        $(\gst[i], \gst'[i]) \in R$ we have that CFA $\CFA$ has a path $q_1,
        g_1, q_2, \dots, q_k$ such that $q_1=q_I$, $q_k=q_F$ and for every
        guard $g_j$ it holds that $v \models g_j$. Moreover, for any $x \in
        \paraset \cup \locset \cup \globset \cup \{\pc\}$ it holds $v(x) =
        \gst[i].x$ and $v(x') = \gst'[i].x$.

        We choose the same path in $abst(\CFA)$ and construct the valuation
        $\absMap(v)$.  From Theorem~\ref{thm:sem-abs-new} we have that for
        every guard $g_j$ it holds that $\absMap(v) \models g_j$. Hence, the
        path $q_1, g_1, q_2, \dots, q_k$ is a path of CFA $abst(\CFA)$ as well.

By the definitions of $\absMap$ and $\absMapSys$ we have that for
     every $x \in \paraset \cup \locset \cup \globset \cup \{\pc\}$ it
     holds $\absMap(v)(x) = \absMapSys(\gst)[i].x$ and $\absMap(v)(x')
     = \absMapSys(\gst')[i].x$.
By the definition of $\SkAAut$ it immediately follows that
     $(\absMapSys(\gst), \absMapSys(\gst')) \in R_{abs}$.
        
        Finally, $(\absMapSys(\gst), \absMapSys(\gst')) \in R_{abs}$ implies
        that $(\absMapSys(\gst), \absMapSys(\gst')) \in R_{\LetterAbstSys}$.
    \qed
    \end{proof}
}

Theorem~\ref{thm:comm} is the first step to prove simulation.
In order to actually do so, we now define the labeling function
     $\labelfun_{\LetterAbstSys}$.
For propositions from~$p \in \PropPC$,
     $\labelfun_{\LetterAbstSys}({\absgst})$ is defined in the same
     way as $\labelfun_{\LetterConcSys}$.
Similarly to~\cite{KP2000} for propositions from $p \in \PropVAR$,
     which are used in justice constraints, we define: 
\begin{multline*}
    \quotes{\exists i.\; x_i + c < y_i} \in \labelfun_{\LetterAbstSys}({\absgst})
\; \text{ iff } \; \\
\Vee{1 \le i \le \syssize(\param)}{}
    \absgst[i] \models \absEx(\{x + c(\param) < y\}) %\label{equ:abstrLabeling1}
   \end{multline*}
\begin{multline*}
    \quotes{\forall i.\; x_i + c \ge y_i} \in \labelfun_{\LetterAbstSys}({\absgst})
\; \text{ iff } \; \\
\Wedge{1 \le i \le \syssize(\param)}{}
    \absgst[i] \models \absEx(\{x + c(\param) \ge y\}) %\label{equ:abstrLabeling2}
\end{multline*}

From Theorem~\ref{thm:comm}, the definition of $\absMapSys$ with
     respect to the variable $\pc$, and the definition of
     $\labelfun_{\LetterAbstSys}$, one immediately obtains the
     following theorems.
Theorem~\ref{thm:justice} ensures that justice constraints $J$ in the
     abstract system $\AbstSys$ are a sound abstraction of justice
     constraints $J$ in $\ConcSys$.

\begin{theorem}
%%[data: simulation] 
\label{thm:simPC1}
For  all $\param\in\AdmP$, and for  all CFA $\CFA$, it holds $\ConcSys \preceq
     \AbstSys$,  with respect to $\PropPC$.
\end{theorem}
%% \begin{corollary}[Simulations for~$\PropPC$] \label{cor:simPC1}
%% For  all finite state process skeletons, and for all $\param\in\AdmP$,
%%      $P$, the set $\{(\gst, \absMapSys(\gst)) \mid \gst \in S_M \}$ is
%%      a simulation relation between $\ConcSys$ and $\AbstSys$ with
%%      respect to $\PropPC$, that is, $\ConcSys \preceq \AbstSys$.
%% \end{corollary}

    % ********************************************************

\newcommand\thmJustice{
    Let $\pi = \{\gst_i\}_{i \ge 1}$ be a $J$-fair path of
    $\ConcSys$. Then
    $\IntAbs{\pi} = \{\absMapSys(\gst_i)\}_{i \ge 1}$ 
    is a $J$-fair path of $\AbstSys$.
}

\begin{theorem}
%%[data: justice preservation]
\label{thm:justice}
    \thmJustice
\end{theorem}

\newcommand\thmJusticeProof{
    \begin{proof}
        By inductively applying Theorem~\ref{thm:comm} to $\pi$
        we conclude that $\IntAbs{\pi}$ is indeed a path of $\AbstSys$.

        Fix an arbitrary justice constraint $q \in J \subseteq \PropVAR$;
        infinitely many states on $\pi$ are labelled with $q$. Fix a
        state $\gst$ on $\pi$ with $q \in \labelfun_{\LetterConcSys}$.  We
        show that $q \in \labelfun_{\LetterAbstSys}(\absMapSys(\gst))$.
        Consider two cases:

        \mypara{Case 1} Proposition $q$ has a form $\quotes{\exists i.\
        \Phi(i)}$, where $\Phi$ has free variables of two types: a vector of
        parameters $\vec{x}^p = x^p_1, \dots, x^p_{|\paraset|}$ from $\paraset$
        and a vector of variables $\vec{x}^v = x^v_1, \dots, x^v_k$. There is a
        process index $i: 1 \le i \le \syssize(\param)$ such that $\gst[i]
        \models \Phi(i)$.  Hence, $\vec{x}^p=\param, x_1^v = \gst[i].x_1,
        \dots, x_k^v = \gst[i].x_k \models \Phi(i)$. From the definition of the
        existential approximation it follows that
        $(\afunc{\param}{\gst[i].x_1}, \dots, \afunc{\param}{\gst[i].x_k}) \in
        ||\Phi(i)||_E$. Thus, $\IntAbs{x}_1^v = \afunc{\param}{\gst[i].x_1},
        \dots, \IntAbs{x}_k^v = \afunc{\param}{\gst[i].x_k} \models
        \absEx(\Phi(i))$.  As for every $x_j: 1 \le j \le k$ the value
        $\absMapSys(\gst)[i].x_j$ is exactly $\IntAbs{x}_j^v$, we arrive at
        $\absMapSys(\gst)[i] \models \absEx(\Phi(i))$. Then by the construction
        of $\labelfun_{\LetterAbstSys}$ it holds that $\quotes{\exists i.\
        \Phi(i)} \in \labelfun_{\LetterAbstSys}(\absMapSys(\gst))$.

        \mypara{Case 2} Proposition $q$ has a form $\quotes{\forall i.\
        \Phi(i)}$, where $\Phi$ has free variables of two types: a vector of
        parameters $\vec{x}^p = x^p_1, \dots, x^p_{|\paraset|}$ from $\paraset$
        and a vector of variables $\vec{x}^v = x^v_1, \dots, x^v_k$.  Then for
        every process index $i: 1 \le i \le \syssize(\param)$ it holds $\gst[i]
        \models \Phi(i)$.  By fixing an arbitrary $i: 1 \le i \le
        \syssize(\param)$ and repeating exactly the same argument as in the
        Case 1, we show that $\absMapSys(\gst)[i] \models \absEx(\Phi(i))$. As $i$
        is chosen arbitrarily, we conclude that $\Wedge{1 \le i \le
        \syssize(\param)}{} \absMapSys(\gst)[i] \models \absEx(\Phi(i))$. By the
        construction of $\labelfun_{\LetterAbstSys}$ it holds that
        $\quotes{\forall i.\ \Phi(i)} \in
        \labelfun_{\LetterAbstSys}(\absMapSys(s))$.

        From Cases 1 and 2 we conclude that
        $q \in \labelfun_{\LetterAbstSys}(\absMapSys(\gst))$. As we chose
        $\gst$ to be an arbitrary state on $\pi$ labelled with $q$ and we know
        that there are infinitely many such states on $\pi$, we have shown
        that there are infinitely many states $\absMapSys(\gst)$ on
        $\IntAbs{\pi}$ labelled with $q$. Finally, as $q$ was chosen to be
        an arbitrary justice constraint from $J$, we conclude that every
        justice constraint $q \in J$ appears infinitely often on $\IntAbs{\pi}$.

        This proves that $\IntAbs{\pi}$ is a fair path.
    \qed
    \end{proof}
}

    % ********************************************************

\subsection{PIA counter abstraction}\label{sec:CountAbs}

In this section, we present a counter abstraction inspired
     by~\cite{PXZ02} which maps a system instance composed of
     \emph{identical finite state} process skeletons to a single
     finite state system.
We use the PIA domain $\absdomain$ along with abstractions
     $\absEx(\{x'=x+1\})$ and $\absEx(\{x'=x-1\})$ for the counters.

Let us consider a process skeleton ${\Sk} = ({S}, {S}_0, {R})$, where
     $S = \PC \times \somedomain^{|\locset|} \times
     \somedomain^{|\globset|} \times \somedomain^{|\paraset|}$ that is
     defined using an arbitrary  finite domain~$\somedomain$.
(Note that we do not require that the skeleton is obtained from a CFA.) Our
     counter abstraction over the abstract domain $\absdomain$
     proceeds in two stages, where the first stage is only a change in
     representation, but not an abstraction.

\mypara{Stage 1: Vector Addition System with States (VASS)}
Let $\localstates = \{ \ell \in \PC \times \somedomain^{|\locset|}
     \mid \exists {s} \in {S}.\; \ell \gleich{\{\pc\}\cup \locset} {s}
     \}$ be the set of \emph{local states} of a process skeleton.
As the domain~$\somedomain$ and the set of local variables $ \locset$
     are finite, $\localstates$ is finite.
We write the elements of $\localstates$ as $\ell_1, \dots,
     \ell_{|\localstates|}$.
We define the counting function $K\colon S_{\LetterConcSys} \times
     \localstates \rightarrow D$ such that $K(\gst,\ell)$ is the
     number of processes $i$ whose local state is $\ell$ in global
     state $\gst$, i.e., ${\gst}[i] \gleich{\{\pc\} \cup \locset}
     \ell$.
Thus, we represent the system state $\gst\in S_{\LetterConcSys}$ as a
     tuple $(g_1, \dots, g_k,K[\gst,\ell_1],\dots,
     K[\gst,\ell_{|\localstates|}]),$ i.e., by the shared global state
     and by the counters for the local states.
If a process moves from local state $\ell_i$ to local state~$\ell_j$,
     the counters of $\ell_i$ and $\ell_j$ will decrement and
     increment, respectively.
\nop{ The parameterized mapping $\vassMap: S_\LetterConcSys
     \rightarrow \NatZero^{|\localstates|} \times
     \somedomain^{|\globset|}$ maps a global state of $\sConcSys$ to
     its VASS representation.}

\mypara{Stage 2: Abstraction of VASS}
We abstract the counters $K$ of the VASS representation using
the PIA domain to obtain a finite state Kripke structure $\CntASys(\Sk)$.
To compute $\CntASys(\Sk)= (S_{\CntASys},
     S_{\CntASys}^0, R_{\CntASys}, \Prop,\labelfun_{\CntASys})$ we proceed as
     follows: %.

A state $w \in S_{\CntASys}$  is given by values of shared variables
     from the set~$\globset$, ranging over $\somedomain^{|\globset|}$,
     and by a vector $(\absCnt{\ell_1}, \dots,
     \absCnt{\ell_{|\localstates|}})$ over the abstract domain
     $\absdomain$ from Section~\ref{sec:AbsDomain}.
More concisely, $S_{\CntASys} = \absdomain^{|\localstates|} \times
     \somedomain^{|\globset|}$.

\begin{definition}
The parameterized  abstraction mapping $\cacMap$ maps
a     global state~${\gst}$ of the system~$\sConcSys$ to a
     state~$w$ of the abstraction~$\CntASys(\Sk)$ such that:
 For all  $\ell \in \localstates$  it holds that
    $w.\absCnt{\ell} = \abst_{\param}(K[{\gst}, \ell])$, and
 $w \gleich{\globset}  {\gst}$.
\end{definition}

From the definition, one can see how to construct the initial states.
Informally, we require (1) that the initial shared states of
     $\CntASys(\Sk)$ correspond to initial shared states of $\Sk$,
     (2) that there are actually $\syssize(\param)$ processes in the
     system, and (3) that initially all processes are in an initial
     state.

The intuition\footnote{A formal definition of the transition relation
     is given in Appendix~\ref{sec:4Igor}.} for the construction of
     the transition relation is as follows:  Like in VASS, a step that
     brings a process from local state $\ell_i$ to $\ell_j$ can be
     modeled by decrementing the (non-zero) counter of $\ell_i$ and
     incrementing the counter of $\ell_j$.
Like  Pnueli, Xu, and Zuck~\cite{PXZ02} we use the idea of
     representing counters in an abstract domain, and performing
     increment and decrement using existential abstraction.
They used a three-valued domain representing 0, 1, or more processes.
As we are interested, e.g., in the fact whether at least $t+1$ or
     $n-t$ processes are in a certain state, the domain
     from~\cite{PXZ02} is too coarse for us.
Therefore, we use counters from $\absdomain$, and we increment and
     decrement counters using the formulas $\absEx(\{x'=x+1\})$ and
     $\absEx(\{x'=x-1\})$.

\newcommand\thmStepTwoGeneral{
For all $\param\in\AdmP$, and all finite state process skeletons
     ${\Sk}$, let system instance $\sConcSys = (S_{\LetterConcSys},
     S^0_{\LetterConcSys}, R_{\LetterConcSys}, \Prop,
     \labelfun_{\LetterConcSys})$, and $\CntASys(\Sk) =
     (S_{\CntASys}, S_{\CntASys}^0, R_{\CntASys},
     \Prop,\labelfun_{\CntASys})$.
Then:
$
\text{if } (\gst, \gst') \in R_{\LetterConcSys} \text{, then }
     \left({\cacMap(\gst)},{\cacMap(\gst')}\right) \in
     R_{\CntASys}.
$
}

\begin{theorem}
%%[counters: transition preservation] 
\label{thm:step2-general}
    \thmStepTwoGeneral
\end{theorem}

\newcommand\thmStepTwoGeneralProof{\begin{proof}
We have to show that if $(\gst, \gst') \in R_{\LetterConcSys}$, then
     $w=\cacMap(\gst)$ and $w'=\cacMap(\gst')$
     satisfy~(\ref{eq:ctr-abs-proc-moves}) to~(\ref{eq:ctr-abs-keep}.
We first note that as $(\gst, \gst') \in R_{\LetterConcSys}$, it
     follows from the \trmove\ property of transition relations that
     there is a process index~$i$ such that $(\gst[i], \gst'[i])
     \in R_{\LetterConcSys}$; we will use the existence of $i$ in the following:

\mypara{(\ref{eq:ctr-abs-proc-moves})}
Abbreviating ${s}=\gst[i]$ and ${s}'= \gst'[i]$,
     (\ref{eq:ctr-abs-proc-moves}) follows.

\mypara{(\ref{abs:existsfrom}) and~(\ref{abs:existsto})}
Follows immediately from the definition of $\localstates$.

\mypara{(\ref{abs:nonzero})}
From the definition of $\cacMap$ it follows that $w.\absCnt{\fromstate} =
    \afunc{\param}{K({\gst}, \fromstate)}$.

From the existence of the index~$i$ it follows that $K({\gst}, \fromstate)\ge
    1$.
Hence, we have $K({\gst}, \fromstate) \ne 0$ and from the definition of
$\abst_\param$ it follows that $\afunc{\param}{K({\gst}, \fromstate)} \ne 0$.
From $\afunc{\param}{1} = \absSym_1$ and Definition~\ref{def:torder} of total
order we conclude (\ref{abs:nonzero}), i.e. $\afunc{\param}{K({\gst},
\fromstate)} \ne \absSym_0$.

\mypara{(\ref{abs:globalfrom}) and (\ref{abs:globalto})}
Follows immediately from the definition of~$\cacMap$.

\mypara{(\ref{abs:nochange})}
Since $\tostate = \fromstate$, it follows from (\ref{abs:existsfrom})
     and~(\ref{abs:existsto}) that  ${s} \gleich{\{ \pc \} \cup
     \locset}  {s}'$.
Thus the process with index~$i$ does not change its local state.
Moreover from the property  \trmaintain\ of transition relations, all
      processes other than $i$ maintain their local state.
It follows that for all $\ell$ in $L$,   $K({\gst}, \ell) =
     K({\gst}', \ell)$, and further that
     $\afunc{\param}{K({\gst}, \ell)} =
     \afunc{\param}{K({\gst}', \ell)}$, and in particular
 $\afunc{\param}{K({\gst}, \fromstate)} =
     \afunc{\param}{K({\gst}', \fromstate)}$.
Then  (\ref{abs:nochange}) follows from the definition of~$\cacMap$.

\mypara{(\ref{abs:inc}) and (\ref{abs:dec})}
From the property  \trmaintain\ of transition relations, all
     processes other than $i$ maintain their local state.
Since $\tostate \ne \fromstate$ it follows that $i$ changes it local
     state.
It follows that
\begin{align}
K({\gst}', \tostate) &= K({\gst}, \tostate) + 1,
\label{eq:bla1}\\
K({\gst}', \fromstate) &= K({\gst}, \fromstate) - 1.
\label{eq:bla2}
\end{align}
From the definition of $\cacMap$ we have
\begin{multline*}
w'.\absCnt{\tostate} = \afunc{\param}{K({\gst}', \tostate)}
 \mbox{ and } \\
w.\absCnt{\tostate} = \afunc{\param}{K({\gst}, \tostate)}
\end{multline*}
\begin{multline*}
w'.\absCnt{\fromstate} = \afunc{\param}{K({\gst}', \fromstate)}
\; \; \mbox{ and } \; \; \\
w.\absCnt{\fromstate} = \afunc{\param}{K({\gst}, \fromstate)}
\end{multline*}
From Proposition~\ref{prop:RussianDolls} follows that
\begin{multline}
K({\gst}', \tostate) \in \cfunc{\param}{w'.\absCnt{\tostate}}
\; \; \mbox{ and } \; \; \\
K({\gst}, \tostate) \in \cfunc{\param}{w.\absCnt{\tostate}}
\label{eq:blu1}
\end{multline}
\begin{multline}
K({\gst}', \fromstate) \in \cfunc{\param}{w'.\absCnt{\fromstate}}
\; \; \mbox{ and } \; \; \\
K({\gst}, \fromstate) \in
     \cfunc{\param}{w.\absCnt{\fromstate}}.
\label{eq:blu2}
\end{multline}

Point (\ref{abs:inc}) follows from (\ref{eq:bla1}), (\ref{eq:blu1}),
     and  the definition of existential abstraction~$\absEx$,
     while  (\ref{abs:dec}) follows from (\ref{eq:bla2}), (\ref{eq:blu2}), and
     the definition of existential abstraction~$\absEx$.

\mypara{(\ref{eq:ctr-abs-keep})}
From property \trmaintain\ processes other than $i$ do not move.
The move of process~$i$ does not change the number of processes in
     states other than $\fromstate$ and $\tostate$.
Consequently, for all local states~$\ell$ different from $\fromstate$
     and $\tostate$ it holds that $K({\gst}', \ell) =
     K({\gst}, \ell)$.
It follows that $\afunc{\param}{K({\gst}', \ell)} =
     \afunc{\param}{K({\gst}, \ell)}$, and
     (\ref{eq:ctr-abs-keep}) follows from the definition of $\cacMap$.
\qed
\end{proof}
}

To prove simulation, we now define the labeling
     function~$\labelfun_{\CntASys}$.
Here we consider propositions from $\PropVAR \cup \PropPC$ in the
      form of $\quotes{\exists i.\ \Phi(i)}$ and
     $\quotes{\forall i.\ \Phi(i)}$.
Formula~$\Phi(i)$ is defined over variables from the
     $|\paraset|$-dimensional vector $\vec{x}^p$ of parameters, a
     $k$-dimensional vector $\vec{x}^\ell$ of local variables and
     $\pc$, an $m$-dimensional vector of global variables $\vec{x}^g$.
Then, the labeling function is defined~by 
        \begin{multline*}
            \quotes{\exists i.\; \Phi(i)}  \in \labelfun_\CntASys(w)
        \text{ if and only if }  \\
         \Vee{\ell \in \localstates}{}
        \left(\vec{x}^\ell \gleich{\locset} \ell, \vec{x}^g \gleich{\globset} w \models abst(\Phi(i))
         \wedge w.k[\ell] \ne \absSym_0 \right)
        \end{multline*}
 \begin{multline*}
        \quotes{\forall i.\; \Phi(i)}  \in \labelfun_\CntASys(w)
        \text{ if and only if }  \\
           \Wedge{\ell \in \localstates}{}
            \left(
            \vec{x}^\ell \gleich{\locset} \ell, \vec{x}^g \gleich{\globset} w  \models abst(\Phi(i))
            \vee w.k[\ell] = \absSym_0
            \right)
                \end{multline*}

%% \begin{align}
%% \left[\forall i.\; \pc_i = \pcval \right] \in \labelfun_\CntASys(w)
%%  \text{ iff }
%% \Wedge{\ell \in \localstates}{}
%% \left(\ell.\pc = \pcval \vee w.\kappa[\ell] = \absZero \right)
%% \label{eq:lfunAllPC}\\
%% \left[\exists i.\; \pc_i = \pcval \right] \in \labelfun_\CntASys(w)
%%  \text{ iff }
%% \Vee{\ell \in \localstates}{}
%% \left(\ell.\pc = \pcval \wedge w.\kappa[\ell] \ne \absZero
%% \right)
%% \label{eq:lfunSomePC}
%% \end{align}

\newcommand\thmSimTwo{
For all $\param\in\AdmP$, and for all finite state process skeletons
     $\Sk$, $\sConcSys \preceq \CntASys(\Sk)$, w.r.t.~$\PropPC$.
}
\begin{theorem}
%[counters: simulation] 
\label{thm:sim2}
    \thmSimTwo
\end{theorem}

\newcommand\thmSimTwoProof{\begin{proof}
Due to Theorem~\ref{thm:step2-general}, it is sufficient to show that
     if a proposition $p\in\PropPC$ holds in state $\gst$ of
     $\sConcSys$ then it also holds in state $\cacMap(\gst)$ and vice versa.
We distinguish two types of propositions.

If $p = \left[\forall i.\; \pc_i = \pcval \right]$ and
    $p\in\labelfun_{\LetterConcSys}(\gst)$, then by the definition of
    $\labelfun_{\LetterConcSys}$ we have $\bigwedge_{1 \le i \le
    \syssize(\param)}{} \left(\gst[i].\pc = \pcval \right)$. 
Thus, in global state $\gst$ all processes are in a local state with
    $\pc=\pcval$. 
In other words, no process is in a local state with $\pc\ne\pcval$. 
It follows that each local state $\ell$ satisfies in $\gst$ that
    $\ell.\pc=\pcval$ or $K(\gst,\ell) = 0$.
From the definition of $\cacMap$ and the definition of $\labelfun_\CntASys$
    this case follows. The same argument works in the opposite direction.

If  $p = \left[\exists i.\; \pc_i = \pcval \right]$ and
    $p\in\labelfun_{\LetterConcSys}(\gst)$, then by the definition of
    $\labelfun_{\LetterConcSys}$ we have $\bigvee_{1
     \le i \le \syssize(\param)}{} \left(\gst[i].\pc = \pcval  \right)$.
Thus, in global state $\gst$ there is a process in a local state
     $\ell$ with $\pc=\pcval$.
It follows that $K(\gst,\ell) > 0$.
From the definition of $\cacMap$ and the definition of~$\labelfun_{\CntASys}$
    the case follows. The same argument works in the opposite direction.

This two cases conclude the proof.    
\qed
\end{proof}}

\newcommand{\thmJusticeForAll}{
    If $\pi = \{\gst_i\}_{i \ge 1}$ is a $J$-fair path of $\sConcSys$,
    then the path $\IntAbs{\pi} = \{\cacMap(\gst_i) \}_{i \ge 1}$ is a $J$-fair path of
        $\CntASys$.
}

\begin{theorem}
%%[counters: justice preservation] 
\label{thm:justice-for-all}
    \thmJusticeForAll
\end{theorem}

\newcommand\thmJusticeForAllProof{
\begin{proof}
        By inductively applying Theorem~\ref{thm:step2-general} to $\pi$ we
            conclude that $\IntAbs{\pi}$ is indeed a path of $\CntASys$.

        Fix an arbitrary justice constraint $q \in J \subseteq \PropVAR$;
            infinitely many states on $\pi$ are labelled with $q$.
        Fix an arbitrary state $\gst$ on $\pi$ such that $q \in
            \labelfun_{\LetterConcSys}$.
        We show that $q \in \labelfun_{\CntASys}(\cacMap(\gst))$.

        Propositions from $\PropVAR$ have the form
            of $\quotes{\exists i.\ \Phi(i)}$ and $\quotes{\forall i.\
            \Phi(i)}$, where each $\Phi(i)$ has free variables of two types: a
            vector of parameters $\vec{x}^p = x^p_1, \dots, x^p_{|\paraset|}$
            from $\paraset$, a vector of local variables $x^\ell = x^\ell_1,
            \dots, x^\ell_k$ from $\locset$, and a vector of global variables
            $x^g = x^g_1, \dots, x^g_m$ from~$\globset$.

        \begin{multline}
            \quotes{\exists i.\; \Phi(i)}  \in \labelfun_\CntASys(w)
        \text{ iff } \\
        \Vee{\ell \in \localstates}{}
        \left(\vec{x}^\ell = \ell, \vec{x}^g \gleich{\globset} w \models \absEx(\Phi(i))
         \wedge w.k[\ell] \ne \absSym_0 \right)
        \label{eq:gen-lfunSomeD} 
\end{multline}
\begin{multline}
        \quotes{\forall i.\; \Phi(i)}  \in \labelfun_\CntASys(w)
        \text{ iff }\\ 
            \Wedge{\ell \in \localstates}{}
            \left(
            \vec{x}^\ell = \ell, \vec{x}^g \gleich{\globset} w  \models \absEx(\Phi(i))
            \vee w.k[\ell] = \absSym_0
            \right)
        \label{eq:gen-lfunAllD}
        \end{multline}

        Consider two cases:
        \mypara{Existential case~(\ref{eq:gen-lfunSomeD})}
        There is a process index $i: 1 \le i \le \syssize(\param)$ such that
            $\absgst[i] \models \absEx(\Phi(i))$.

        Consider a local state $\ell \in \localstates$ with
            $\ell \gleich{\localstates} \absgst[i]$.
        As $\absgst[i] \models \absEx(\Phi(i))$ it follows that
            $x^\ell_1=\ell.x^\ell_1,\dots,x^\ell_k=\ell.x^\ell_k,
               x^g_1=w.x^g_1,\dots,x^g_m=w.x^g_m \models \absEx(\Phi(i))$.
        As $i$ is the index of a process with $\ell \gleich{\localstates}
            \absgst[i]$, it immediately follows that $K(w, \ell) \ne 0$.
        From the definition of $\abst$ it follows that for every $\param \in
            \AdmP$ it holds $\afunc{\param}{K(w, \ell)} \ne \absSym_0$.
        Thus, by the definition of $\cacMap$ we have
            $w.\absCnt{\ell} \ne \absSym_0$.

        Hence, both requirements of equation~(\ref{eq:gen-lfunSomeD}) are met for
            $\ell$ and from the property of disjunction we have $q \in
            \labelfun_\CntASys(w)$.

        \mypara{Universal case~(\ref{eq:gen-lfunAllD})}
        Then for every process index $i: 1 \le i \le \syssize(\param)$ it holds
            $\absgst[i] \models \absEx(\Phi(i))$.

        By fixing an arbitrary $i: 1 \le i \le \syssize(\param)$,
            choosing $\ell \in \localstates$ with
            $\ell \gleich{\localstates} w$ and by repeating
            exactly the same argument as in the existential case, we show that
            $x^\ell_1=\ell.x^\ell_1,\dots,x^\ell_k=\ell.x^\ell_k,
            x^g_1=w.x^g_1,\dots,x^g_m=w.x^g_m \models \absEx(\Phi(i))$.
        Thus, for every $\ell \in \localstates$ such that there exists
            $i: 1 \le i \le \syssize(\param)$ with $\ell 
            \gleich{\localstates} w$ the disjunct for $\ell$
            in~\ref{eq:gen-lfunAllD} holds true.

        Consider $\ell' \in \localstates$ such that for every 
            $i: 1 \le i \le \syssize(\param)$ it holds
            $\ell' \nichtgleich{\localstates} w$.
        It immediately follows that $K(w, \ell') = 0$; from the definition of
            $\abst_\param$ we have that
            $\afunc{\param}{K(w, \ell')} = \absSym_0$
            and thus $\absCnt{\ell'} = \absSym_0$.
        Then for $\ell'$ the disjunct in ~\ref{eq:gen-lfunAllD} holds true as well.

        Thus, we conclude that the conjunction in the right-hand side of
            the equation~(\ref{eq:gen-lfunAllD}) holds, which immediately results
            in $q \in \labelfun_\CntASys(w)$.

        From Universal case and Existential Case we conclude that $q \in
            \labelfun_{\CntASys}(w)$.
        As we chose $\absgst$ to be an arbitrary state on $\pi$ labelled with
            $q$ and we know that there are infinitely many such states on
            $\pi$, we have shown that there are infinitely many states
            $\cacMap(\absgst)$ on $\IntAbs{\pi}$ labelled with $q$.
        Finally, as $q$ was chosen to be an arbitrary justice constraint from
            $J$, we conclude that every justice constraint $q \in J$ appears
            infinitely often on $\IntAbs{\pi}$.

        This proves that $\IntAbs{\pi}$ is a fair path.
\qed
\end{proof}
}

From Theorems~\ref{thm:simPC1}, \ref{thm:justice}, \ref{thm:sim2},
     ~\ref{thm:justice-for-all}, and \cite[Thm.~16]{CGP1999} we obtain
     the following central corollary in the form necessary for our
     parameterized model checking problem.

\begin{corollary}[Soundness of data \& counter abstraction]
\label{cor:completeSound}
For all CFA $\CFA$, and for all formulas
     $\varphi$ from \LTLX\ over $\PropPC$ and justice constraints
     $J \subseteq \PropVAR$:
      if
     $\CntASys(\SkAAut) \models_J \varphi$,
     then for all $\param\in\AdmP$ it holds $\ConcSys \models_J \varphi$.
\end{corollary}

%%\jw{there was some discussion whether this actually is a
%%  corollary. What should we write here?}

%% As discussed in Section~\ref{sec:APs}, for the verification of
%%      asynchronous distributed algorithms, for specifying fairness, we
%%      are interested in a special combination of formulas over
%%      $\PropPC$ and $\PropVAR$.
%% For this form we directly obtain:

%% \begin{corollary}\label{cor:casestudy}
%% For each parameterized CFA $\pCFA$, for each $\param\in\AdmP$, for
%%      each formula $\varphi$ from \ACTL\ over $\PropPC$ and
%%      any negation-free formula $\psi$ from \ACTL\ over $\PropVAR$,
%%      if $\CntASys(P(\aCFA)) \models
%%      \psi \vee \varphi$, then $\ConcSys \models \psi \vee \varphi$.
%% \end{corollary}

\section{Abstraction Refinement}
    \label{sec:Refine} \label{sec:RefineSmall}

Due to our parametric existential abstraction of comparison guards,
     which may be imprecise, we have to deal with several kinds of
     spurious behavior.

The first one is caused by spurious transitions.
Consider a transition $\tau$ of $\CntSkAut$.
We say that the transition~$\tau$ is spurious w.r.t.\ $\param \in
     \AdmP$, if there is no transition in $\ConcSys$ that is a
     concretization of $\tau$.
This situation can be detected by known techniques~\cite{Clarke2003}
     for a fixed $\param$.
However, it is not sound to remove $\tau$ from $\CntSkAut$,
     unless~$\tau$ is spurious w.r.t.\ \emph{all} $\param \in \AdmP$.
We call transitions that are spurious w.r.t.\ all admissible
     parameters \emph{uniformly spurious}.
Detecting such transitions is a challenge and to the best of our
     knowledge, this problem has not been investigated before.
To detect such transitions, we use one more intermediate abstraction
     in the form of VASS that abstracts local variables as  in
     Section~\ref{sec:FirstAbs} and keeps concrete shared variables
     and process counters.

Independently of uniformly spurious transitions, parametric
     abstraction leads to the second interesting problem.
Consider transitions $\tau_1$ and $\tau_2$ of $\CntSkAut$ that are not
     spurious w.r.t.\ $\param_1$ and $\param_2$ in $\AdmP$,
     respectively, for  $\param_1 \ne \param_2$.
It can be the case that a path $\tau_1, \tau_2$ is in $\CntSkAut$ and
     there is no $\param_3 \in \AdmP$ such that the concretization of
     $\tau_1, \tau_2$ is a path in $\ConcSysP{\param_3}$, i.e., the
     path $\tau_1, \tau_2$ is uniformly spurious.
We detect such spurious behavior by invariants.
These invariants are provided by the user as invariant candidates, and
     then automatically checked to actually be invariant using an
     SMT~solver.

As observed by~\cite{PXZ02}, counter abstraction may lead to justice
     suppression.
Given a counter-example in the form of a lasso, we detect whether its
     loop contains only unjust states.
If this is the case, we refine $\CntSkAut$ by adding a justice
     requirement, which is consistent with existing requirements in
     all concrete instances $\ConcSys$.
This refinement is similar to an idea from~\cite{PXZ02}.

Below, we give a general framework for a sound refinement
     of $\CntSkAut$.\footnote{In the appendix, we provide techniques
     that allow us to do refinement in practice in
     Sections~\ref{sec:cegar-detect} and~\ref{sec:cegar-inv}.} 

%% \subsection{Refinement Framework for Parameterized Systems}
%% \label{sec:cegar-framework}

\smallskip

To simplify presentation, we define a \emph{monster system} as a
     (possibly infinite) Kripke structure 
%\footnote{We use a name from stability theory to emphasize that
%the system is huge and infinite and includes behavior of all parameterized
%system instances.}
 $\WrapSys = (S_\LetterWrapSys, S^0_\LetterWrapSys,
R_\LetterWrapSys, \Prop, \labelfun_\LetterWrapSys)$, whose state space and
transition relation are disjoint unions of state spaces and transition
relations of system instances $\ConcSys = (S_\param, S^0_\param, R_\param,
\Prop, \labelfun_\param)$ over all admissible parameters:
\begin{align*}
    S_\LetterWrapSys & = \Union{\param \in \AdmP}{} S_\param, \qquad
  S^0_\LetterWrapSys = \Union{\param \in \AdmP}{} S^0_\param, \qquad
    R_\LetterWrapSys = \Union{\param \in \AdmP}{} R_\param \\
    \labelfun_\LetterWrapSys & : S_\LetterWrapSys \rightarrow 2^{\Prop} \ 
    \mbox{and } \forall \param \in \AdmP, \forall s \in S_\param.\;
    \labelfun_\LetterWrapSys(s) = \labelfun_\param(s)
\end{align*}

Using abstraction mappings $\absMapSys$ and $\cacMap$ we define an
abstraction mapping $\wrapMap : S_\LetterWrapSys \rightarrow S_\CntASys$
from $\WrapSys$ to $\CntSkAut$:
    $\mbox {If } \gst \in S_\param \mbox{, then }
    \wrapMap(\gst) = \cacMap(\absMapSys(\gst))$.

\begin{definition}
    A sequence $T=\{\gst_i\}_{i \ge 1}$ is a \emph{concretization} of path
        $\IntAbs{T} = \{w_i\}_{i \ge 1}$ from $\CntSkAut$ if and only if $
        \gst_1 \in S^0_\LetterWrapSys$ and for all $i \ge 1$ it holds
        $\wrapMap(\gst_i) = w_i$.
\end{definition}

\begin{definition}\label{def:spurious-path}
    A path $\IntAbs{T}$ of $\CntSkAut$
        is a \emph{spurious path} iff every concretization $T$ of $\IntAbs{T}$
        is not a path in $\WrapSys$.
\end{definition}

While for finite state systems there are methods to detect whether a
     path is spurious~\cite{Clarke2003}, we are not aware of a method
     to detect whether a path $\IntAbs{T}$ in $\CntSkAut$ corresponds
     to a path in the (concrete) infinite monster system~$\WrapSys$.
Therefore, we limit ourselves to detecting and refining uniformly
     spurious transitions and unjust states.

\nop{
It is easy to see that the definition~\ref{def:spurious-path} does not give an
immediate way to check if a path $\IntAbs{T}$ is spurious. Consider our example
in the Figure~\ref{Fig:CFA_SSA}. If we follow the algorithm \textsf{SplitLOOP}
from~\cite{Clarke2003}, then for an arbitrarily large number $C$ one can give a
resilience condition $n > 3t \wedge t \ge f \wedge t \ge C$,
causing at least $C$ processes to send a message until surpassing the threshold $t+1$
and thus to unwind a spurious loop at least $C$ times.
}

\begin{definition}\label{def:uni-spurious}
    An abstract transition $(w, w') \in R_\CntASys$ is \emph{uniformly
    spurious} iff there is no transition $(\gst, \gst') \in R_\LetterWrapSys$
    with $w = \wrapMap(\gst)$ and $w' = \wrapMap(\gst')$.
\end{definition}

\begin{definition}\label{def:unjust}
    An abstract state $w \in S_\CntASys$ is \emph{unjust
    under} $q \in \PropVAR$ iff there is no concrete state $\gst \in
    S_\LetterWrapSys$ with $w = \wrapMap(\gst)$ and $q \in
    \labelfun_\LetterWrapSys(\gst)$.
\end{definition}

We give a general criterion that ensures soundness of abstraction, when
    removing uniformly spurious transitions.
In other words, removing a transition does not affect the property of
    transition preservation.

\newcommand\thmRemoveSpur {    
    Let $T \subseteq R_\CntASys$ be a set of spurious transitions.
    Then for every transition $(\gst, \gst') \in R_\LetterWrapSys$ there is a
        transition $(\wrapMap(\gst), \wrapMap(\gst'))$ in $R_\CntASys \setminus
        T$.
}

\begin{theorem}\label{thm:remove-spur}
  \thmRemoveSpur
\end{theorem}

\newcommand\thmRemoveSpurProof {
    \begin{proof}
        Assume that there is transition $(\gst, \gst') \in R_\LetterWrapSys$
            with $w=\wrapMap(\gst)$, $w'=\wrapMap(\gst')$, and
            $(w, w') \in R_\CntASys \cap T$.
        As $T$ is a set of uniformly spurious transitions, we have that
            the transition $(w, w')$ is uniformly spurious.
        Consider a pair of states $\rho, \rho' \in S_\LetterWrapSys$ with
            the property $\wrapMap(\rho) = w$ and $\wrapMap(\rho') = w'$.
        From Definition~\ref{def:uni-spurious} it follows that
            $(\rho, \rho') \not \in R_\LetterWrapSys$.
        This contradicts the assumption $(\gst, \gst') \in R_\LetterWrapSys$
        as we can take $\rho=\gst$ and $\rho'=\gst'$.
        \qed
    \end{proof}
}

From the theorem it follows that the system
    $\CntASys_{ref} = (S_\CntASys, S^0_\CntASys, R_\CntASys \setminus T, \Prop,
    \labelfun_\CntASys)$ still simulates $\WrapSys$.

After the criterion of removing individual transitions, we now consider
    infinite counterexamples of $\CntSkAut$, which have a form of lassos
    $w_1 \dots w_k (w_{k+1} \dots w_{m})^\omega$. 
For such a counterexample $\IntAbs{T}$ we denote the set of states in the
    lasso's loop by $U$. 
We then check, whether all states of $U$ are unjust under some justice
    constraint $q \in J$. 
If this is the case, $\IntAbs{T}$ is a spurious counterexample, because
    the justice constraint $q$ is violated. 
Note that it is sound to only consider infinite paths, where states
    outside of $U$ appear infinitely often; in fact, this is a justice
    requirement. 
To refine $\CntASys$'s unjust behavior we add a corresponding justice
    requirement. 
Formally, we augment $J$ (and $\PropVAR$) with a propositional
    symbol~$\offu$. 
Further, we augment the labelling function $\labelfun_\CntASys$ such that
    every $w \in S_\CntASys$ is labelled with $\offu$ if and only if $w
    \in U$.

\newcommand\thmRemoveUnjustice {
Let $J \subseteq \PropVAR$ be a set of justice requirements,
    $q \in J$, and $U \subseteq S_\CntASys$ be a set of unjust states under $q$.
Let $\pi = \{\sigma_i\}_{i \ge 1}$ be an arbitrary fair path of $\WrapSys$
    under $J$.
The path $\IntAbs{\pi} = \{\wrapMap(\sigma_i)\}_{i\ge 1}$ is a fair path
    in $\CntSkAut$ under
    $ J  \cup \{ \offu \}$.
}

\begin{theorem}\label{thm:remove-unjustice}
    \thmRemoveUnjustice
\end{theorem}

\newcommand\thmRemoveUnjusticeProof {
    \begin{proof}
        Consider an arbitrary fair path $\pi = \{\sigma_i\}_{i \ge 1}$ of
            $\WrapSys$ under $J$.
        Assume that
            $\IntAbs{\pi} = \{\wrapMap(\sigma_i)\}_{i\ge 1}$ is fair under $J$,
            but it becomes unfair under $J \cup \{ \offu \}$.

        If $\IntAbs{\pi}$ is unfair under $\{ \offu \}$, then $\IntAbs{\pi}$
            does not have infinitely many states labelled with $\offu$.
        Thus, $\IntAbs{\pi}$ must have an infinite suffix
            $\mathit{suf}(\IntAbs{\pi})$, where each
            $w \in \mathit{suf}(\IntAbs{\pi})$ has the property
            $\offu \notin \labelfun_\CntASys$.
        From the definition of $\offu$ we immediately conclude that every
            state $w \in \mathit{suf}(\IntAbs{\pi})$ belongs to $U$, i.e.,
            $w$ is unjust under $q \in J$.
        
        Using the suffix $\mathit{suf}(\IntAbs{\pi})$ we reconstruct a
            corresponding suffix $\mathit{suf}(\pi)$ of $\pi$ (by skipping the
            prefix of the same length as in $\IntAbs{\pi}$).
        From the fact that every state of $\mathit{suf}(\IntAbs{\pi})$ is unjust
            under $q$ we know that every state $\sigma \in \mathit{suf}(\pi)$
            violates the constraint $q$ as well, namely,
            $q \not \in \labelfun_\LetterWrapSys(\sigma)$.
        Thus, $\pi$ has at most finitely many states labelled with $q \in J$.
        It immediately follows from the definition of fairness that $\pi$ is
            not fair under $J$. This contradicts the assumption of the theorem.
        \qed
    \end{proof}
}

From this we derive that loops containing only unjust states can be
     eliminated, and thus $\CntSkAut$ be refined.

% unmoved to the conclusion

\section{Experimental Evaluation}\label{sec:exp}

\begin{table}
  \begin{tabular}{rrrrrrrr}
%   \hline
   \textbf{\scriptsize{$M \models \varphi?$}} & \textbf{\scriptsize{RC}} &\textbf{\scriptsize{Spin}} & \textbf{\scriptsize{Spin}} & \textbf{\scriptsize{Spin}} & \textbf{\scriptsize{Spin}} & \textbf{\scriptsize{\#R}} & \textbf{\scriptsize{Total}} \\
   & & \textbf{\scriptsize{Time}} & \textbf{\scriptsize{Memory}} & \textbf{\scriptsize{States}} & \textbf{\scriptsize{Depth}} &  & \textbf{\scriptsize{Time}} \\
   \hline

 $\mathit{Byz} \models U$ & \textsc{(a)} & 2.2 s& 83 MB & 483k & 9154 & 0 & 3 s\\
 $\mathit{Byz} \models C$ & \textsc{(a)} & 2.6 s& 87 MB & 614k & 12966 & 6 & 18 s\\
 $\mathit{Byz} \models R$ & \textsc{(a)} & 6.6 s& 101 MB & 1122k & 14820 & 8 & 21 s\\
 $\mathit{Sym} \models U$ & \textsc{(a)} & 0.1 s& 68 MB & 18k & 897 & 0 & 1 s\\
 $\mathit{Sym} \models C$ & \textsc{(a)} & 0.1 s& 68 MB & 19k & 1221 & 3 & 5 s\\
 $\mathit{Sym} \models R$ & \textsc{(a)} & 0.2 s& 69 MB & 40k & 1669 & 8 & 12 s\\
 $\mathit{Omt} \models U$ & \textsc{(a)} & 0.1 s& 68 MB & 4k & 487 & 0 & 1 s\\
 $\mathit{Omt} \models C$ & \textsc{(a)} & 0.1 s& 68 MB & 6k & 627 & 3 & 9 s\\
 $\mathit{Omt} \models R$ & \textsc{(a)} & 0.1 s& 68 MB & 8k & 704 & 5 & 9 s\\
 $\mathit{Cln} \models U$ & \textsc{(a)} & 0.3 s& 68 MB & 30k & 1371 & 0 & 2 s\\
 $\mathit{Cln} \models C$ & \textsc{(a)} & 0.3 s& 68 MB & 35k & 2043 & 6 & 10 s\\
 $\mathit{Cln} \models R$ & \textsc{(a)} & 1.1 s& 69 MB & 51k & 2647 & 20 & 60 s\\
   \hline
 $\mathit{RBC} \models U$      & --- & 0.1 s & 68 MB & 0.8k & 232 & 0 & 1 s\\
 $\mathit{RBC} \models A$      & --- & 0.1 s & 68 MB & 1.7k & 333 & 0 & 1 s \\
 $\mathit{RBC} \models R$      & --- & 0.1 s & 68 MB & 1.2k & 259 & 0 & 1 s \\
 $\mathit{RBC} \not \models C$ & --- & 0.1 s & 68 MB & 0.8k & 232 & 0 & 2 s\\
   \hline
% incorrect resilience conditions
$\mathit{Byz} \not \models U$ & \textsc{(b)} & 7.2 s & 168 MB & 3160k & 23120 & 12 & 85 s\\
$\mathit{Byz} \not \models C$ & \textsc{(b)} & 3.2 s & 100 MB & 1123k & 17829 & 6 & 28 s\\
$\mathit{Byz} \not \models R$ & \textsc{(b)} & 0.4 s & 71 MB & 136k & 6387 & 9 & 23 s\\
$\mathit{Byz} \models U$ & \textsc{(c)}& 4.0 s & 103 MB & 1192k & 11073 & 0 & 5 s\\
$\mathit{Byz} \models C$ &  \textsc{(c)} & 8.9 s & 175 MB & 3781k & 36757 & 19 & 104 s\\
$\mathit{Byz} \not \models R$ & \textsc{(c)} & 3.0 s & 110 MB & 1451k & 24656 & 43 & 165 s\\
$\mathit{Sym} \not \models U$ & \textsc{(b)} & 0.1 s & 68 MB & 31k & 1503 & 1 & 1 s\\
$\mathit{Sym} \not \models C$ & \textsc{(b)} & 0.1 s & 68 MB & 32k & 1837 & 3 & 3 s\\
$\mathit{Sym} \models R$ & \textsc{(b)} & 0.2 s & 69 MB & 54k & 2161 & 6 & 16 s\\
$\mathit{Omt} \models U$ & \textsc{(d)} & 0.1 s & 68 MB & 6k &544 & 0 & 1 s\\
$\mathit{Omt} \not \models C$ & \textsc{(d)} & 0.1 s & 68 MB & 6k & 544 & 0 & 2 s\\
$\mathit{Omt} \not \models R$ & \textsc{(d)} & 0.1 s & 68 MB & 0.1k & 481 & 0 & 4 s\\
    \hline
  \end{tabular}
     \caption{
Experimental data on abstraction
     of algorithms tolerant to faults:
     Byzantine, symmetric, omission, clean crashes. RBC corresponds to
     the reliable broadcast algorithm also considered
     in~\cite{FismanKL08}. 
     Checked under different resilience conditions (RC): 
     (a) $n > 3t \wedge f \le t$;
     (b) $n > 3t \wedge f \le t +1$; (c) $n \ge 3t \wedge f \le t$;
     (d) $n \ge 2t \wedge f \le t$.
RBC works for $n \ge t \ge f$.
    Run on a 3.3GHz Intel\textregistered{} Core\texttrademark{} 4GB
    machine.}
\label{tab:exp}

\end{table}

To show practicability of the abstraction, we have implemented the PIA
     abstractions and the refinement loop in OCaml as a prototype
     tool~\textsc{ByMC}.
We evaluated it on the algorithms and the specifications discussed in
     Section~\ref{sec:CaseStudy}, and  conducted experiments that are
     summarized in Table~\ref{tab:exp}.

We extended the \textsc{Promela} language~\cite{H2003} with
     constructs to express $\paraset$,  $\Prop$,  $\ResCond$, and
     $\syssize$.
\textsc{ByMC} receives a description of a CFA $\CFA$ in this
     extended~\textsc{Promela}, and then syntactically extracts the
     thresholds.
The tool chain uses the  Yices SMT solver for existential abstraction,
     and generates the counter abstraction~$\CntSkAut$ in standard
     Promela, such that we can use Spin to do finite state model
     checking.
Finally, \textsc{ByMC} also implements the refinements introduced in
     Section~\ref{sec:RefineSmall} and refines the Promela code
     for~$\CntSkAut$ by introducing predicates capturing spurious
     transitions and unjust states.

The column ``\#R'' gives the numbers of refinement steps.
In the cases where it is greater than zero, refinement was necessary,
     and ``Spin Time'' refers to the \textsc{Spin} running time after
     the last refinement step.

In the cases (A) we used resilience conditions as provided by the
     literature, and verified the specification.
The model RBC represents the reliable broadcast algorithm also
     considered in~\cite{FismanKL08}.

In the bottom part of Table~\ref{tab:exp} we used different resilience
     conditions under which we expected the algorithms to fail.
The cases (B) capture the case where more faults occur than expected
     by the algorithm designer, while the cases (C) and (D) capture
     the cases where the algorithms were designed by assuming wrong
     resilience conditions.
We omit $(\textsc{clean})$ as the only sensible case $n=t= f$ (all
     processes are faulty) results into a trivial abstract domain of
     one interval $[0, \infty)$.

\section{Conclusions}

We presented a novel technique to model check fault-tolerant
     distributed algorithms.
To this end, we extended the standard setting of parameterized model
     checking to processes which use threshold guards, and are
     parameterized with a resilience condition.
As a case study we have chosen the core of several broadcasting
     algorithms under different failure models, including the one by
     Srikanth and Toueg~\cite{ST87:abc} that tolerates Byzantine
     faults.
These algorithms are widely applied in the literature: typically,
     multiple (possibly an unbounded number of) instances are used in
     combination.
As future work, we plan to use compositional model checking
     techniques~\cite{McMillan01} for parameterized verification of
     such algorithms.
%% Moreover, we will consider parameterized processes with process IDs,
%%      and investigate verification of distributed algorithm with
%%      communication and failure models including continuous
%%      time~\cite{FSFK06:EDCC}.

\mypara{Acknowledgments}
We are grateful to Javier Esparza for
     valuable discussions on decidability of VASS.

\bibliographystyle{IEEEtranS}

\bibliography{lit}

% Generated by IEEEtranS.bst, version: 1.13 (2008/09/30)
\begin{thebibliography}{10}
\providecommand{\url}[1]{#1}
\csname url@samestyle\endcsname
\providecommand{\newblock}{\relax}
\providecommand{\bibinfo}[2]{#2}
\providecommand{\BIBentrySTDinterwordspacing}{\spaceskip=0pt\relax}
\providecommand{\BIBentryALTinterwordstretchfactor}{4}
\providecommand{\BIBentryALTinterwordspacing}{\spaceskip=\fontdimen2\font plus
\BIBentryALTinterwordstretchfactor\fontdimen3\font minus
  \fontdimen4\font\relax}
\providecommand{\BIBforeignlanguage}[2]{{%
\expandafter\ifx\csname l@#1\endcsname\relax
\typeout{** WARNING: IEEEtranS.bst: No hyphenation pattern has been}%
\typeout{** loaded for the language `#1'. Using the pattern for}%
\typeout{** the default language instead.}%
\else
\language=\csname l@#1\endcsname
\fi
#2}}
\providecommand{\BIBdecl}{\relax}
\BIBdecl

\bibitem{AJ93}
P.~Abdulla and B.~Jonsson, ``Verifying programs with unreliable channels,'' in
  \emph{LICS}, jun 1993, pp. 160--170.

\bibitem{A12}
P.~A. Abdulla, ``Regular model checking,'' \emph{International Journal on
  Software Tools for Technology Transfer}, vol.~14, pp. 109--118, 2012.

\bibitem{AJNOS12}
P.~A. Abdulla, B.~Jonsson, M.~Nilsson, J.~d'Orso, and M.~Saksena, ``Regular
  model checking for {LTL}({MSO}),'' \emph{STTT}, vol.~14, no.~2, pp. 223--241,
  2012.

\bibitem{ADFT06:DSN}
M.~K. Aguilera, C.~Delporte-Gallet, H.~Fauconnier, and S.~Toueg, ``Consensus
  with {B}yzantine failures and little system synchrony,'' in \emph{DSN}.\hskip
  1em plus 0.5em minus 0.4em\relax IEEE Computer Society, 2006, pp. 147--155.

\bibitem{AK86}
K.~Apt and D.~Kozen, ``Limits for automatic verification of finite-state
  concurrent systems,'' \emph{IPL}, vol.~15, pp. 307--309, 1986.

\bibitem{AW04}
H.~Attiya and J.~Welch, \emph{Distributed Computing}, 2nd~ed.\hskip 1em plus
  0.5em minus 0.4em\relax Wiley, 2004.

\bibitem{BallMMR01}
T.~Ball, R.~Majumdar, T.~D. Millstein, and S.~K. Rajamani, ``Automatic
  predicate abstraction of c programs,'' in \emph{PLDI}, 2001, pp. 203--213.

\bibitem{BiereCCFZ99}
A.~Biere, A.~Cimatti, E.~M. Clarke, M.~Fujita, and Y.~Zhu, ``Symbolic model
  checking using {SAT} procedures instead of {BDDs},'' in \emph{DAC}, 1999, pp.
  317--320.

\bibitem{BKA12}
B.~Bonakdarpour, S.~S. Kulkarni, and F.~Abujarad, ``Symbolic synthesis of
  masking fault-tolerant distributed programs,'' \emph{Distributed Computing},
  vol.~25, no.~1, pp. 83--108, 2012.

\bibitem{BCG1989}
M.~C. Browne, E.~M. Clarke, and O.~Grumberg, ``Reasoning about networks with
  many identical finite state processes,'' \emph{Inf. Comput.}, vol.~81, pp.
  13--31, 1989.

\bibitem{CT96}
T.~D. Chandra and S.~Toueg, ``Unreliable failure detectors for reliable
  distributed systems,'' \emph{JACM}, vol.~43, no.~2, pp. 225--267, March 1996.

\bibitem{Charron-BostM09}
B.~Charron-Bost and S.~Merz, ``Formal verification of a consensus algorithm in
  the heard-of model,'' \emph{IJSI}, vol.~3, no. 2--3, pp. 273--303, 2009.

\bibitem{CGP1999}
E.~Clarke, O.~Grumberg, and D.~Peled, \emph{Model Checking}.\hskip 1em plus
  0.5em minus 0.4em\relax MIT Press, 1999.

\bibitem{Clarke2003}
E.~Clarke, O.~Grumberg, S.~Jha, Y.~Lu, and H.~Veith, ``Counterexample-guided
  abstraction refinement for symbolic model checking,'' \emph{J. ACM}, vol.~50,
  no.~5, pp. 752--794, 2003.

\bibitem{CTV2008}
E.~Clarke, M.~Talupur, and H.~Veith, ``Proving {Ptolemy} right: the environment
  abstraction framework for model checking concurrent systems,'' in
  \emph{TACAS'08/ETAPS'08}.\hskip 1em plus 0.5em minus 0.4em\relax Springer,
  2008, pp. 33--47.

\bibitem{ClarkeE81}
E.~M. Clarke and E.~A. Emerson, ``Design and synthesis of synchronization
  skeletons using branching-time temporal logic,'' in \emph{Logic of Programs},
  ser. LNCS, vol. 131, 1981, pp. 52--71.

\bibitem{Clarke1994}
E.~M. Clarke, O.~Grumberg, and D.~E. Long, ``Model checking and abstraction,''
  \emph{ACM TOPLAS}, vol.~16, no.~5, pp. 1512--1542, 1994.

\bibitem{CC1977}
P.~Cousot and R.~Cousot, ``Abstract interpretation: a unified lattice model for
  static analysis of programs by construction or approximation of fixpoints,''
  in \emph{POPL}.\hskip 1em plus 0.5em minus 0.4em\relax ACM, 1977, pp.
  238--252.

\bibitem{Cytron1991}
R.~Cytron, J.~Ferrante, B.~K. Rosen, M.~N. Wegman, and F.~K. Zadeck,
  ``Efficiently computing static single assignment form and the control
  dependence graph,'' \emph{ACM TOPLAS}, vol.~13, no.~4, pp. 451--490, 1991.

\bibitem{DGG1997}
D.~Dams, R.~Gerth, and O.~Grumberg, ``Abstract interpretation of reactive
  systems,'' \emph{ACM TOPLAS}, vol.~19, no.~2, pp. 253--291, 1997.

\bibitem{PriscoMR01}
R.~De~Prisco, D.~Malkhi, and M.~K. Reiter, ``On k-set consensus problems in
  asynchronous systems,'' \emph{IEEE Trans. Parallel Distrib. Syst.}, vol.~12,
  no.~1, pp. 7--21, 2001.

\bibitem{DolevLPSW86}
D.~Dolev, N.~A. Lynch, S.~S. Pinter, E.~W. Stark, and W.~E. Weihl, ``Reaching
  approximate agreement in the presence of faults,'' \emph{J. ACM}, vol.~33,
  no.~3, pp. 499--516, 1986.

\bibitem{DLS88}
C.~Dwork, N.~Lynch, and L.~Stockmeyer, ``Consensus in the presence of partial
  synchrony,'' \emph{J.ACM}, vol.~35, no.~2, pp. 288--323, 1988.

\bibitem{EmersonK00}
E.~A. Emerson and V.~Kahlon, ``Reducing model checking of the many to the
  few,'' in \emph{CADE}, ser. LNCS.\hskip 1em plus 0.5em minus 0.4em\relax
  Springer, 2000, vol. 1831, pp. 236--254.

\bibitem{EmersonK03}
------, ``Exact and efficient verification of parameterized cache coherence
  protocols,'' in \emph{CHARME}, ser. LNCS, vol. 2860, 2003, pp. 247--262.

\bibitem{EN95}
E.~Emerson and K.~Namjoshi, ``Reasoning about rings,'' in \emph{POPL}, 1995,
  pp. 85--94.

\bibitem{E97}
J.~Esparza, ``Decidability of model checking for infinite-state concurrent
  systems,'' \emph{Acta Informatica}, vol.~34, no.~2, pp. 85--107, 1997.

\bibitem{Esparza99}
J.~Esparza, A.~Finkel, and R.~Mayr, ``On the verification of broadcast
  protocols,'' in \emph{LICS}.\hskip 1em plus 0.5em minus 0.4em\relax IEEE
  Computer Society, 1999, pp. 352--359.

\bibitem{FismanKL08}
D.~Fisman, O.~Kupferman, and Y.~Lustig, ``On verifying fault tolerance of
  distributed protocols,'' in \emph{TACAS}, ser. LNCS, vol. 4963.\hskip 1em
  plus 0.5em minus 0.4em\relax Springer, 2008, pp. 315--331.

\bibitem{FSFK06:EDCC}
M.~Fuegger, U.~Schmid, G.~Fuchs, and G.~Kempf, ``{F}ault-{T}olerant
  {D}istributed {C}lock {G}eneration in {VLSI} {S}ystems-on-{C}hip,'' in
  \emph{EDCC-6}.\hskip 1em plus 0.5em minus 0.4em\relax IEEE Computer Society
  Press, 2006, pp. 87--96.

\bibitem{GS1992}
S.~M. German and A.~P. Sistla, ``Reasoning about systems with many processes,''
  \emph{J. ACM}, vol.~39, pp. 675--735, 1992.

\bibitem{GrafS97}
S.~Graf and H.~Sa\"{\i}di, ``Construction of abstract state graphs with pvs,''
  in \emph{CAV}, ser. LNCS, vol. 1254, 1997, pp. 72--83.

\bibitem{HenzingerJMS02}
T.~A. Henzinger, R.~Jhala, R.~Majumdar, and G.~Sutre, ``Lazy abstraction,'' in
  \emph{POPL}.\hskip 1em plus 0.5em minus 0.4em\relax ACM, 2002, pp. 58--70.

\bibitem{H2003}
G.~Holzmann, \emph{The SPIN Model Checker}.\hskip 1em plus 0.5em minus
  0.4em\relax Addison-Wesley, 2003.

\bibitem{Ip1996}
C.~Ip and D.~Dill, ``Verifying systems with replicated components in
  mur$\phi$,'' in \emph{CAV}, ser. LNCS.\hskip 1em plus 0.5em minus 0.4em\relax
  Springer, 1996, vol. 1102, pp. 147--158.

\bibitem{JKSVW12b}
A.~John, I.~Konnov, U.~Schmid, H.~Veith, and J.~Widder, ``Starting a dialog
  between model checking and fault-tolerant distributed algorithms,''
  \emph{arXiv CoRR}, vol. abs/1210.3839, 2012.

\bibitem{KP2000}
Y.~Kesten and A.~Pnueli, ``Control and data abstraction: the cornerstones of
  practical formal verification,'' \emph{STTT}, vol.~2, pp. 328--342, 2000.

\bibitem{KVW12}
I.~Konnov, H.~Veith, and J.~Widder, ``{Who is afraid of Model Checking
  Distributed Algorithms?}'' 2012, unpublished contribution to: CAV Workshop
  {$(EC)^2$}. \url{http://forsyte.at/download/ec2-konnov.pdf}.

\bibitem{Lamport11a}
L.~Lamport, ``Byzantizing paxos by refinement,'' in \emph{DISC}, ser. LNCS,
  vol. 6950.\hskip 1em plus 0.5em minus 0.4em\relax Springer, 2011, pp.
  211--224.

\bibitem{LR93}
P.~Lincoln and J.~Rushby, ``A formally verified algorithm for interactive
  consistency under a hybrid fault model,'' in \emph{FTCS}, 1993, pp. 402--411.

\bibitem{Lyn96}
N.~Lynch, \emph{Distributed Algorithms}.\hskip 1em plus 0.5em minus 0.4em\relax
  Morgan Kaufman, 1996.

\bibitem{Mayr03}
R.~Mayr, ``Undecidable problems in unreliable computations,'' \emph{Theoretical
  Computer Science}, vol. 297, no. 1-3, pp. 337--354, 2003.

\bibitem{McM93}
K.~McMillan, \emph{Symbolic model checking}.\hskip 1em plus 0.5em minus
  0.4em\relax Kluwer, 1993.

\bibitem{McMillan01}
K.~L. McMillan, ``Parameterized verification of the flash cache coherence
  protocol by compositional model checking,'' in \emph{CHARME}, ser. LNCS, vol.
  2144, 2001, pp. 179--195.

\bibitem{PXZ02}
A.~Pnueli, J.~Xu, and L.~Zuck, ``Liveness with (0,1,{$\infty$})- counter
  abstraction,'' in \emph{CAV}, ser. LNCS.\hskip 1em plus 0.5em minus
  0.4em\relax Springer, 2002, vol. 2404, pp. 93--111.

\bibitem{SIG07}
S.~Sankaranarayanan, F.~Ivancic, and A.~Gupta, ``Program analysis using
  symbolic ranges,'' in \emph{SAS}, ser. LNCS, vol. 4634, 2007, pp. 366--383.

\bibitem{SWR02:hom}
U.~Schmid, B.~Weiss, and J.~Rushby, ``Formally verified {B}yzantine agreement
  in presence of link faults,'' in \emph{ICDCS}, 2002, pp. 608--616.

\bibitem{ST87}
T.~K. Srikanth and S.~Toueg, ``Optimal clock synchronization,'' \emph{Journal
  of the ACM}, vol.~34, no.~3, pp. 626--645, 1987.

\bibitem{ST87:abc}
T.~Srikanth and S.~Toueg, ``Simulating authenticated broadcasts to derive
  simple fault-tolerant algorithms,'' \emph{Dist. Comp.}, vol.~2, pp. 80--94,
  1987.

\bibitem{SteinerRSP04}
W.~Steiner, J.~M. Rushby, M.~Sorea, and H.~Pfeifer, ``Model checking a
  fault-tolerant startup algorithm: From design exploration to exhaustive fault
  simulation,'' in \emph{DSN}, 2004, pp. 189--198.

\bibitem{S88}
I.~Suzuki, ``Proving properties of a ring of finite-state machines,''
  \emph{Inf. Process. Lett.}, vol.~28, no.~4, pp. 213--214, 1988.

\bibitem{TalupurT08}
M.~Talupur and M.~R. Tuttle, ``Going with the flow: Parameterized verification
  using message flows,'' in \emph{FMCAD}, 2008, pp. 1--8.

\bibitem{TS11}
T.~Tsuchiya and A.~Schiper, ``Verification of consensus algorithms using
  satisfiability solving,'' \emph{Dist. Comp.}, vol.~23, no. 5--6, pp.
  341--358, 2011.

\bibitem{WT07}
S.~W{\"o}hrle and W.~Thomas, ``Model checking synchronized products of infinite
  transition systems,'' \emph{LMCS}, vol.~3, no.~4, 2007.

\end{thebibliography}

\clearpage

%\appendix
%\section*{APPENDIX}
\appendices

\section{Undecidability of Liveness Properties}
\label{sec:undec}

\newcommand\pnet{{\cal N}}
\newcommand\ppnet{\Sigma}
\newcommand\ppl{{\cal P}}
\newcommand\ptr{{\cal T}}
\newcommand\pfl{{\cal F}}
\newcommand\pact{\mathit{Act}}
\newcommand\plb{l}
\newcommand\ppost[1]{#1^{\bullet}}
\newcommand\ppre[1]{{}^{\bullet}#1}
\newcommand\pmark{M}

\newcommand\pacc{p_{\mathrm{\AC}}}
\newcommand\sacc{Y_{\mathrm{\AC}}}
\newcommand\pnacc{p_{\mathrm{N\AC}}}
\newcommand\tstart{t^{\theta}}

In this section we show that the non-halting problem of a two counter
     machine (2CM) is reducible to the parameterized model checking
     problem as defined above, that is, using a parameter $n$, a CFA
     $\CFA$ to construct an instance ${\instBLA(n+1,\SkAut)}$, an
     \LTLX{} formula that uses $\ltlG$, $\ltlF$, and atomic
     propositions $\quotes{\exists i.\ \pc_i = Z}$.
We show that ${\instBLA(n+1,\SkAut)}$ simulates at least $n$ steps of
     the 2CM, and that therefore the general parameterized model
     checking problem as formulated in Section~\ref{sec:prelim} is
     undecideable.
Note that $\ltlG$, $\ltlF$, and $\quotes{\exists i.\ \pc_i = Z}$ are
     required to express the liveness property (\ref{ST:rel}) of
     our~FTDA.

The CFA $\CFA$ contains the functionalities of a \emph{control
     process} that simulates the program of the 2CM, as well as of
     \emph{data processes} that each store at most one digit of one of
     the two counters $B$ and $C$ encoded in unary representation.
For simplicity of presentation, we say that if a data process does not
     store a digit for $B$ or $C$, then it stores a digit for a
     counter~$D$.
Counter~$D$ thus serves as a capacity (initially set to $n$), from
     which $B$ and $C$ can borrow (and return) digits, that is,
     initially all data processes' status variable corresponds to $D$.
Our CFA $\CFA$ uses only the status variable $\pc$, while the sets of
     local and global variables can be empty.
We consider paths where exactly one process plays the role of the
     control process, and the remaining $n$ processes are data
     processes.
This can be encoded using $\ltlG$, $\ltlF$, and  $\quotes{\exists i.\
     \pc_i = Z}$.

Intuitively, whenever the control process has to increase or decrease
     the value of a counter, this is done by a handshake of the
     control with a data process; up to this point, our proof follows
     ideas from~\cite{GS1992,EN95,E97}.
In contrast to these papers, however, our system model does not
     provide primitives for such a handshake, which leads to the
     central contribution for our proof: we ``move'' this handshake
     into the specification without using the ``next time'' operator
     not present in \LTLX{}.
In addition, as in \cite{E97}, also the test for zero is moved into
     the specification using our propositions $\neg \quotes{\exists
     i.\ \pc_i = Z}$.
We start with some preliminary definitions.

%% Such atomic propositions play the role of a test for zero.
%% We are reusing two ideas: (a) the number of processes in a certain
%%      location represents a counter~\cite{GS92}, i.e., a place in Petri
%%      net; (b) model checking a Petri net against a formula of linear
%%      $\mu$-calculus is undecidable if the formula can test places for
%%      zero~\cite{E97}.

\newcommand\cif{\mathbf{if\ }}
\newcommand\celse{\mathbf{else\ }}
\newcommand\cthen{\mathbf{then\ }}
\newcommand\cgoto{\mathbf{goto\ }}
\newcommand\cinc[1]{\mathbf{inc}\ #1}
\newcommand\cdec[1]{\mathbf{dec}\ #1}
\newcommand\chalt{\mathbf{halt}}
\newcommand\cthresh{a_0 + \sum_{1 \le i \le \numparam} a_i \cdot p_i}
\newcommand\cloc[1]{#1}
\newcommand\cv{{\cal V}}
\newcommand\ce{{\cal E}}
\newcommand\cc{{\cal C}}

A \emph{two counter machine} (2CM) $\cal M$ is a list of $m+1$
     statements over two counters $B$ and $C$.
A statement at location~$\cloc{v}$ uses a counter $\cc(v) \in \{B,
     C\}$ and has one of the following forms (note, that the machine
     halts at location $m$): 
\begin{align}
    \cloc{v}:\ & \cinc{\cc(v)};\ \cgoto{\cloc{w}} \label{eq:CMinc} \\
    \cloc{v}:\ & \cif \cc(v) = 0\ \cthen\cgoto \cloc{w'} \label{eq:CMzero}\\
        & \celse \cdec{\cc(v)};\ \cgoto \cloc{w''} \label{eq:CMdec} \\
    \cloc{m}:\ & \chalt \label{eq:CMhalt}
\end{align}

The control flow of the machine $\cal M$ is defined by the labelled
     graph $(\cv,\ce,\cc)$, where $\cv = \{ \cloc{v} \mid 0 \le v \le
     m \}$ is the set of locations, $\ce = \ce_{+} \cup \ce_0 \cup
     \ce_{-} \cup \{m, m\}$ is the set of edges, and $\cc: \cv
     \rightarrow \{B, C\}$ is the labelling function which maps a
     location to the counter used in this location.
The sets $\ce_{+}$, $\ce_{0}$, $\ce_{-}$ are defined as follows: 
\begin{itemize}
  \item $\ce_{+} = \{ (v, w) \mid \mbox{ statement at } v \mbox{ goes to } w
    \mbox{ as in } (\ref{eq:CMinc}) \}$;
  \item $\ce_{0} = \{ (v, w') \mid \mbox{ statement at } v \mbox{ goes to } w'
    \mbox{ as in } (\ref{eq:CMzero}) \}$;
  \item $\ce_{-} = \{ (v, w'') \mid \mbox{ statement at } v \mbox{ goes to } w''
    \mbox{ as in } (\ref{eq:CMdec}) \}$.
\end{itemize}

%% Petri nets~\cite{E97} and parameterized systems with
%%      rendezvous~\cite{GS92} have powerful primitives.
%% In the former a program can increment and decrement counters directly.
%% In the latter the control process may pick one process to send a
%%      message to.
%% On the other hand, our CFAs support very limited means of scheduling
%%      and synchronization: Every process can be scheduled at every
%%      point of time if it not blocked by a guard; Processes send
%%      messages to all other processes and can only count the number of
%%      messages received.
In what follows, we model a handshake between the control process and
     a data process in order to implement an increment as in
     (\ref{eq:CMinc}) or decrement as in (\ref{eq:CMdec}).
The handshake is guaranteed by a combination of steps both in the
     control and the data processes and by a constraint formulated in
     \LTLX{} as follows.

\newcommand\cidleC{\mathit{IdlC}}
\newcommand\csynC{\mathit{SynC}}
\newcommand\cackC{\mathit{AckC}}
\newcommand\cidleD{\mathit{IdlD}}
\newcommand\csynD{\mathit{SynD}}
\newcommand\cackD{\mathit{AckD}}
\newcommand\cfaJump{J}
\newcommand\cfaInc{I}
\newcommand\cfaSync{\mathit{HS}}
\newcommand\cfaTest{\mathit{EQ_0}}

We define the set $\PC_C$ of status values of the control process and 
the set $\PC_D$ be the set of status values of a data process:
\begin{align*}
  \PC_C &= \bigcup_{v,w \in \cv } \{(v,v,\cidleC), (v, w, \csynC),
    (v, w, \cackC) \} \\
  \PC_D &= \bigcup_{x,y \in \{ B,C,D \}} \{ (x,x,\cidleD ),
    (x,y,\csynD ), (x,y,\cackD)
\end{align*}

For each $(f, t, h) \in \PC_C \cup \PC_D$, $f$ is the state before a handshake,
    $t$ is the scheduled state after a handshake, and $h$ is the status of the
    handshake.

\begin{figure}
    %\hskip -.5cm
    \scalebox{0.65}{
      \tikzstyle{node}=[circle,draw=black,thick,minimum size=4mm,font=\normalsize]
\tikzstyle{init}=[circle,draw=black!90,fill=white!20,thick,minimum size=4mm,font=\normalsize]
\tikzstyle{dest}=[draw=black!90,circle,double,fill=white!20,thick,minimum size=4mm,font=\normalsize]
\tikzstyle{post}=[->,thick]
\tikzstyle{pre}=[<-,thick]
\tikzstyle{cond}=[rounded
  corners,rectangle,minimum
  width=0.8cm,draw=black,fill=white,font=\normalsize]
\tikzstyle{asign}=[rectangle,minimum
  width=0.8cm,draw=black,fill=gray!5,font=\normalsize]

\begin{tikzpicture}[>=latex,rotate=0]
  %\draw[step=0.5,black,thin] (-1,0) grid (12.5,6);

  \node at ( 1.5,6) [init] (I) {$q_I$};
  \draw [thick,->] (1,6.5) --(I);
  \node at ( 7.5,6) [init] (I2) {$q_I$};
  \draw [thick,->] (7,6.5) --(I2);

  \node at ( 0,3) [node] (1) {$q^{v,w}_1$};
  \node at ( 1.5,3) [node] (2) {$q^{v,w}_2$} ;
  \node at ( 3,3) [node] (3) {$q^{v,w}_3$};
  \node at ( 6,3) [node] (4) {$r^{x,y}_1$};
  \node at ( 7.5,3) [node] (5) {$r^{x,y}_2$} ;
  \node at ( 9,3) [node] (6) {$r^{x,y}_3$};

  \node at ( 1.5,0) [dest] (F) {$q_F$};
  \node at ( 7.5,0) [dest] (F2) {$q_F$};

\draw [post] (I) .. controls (0,6) .. (0,5.5) -- node[cond,anchor=south]
    {$\CFApc = (v,v,\cidleC)$} (0,4) -- (1);
\draw [post] (I) -- (1.5,5) -- node[cond] {$\CFApc = (v,w,\csynC)$} (2);
\draw [post] (I) .. controls (3,6) .. (3,5.5) -- node[cond,anchor=south]
    {$\CFApc = (v, w, \cackC)$} (3,4) -- (3);

\draw [post] (1) -- node[asign,anchor=north]
    {$\CFApc' = (v,w,\csynC)$} (0,0.5) .. controls (0,0) .. (F);
\draw [post] (2) -- node[asign] {$\CFApc' = (v,w,\cackC)$} (1.5,1.5) -- (F);
\draw [post] (3) -- node[asign,anchor=north]
    {$\CFApc' = (w,w,\cidleC)$} (3,0.5) .. controls (3,0) .. (F);

\draw [post] (I2) .. controls (6,6) .. (6,5.5) -- node[cond,anchor=south]
    {$\CFApc = (x,x,\cidleD)$} (6,4) --(4);
\draw [post] (I2) -- (7.5,5) -- node[cond] {$\CFApc = (x,y,\csynD)$} (5);
\draw [post] (I2) .. controls (9,6) .. (9,5.5) -- node[cond,anchor=south]
    {$\CFApc = (x, y, \cackD)$} (9,4) --(6);

\draw [post] (4) -- node[asign,anchor=north]
    {$\CFApc' = (x,y,\csynD)$} (6,0.5) .. controls (6,0) .. (F2);
\draw [post] (5) -- node[asign] {$\CFApc' = (x,y,\cackD)$} (7.5,1.5) -- (F2);
\draw [post] (6) -- node[asign,anchor=north]
    {$\CFApc' = (y,y,\cidleD)$} (9,0.5) .. controls (9,0) .. (F2);

\end{tikzpicture}
    }
    \caption{CFA $\cfaJump(v,w)$ for $(v,w) \in \ce_{+}$ and
        $\cfaInc(x,y)$ for $\cinc{y}$ (and $\cdec x$).}
    \label{Fig:CFAJumpInc}
\end{figure}

Consider an edge $(v, w) \in \ce_{+}$ and $x=D$ and $y=\cc(v)$, that
     is, in location $v$ the counter $\cc(v)$ is incremented, and
     then the control goes to location $w$.
Incrementing the counter is done by a handshake during which the
     control process goes from $v$ to $w$, while a data process goes
     from $D$ (the capacity) to~$\cc(v)$.

To do so, we construct two CFAs $\cfaJump(v,w)$ and $\cfaInc(x,y)$
     shown in Figure~\ref{Fig:CFAJumpInc}: $\cfaJump(v,w)$ goes from
     location $v$ of $\cal M$ to location~$w$ in three steps $\csynC
     \rightarrow \cackC \rightarrow \cidleC$, whereas $\cfaInc(x, y)$
     transfers one digit from counter $x$ to counter $y$ in steps
     $\csynD \rightarrow \cackD \rightarrow \cidleD$.
%% Suppose, the control process is following $\cfaJump(v,w)$ and one or more data
%%     processes are following $\cfaInc(x,y)$.
%% The steps of the processes are not synchronized yet, i.e., two data processes
%%     can model an increment of $y$, or no process make an increment, although
%%     the control process changed its state from $\pc = (v, v, \cidleC)$ to $\pc
%%     = (w, w, \cidleC)$. 
To actually enforce the handshake synchronization, we add the following
     formula $\cfaSync(v,w,x,y)$ that must hold in every state of a
     system instance:  
\begin{align}
    \quotes{\exists k. \pc_k = (x, y, \csynD)} 
        \rightarrow &\quotes{\exists k. \pc_k = (\cloc{v}, \cloc{w}, \csynC)}
        \label{eq:syn-syn} \ \wedge \\
    \quotes{\exists k. \pc_k = (x, y, \cackD)}
        \rightarrow & \neg \quotes{\exists k. \pc_k = (\cloc{x}, \cloc{y}, \csynD)}
        \label{eq:ack-syn}\, \wedge \\
    \quotes{\exists k. \pc_k = (x, y, \cackD)}
        \rightarrow &\quotes{\exists k. \pc_k = (\cloc{v}, \cloc{w}, \cackC)}
        \label{eq:ack-ack} \ \wedge \\
    \quotes{\exists k. \pc_k = (\cloc{w}, \cloc{w}, \cidleC}
        \rightarrow &( \neg \quotes{\exists k. \pc_k = (x, y, \csynD)}
\nonumber\\
 &\neg \quotes{\exists k. \pc_k = (x, y, \cackD)} )
        \label{eq:idle-ack}
\end{align}

In what follows, we will consider the union of CFAs, where
     union is defined naturally as the union of the sets of nodes and
     the union of the sets of edges (note that the CFAs are joint at
     the initial node $q_I$, and the final node $q_F$).

We are now ready to prove the central result: Let $M$ be a system of
     $K$ processes ${{\instBLA(K,\Sk(\cfaJump(v,w) \cup \cfaInc(x,
     y)))}}$ and~$\gst_0$ be a global state of $M$ such that
\begin{align*}    
     \gst_1[1].\pc &= (v,v,\cidleC) \mbox{ and } \\
     \gst_1[k].\pc &= (x,x,\cidleD)\mbox{ for all $2 \le k \le K$.}
\end{align*}
 The constraints (\ref{eq:syn-syn})-(\ref{eq:idle-ack}) impose
     a synchronization behavior:

\newcommand{\propHandshake}{
Let $\pi$ be an infinite path $\{\sigma_i\}_{i \ge 1}$ of $M$ starting with
    $\sigma_1$.
If $\pi \models \ltlG \cfaSync(v,w,x,y)$, then there exists an
    index~$\ell$ such that $2 \le \ell \le K$ and
\begin{align*}    
    \sigma_7[1].\pc &= (w,w,\cidleC) \\
    \sigma_7[\ell].\pc &= (y,y,\cidleD) \\
    \sigma_7[k].\pc &= (x,x,\cidleD) \mbox{ for } 2 \le k \le K, k \ne \ell.
\end{align*}
}

\begin{proposition}\label{prop:handshake}
\propHandshake
\end{proposition}

\newcommand\propHandshakeProof{
\begin{proof} We consider all states in the prefix $\sigma_1, \dots
\sigma_7$.
To prove the proposition, it is sufficient to show that the prefix
     $\sigma_1, \dots \sigma_7$ is as follows for some
     process~$\ell$ such that $2 \le \ell \le K$: 

\smallskip

\begin{tabular}{lllll}
    & step & $\sigma_i[1]$ & $\sigma_i[\ell]$ & $\sigma_i[k]$, $k \ne \ell$ \\
    \hline
    $\sigma_1$& 1& $(v,v,\cidleC)$ & $(x,x,\cidleD)$ & $(x,x,\cidleD)$ \\
    $\sigma_2$& $\ell$ & $(v,w,\csynC)$ & $(x,x,\cidleD)$ & $(x,x,\cidleD)$ \\
    $\sigma_3$& 1& $(v,w,\csynC)$ & $(x,y,\csynD)$ & $(x,x,\cidleD)$ \\
    $\sigma_4$& $\ell$ & $(v,w,\cackC)$ & $(x,y,\csynD)$ & $(x,x,\cidleD)$ \\
    $\sigma_5$& $\ell$ & $(v,w,\cackC)$ & $(x,y,\cackD)$ & $(x,x,\cidleD)$ \\
    $\sigma_6$& 1 & $(v,w,\cackC)$ & $(y,y,\cidleD)$ & $(x,x,\cidleD)$ \\
    $\sigma_7$&  & $(w,w,\cidleC)$ & $(y,y,\cidleD)$ & $(x,x,\cidleD)$ 
\end{tabular}

\smallskip

Now we show that other possible executions contradict to the lemma's hypothesis.
Recall that a single step of a process corresponds to a path of its CFA from
    $q_I$ to $q_F$.

\mypara{State $\sigma_1$} By the hypothesis of the lemma.

\mypara{State $\sigma_2$} As data processes are blocked in $\sigma_1$ by
    (\ref{eq:syn-syn}), only process~1 can move.

\mypara{State $\sigma_3$} Suppose by contradiction that process 1
     makes a step in~$\sigma_2$.
Then $\sigma'_3[1] = (v,w,\cackC)$.
In such a $\sigma'_3$, process 1 is blocked by~(\ref{eq:idle-ack}), and
     the other processes are blocked by~(\ref{eq:syn-syn}).
It follows that no process can make a further step, and $\pi$ cannot
     be infinite, which provides the required contradiction.
Hence, there is some data process $\ell$ which makes the step.

\mypara{State $\sigma_4$} In $\sigma_3$ process $\ell$ is blocked by
     (\ref{eq:ack-ack}).
Further, suppose by contradiction that data process $k \neq \ell$
     makes a step resulting in $\sigma'_4[k] = (x,y,\csynD)$.
Then due to (\ref{eq:syn-syn}), in  $\sigma'_4$, process 1 is blocked.
Moreover, due to (\ref{eq:ack-syn}), processes~$\ell$ and $k$ are also
     blocked in $\sigma'_4$.
It follows that in the path starting with $\sigma'_4$, all data
     processes move from $(x,x,\cidleD)$ to $(x,y,\csynD)$.
As we have finitely many processes, eventually all data processes will
     stop at $(x,y,\csynD)$.
In this state, every process will be blocked by (\ref{eq:syn-syn}) and
     (\ref{eq:ack-syn}).
This contradicts the assumption that $\pi$ is infinite.

\mypara{State $\sigma_5$} In $\sigma_4$ every process $k > 1$
     different from $\ell$ is blocked by the same argument as in
     state~$\sigma_3$ (they all eventually group and are blocked in
     $(x,y,\csynD)$).
Process 1 cannot move due to~(\ref{eq:idle-ack}).

\mypara{State $\sigma_6$} In $\sigma_5$ process 1 cannot move by
     (\ref{eq:idle-ack}).
As process 1 resides in $(v,w,\cackC)$\dash---and thus not in
     $(v,w,\csynC)$\dash---by (\ref{eq:syn-syn}), every data process
     $k > 1$ different from $\ell$ is blocked.

\mypara{State $\sigma_7$} In $\sigma_6$, every data process $k > 1$
     different from $\ell$ is blocked due to (\ref{eq:syn-syn}), as
     process 1 resides at $(v,w,\cackC)$.
Process~$\ell$ cannot move because\footnote{In a proof for the
     composition of several such automata, one would apply $\cfaSync$
     and (\ref{eq:syn-syn}) for different counters to obtain the same
     conclusion.} its CFA contains no guard $\pc = (y,y,\cidleD)$.
\end{proof}
}

\propHandshakeProof

Using $\cfaJump(v,w)$ and $\cfaInc(x,y)$ we can
     simulate~(\ref{eq:CMinc}) for $(v,w) \in \ce_{+}$ by
     instantiating $\cfaJump(v,w)$ and $\cfaInc(D, \cc(v))$.
Moreover, we can simulate~(\ref{eq:CMdec}) for $(v, w'') \in \ce_{-}$
     by instantiating $\cfaJump(v, w'')$ and $\cfaInc(\cc(v), D)$.
Finally, we can simulate~(\ref{eq:CMzero}) for $(v,w') \in \ce_0$
     (that is, the test for zero) by instantiating $\cfaJump(v,w')$
     and adding one more temporal constraint $\cfaTest(v,w')$: 
$$ \quotes{\exists k.
\pc_k = (v,w',\csynC)} \rightarrow \neg \quotes{\exists k.
\pc_k = (\cc(v), \cc(v), \cidleD)}
$$

Now we can construct the CFA $\CFA({\cal M})$ that simulates~$\cal M$.
This CFA is a union of CFAs constructed for edges of $\ce$. If
$$
Q_J = \bigcup_{(v,w) \in \ce} \cfaJump(v,w),
$$ 
then
\begin{align*}
    \CFA({\cal M}) = Q_J \cup 
    \bigcup_{(v,w) \in \ce_{+}} \cfaInc(D, \cc(v))
    \cup \bigcup_{(v,w) \in \ce_{-}} \cfaInc(\cc(v), D).
\end{align*}

Further we define a specification which ensures that there is always
exactly one control process, $\mathit{CP}$, as 
\begin{align*}
  \bigwedge_{q \in Q_J} \bigwedge_{q' \in Q_J \setminus
    \{q\}} \neg \quotes{\exists k.\ \pc_k = q}
  \vee \neg \quotes{\exists k.\ \pc_k = q'}
\end{align*}

We specify the non-halting property $\varphi_{\mathrm{nonhalt}}$ as
follows:
\begin{multline*}
    \ltlG \Big(\neg \quotes{ \exists k. \pc_k =(m,m,\cidleC) }\ \Big) \; \vee \\
    \ltlF \neg \Big(\mathit{CP} \wedge \bigwedge_{(v,w) \in \ce_0}
  \cfaTest(v,w) \; \wedge \\
     \bigwedge_{(v,w) \in \ce_{+} \cup \ce_{-}} \cfaSync(v,w,\cc(v),\cc(v))
         \Big)
\end{multline*}

We can specify two initial process states: One is where the process
stays in $\pc_C=(\ell_0, \ell_0, \cidleC)$; Another one is where the
process $\pc_D=(D, D, \cidleD)$. Then $\PC_0 = \{ \pc_C, \pc_D \}$.

\recallthm{thm:thmUndecBLA}{\thmUndecBLA}

\tikzstyle{node}=[circle,draw=black,thick,minimum size=4mm,font=\normalsize]
\tikzstyle{init}=[circle,draw=black!90,fill=white!20,thick,minimum size=4mm,font=\normalsize]
\tikzstyle{dest}=[draw=black!90,circle,double,fill=white!20,thick,minimum size=4mm,font=\normalsize]
\tikzstyle{post}=[->,thick]
\tikzstyle{pre}=[<-,thick]
\tikzstyle{cond}=[rounded
  corners,rectangle,minimum
  width=1cm,draw=black,fill=white,font=\normalsize]
\tikzstyle{asign}=[rectangle,minimum
  width=1cm,draw=black,fill=gray!5,font=\normalsize]

\section{CFAs of FTDA}

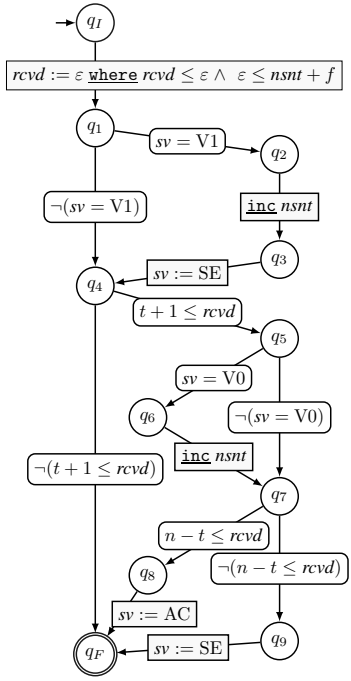
\begin{figure}[t]
\begin{center}
     \scalebox{0.7}{
      \tikzstyle{node}=[circle,draw=black,thick,minimum size=4mm,font=\normalsize]
\tikzstyle{init}=[circle,draw=black!90,fill=white!20,thick,minimum size=4mm,font=\normalsize]
\tikzstyle{dest}=[draw=black!90,circle,double,fill=white!20,thick,minimum size=4mm,font=\normalsize]
\tikzstyle{post}=[->,thick]
\tikzstyle{pre}=[<-,thick]
\tikzstyle{cond}=[rounded
  corners,rectangle,minimum
  width=1cm,draw=black,fill=white,font=\normalsize]
\tikzstyle{asign}=[rectangle,minimum
  width=1cm,draw=black,fill=gray!5,font=\normalsize]

\begin{tikzpicture}[>=latex,rotate=0]

  \node at ( 0,12) [init] (I) {$q_I$};
  \draw [thick,->] (-.75,12) --(I);

  \node at ( 0,10) [node] (D) {$q_1$};
  \node at ( 3.5,9.5) [node] (1) {$q_2$} ;
  \node at ( 3.5,7.5) [node] (2) {$q_3$};
  \node at ( 0,7) [node] (3) {$q_4$};  
\draw [post] (D) -- node[cond] {$\CFApc = \RI$} (1);
\draw [post] (D) -- node[cond] {$\neg(\CFApc = \RI)$} (3);
\draw [post] (1) -- node[asign] {$\CFAinc \; \sent$} (2);
\draw [post] (2) -- node[asign] {$\CFApc := \SE$} (3);

  \node at ( 3.5,6) [node] (4) {$q_{5}$};
  \node at ( 1,4.5) [node] (5) {$q_6$};
  \node at ( 3.5,3) [node] (6) {$q_7$} ;

  \node at ( 1,1.5) [node] (7) {$q_8$};

  \node at ( 3.5,0.25) [node] (8) {$q_9$};
  \node at ( 0,0) [dest] (9) {$q_F$};

\draw [post] (I) --  (D);
\node at ( 1.5,11) [asign] {$\rcvd := \CFApickOp{\CFAdummy}{\CFApick}{\rcvd\le\CFAdummy  \; \wedge \;$ $\CFAdummy \le\sent+f}$};

\draw [post] (3) -- node[cond] {$\neg(t+1 \le \rcvd)$}  (9);  
%%\node[below=.75 of 3,cond] {$\neg(t+1 \le \rcvd)$};
\draw [post] (3) -- node[cond] {$t+1\le\rcvd$} (4);

\draw [post] (4) -- node[cond] {$\CFApc = \IT$} (5);
\draw [post] (4) -- node[cond] {$\neg(\CFApc = \IT)$} (6);
\draw [post] (5) -- node[asign] {$\CFAinc \; \sent$} (6);

\draw [post] (6) -- node[cond] {$n-t \le \rcvd$} (7); 
\draw [post] (6) -- node[cond] {$\neg(n-t \le \rcvd)$} (8); 

\draw [post] (8) -- node[asign] {$\CFApc := \SE$} (9);
\draw [post] (7) -- node[asign,right=-0.2cm] {$\CFApc := \AC$} (9);

\end{tikzpicture}
    }
\end{center}
    \caption{Example CFA of a FTDA as formalized in \cite{JKSVW12b}.}
    \label{Fig:STCFA}
\end{figure}

\begin{figure*}[t]
\begin{minipage}{.33\linewidth}
\begin{center}
\scalebox{.6}{
\begin{tikzpicture}[>=latex,rotate=0]

  \node at ( 0,11.6) [dest] (I) {$q_I$};

  \node at ( 0,10) [node] (D) {$q_1$};
  \node at ( 3,9.5) [node] (1) {$q_2$} ;
  \node at ( 3,7.5) [node] (2) {$q_3$};
  \node at ( 0,7) [node] (3) {$q_4$};  
\draw [post] (D) -- node[cond] {$\CFApc = \RI$} (1);
\draw [post] (D) -- node[cond] {$\neg(\CFApc = \RI$)} (3);
\draw [post] (1) -- node[asign] {$\CFAinc \; \sent$} (2);
\draw [post] (2) -- node[asign] {$\CFApc := \SE$} (3);

  \node at ( 3,6) [node] (4) {$q_{5}$};
  \node at ( 1,4.5) [node] (5) {$q_6$};
  \node at ( 3,3) [node] (6) {$q_7$} ;

  %% \node at ( 1,1.5) [node] (7) {$q_8$};

  %% \node at ( 3,0.25) [node] (8) {$q_9$};
  \node at ( 0,0) [dest] (9) {$q_F$};

\draw [post] (I) --  (D);
\node at ( 1.5,10.9) [asign] {$\rcvd := \CFApickOp{\CFAdummy}{\CFApick}{\rcvd\le\CFAdummy  \; \wedge \;$ $\CFAdummy \le\sent+f_s}$};

\draw [post] (3) -- node[cond] {$\neg(t_s+1 \le \rcvd)$}  (9);  
%%\node[below=.75 of 3,cond] {$\neg(t+1 \le \rcvd)$};
\draw [post] (3) -- node[cond] {$t_s+1\le\rcvd$} (4);

\draw [post] (4) -- node[cond] {$\CFApc = \IT$} (5);
\draw [post] (4) -- node[cond] {$\neg(\CFApc = \IT)$} (6);
\draw [post] (5) -- node[asign] {$\CFAinc \; \sent$} (6);

%% \draw [post] (6) -- node[cond] {$n-t_s \le \rcvd$} (7); 
%% \draw [post] (6) -- node[cond] {$\neg(n-t_s \le \rcvd)$} (8); 

%% \draw [post] (8) -- node[asign] {$\CFApc := \SE$} (9);
%% \draw [post] (7) -- node[asign,right=-0.25cm] {$\CFApc := \AC$} (9);

\draw [post] (6) -- node[asign] {$\CFApc := \AC$} (9);

\end{tikzpicture}
}
\end{center}
\end{minipage}
\begin{minipage}{.33\linewidth}
\begin{center}
\scalebox{.6}{
\begin{tikzpicture}[>=latex,rotate=0]

  \node at ( 0,11.6) [dest] (I) {$q_I$};

  \node at ( 0,10) [node] (D) {$q_1$};
  \node at ( 3,9.5) [node] (1) {$q_2$} ;
  \node at ( 3,7.5) [node] (2) {$q_3$};
  \node at ( 0,7) [node] (3) {$q_4$};  
\draw [post] (D) -- node[cond] {$\CFApc = \RI$} (1);
\draw [post] (D) -- node[cond] {$\neg(\CFApc = \RI$)} (3);
\draw [post] (1) -- node[asign] {$\CFAinc \; \sent$} (2);
\draw [post] (2) -- node[asign] {$\CFApc := \SE$} (3);

  \node at ( 3,6) [node] (4) {$q_{5}$};
  \node at ( 1,4.5) [node] (5) {$q_6$};
  \node at ( 3,3) [node] (6) {$q_7$} ;

  \node at ( 1,1.5) [node] (7) {$q_8$};

  \node at ( 3,0.25) [node] (8) {$q_9$};
  \node at ( 0,0) [dest] (9) {$q_F$};

\draw [post] (I) --  (D);
\node at ( 1.5,10.9) [asign] {$\rcvd := \CFApickOp{\CFAdummy}{\CFApick}{\rcvd\le\CFAdummy  \; \wedge \;$ $\CFAdummy \le\sent}$};

\draw [post] (3) -- node[cond] {$\neg(1 \le \rcvd)$}  (9);  
%%\node[below=.75 of 3,cond] {$\neg(t+1 \le \rcvd)$};
\draw [post] (3) -- node[cond] {$1\le\rcvd$} (4);

\draw [post] (4) -- node[cond] {$\CFApc = \IT$} (5);
\draw [post] (4) -- node[cond] {$\neg(\CFApc = \IT)$} (6);
\draw [post] (5) -- node[asign] {$\CFAinc \; \sent$} (6);

\draw [post] (6) -- node[cond] {$\neg(t_o+1 \le \rcvd)$} (8); 

\draw [post] (8) -- node[asign] {$\CFApc := \SE$} (9);
\draw [post] (6) -- node[cond] {$t_o+1 \le \rcvd$} (7); 
\draw [post] (7) -- node[asign,right=-0.25cm] {$\CFApc := \AC$} (9);

\end{tikzpicture}
}
\end{center}
\end{minipage}
\begin{minipage}{.33\linewidth}
\begin{center}
\scalebox{0.6}{
\begin{tikzpicture}[>=latex,rotate=0]

  \node at ( 0,11.6) [dest] (I) {$q_I$};

  \node at ( 0,10) [node] (D) {$q_1$};
  \node at ( 3,9.5) [node] (1) {$q_2$} ;
  \node at ( 3,7.5) [node] (2) {$q_3$};
  \node at ( 0,7) [node] (3) {$q_4$};  
\draw [post] (D) -- node[cond] {$\CFApc = \RI$} (1);
\draw [post] (D) -- node[cond] {$\neg(\CFApc = \RI$)} (3);
\draw [post] (1) -- node[asign] {$\CFAinc \; \sent$} (2);
\draw [post] (2) -- node[asign] {$\CFApc := \SE$} (3);

  \node at ( 3,3.5) [node] (4) {$q_{5}$};

  \node at ( 0,0) [dest] (9) {$q_F$};

\draw [post] (I) --  (D);
\node at ( 1.5,10.9) [asign] {$\rcvd := \CFApickOp{\CFAdummy}{\CFApick}{\rcvd\le\CFAdummy  \; \wedge \;$ $\CFAdummy \le\sent}$};

\draw [post] (3) -- node[cond] {$\neg(n-t \le \rcvd)$}  (9);  
%%\node[below=.75 of 3,cond] {$\neg(t+1 \le \rcvd)$};
\draw [post] (3) -- node[cond] {$n-t\le\rcvd$} (4);

\draw [post] (4) -- node[asign] {$\CFApc := \AC$} (9);

\end{tikzpicture}
}
\end{center}
\end{minipage}
\caption{Control flow automata for the steps of the FTDA from left to
     right tolerating symmetric faults, omission faults, and clean
     crashes; faults are modeled by fairness constraints not shown.}
%% The algorithm to the right is the folklore broadcasting algorithm also
%%      considered in~\cite{FismanKL08}, with explicit fault modeling
%%      where $\DD$ is a special statues denoting the process has
%%      crashed.} 
\label{Fig:manyCFA}
\end{figure*}
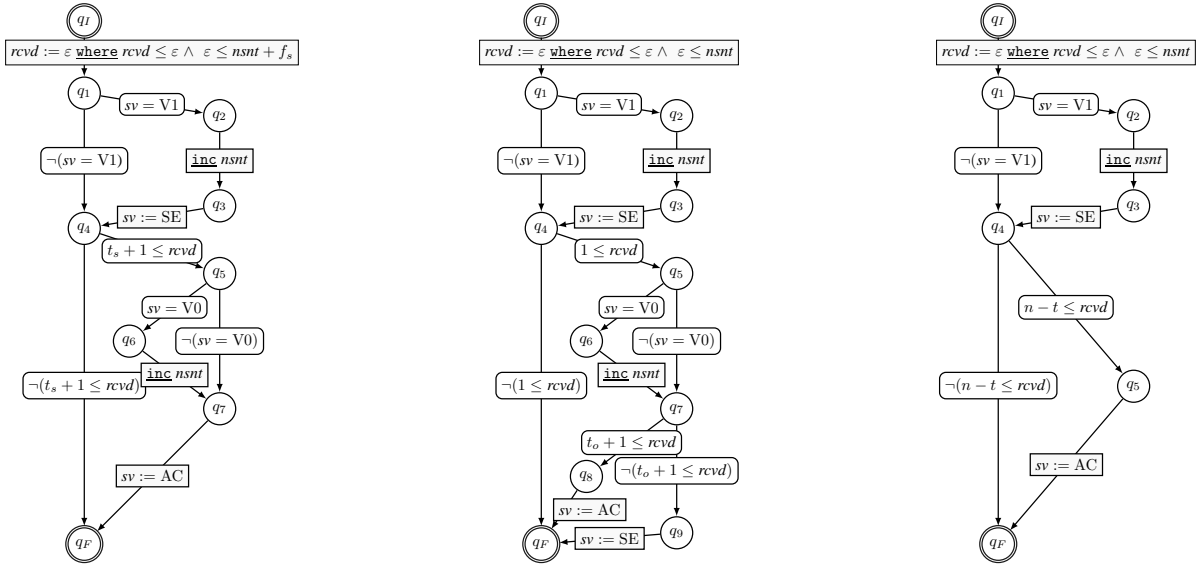

In this section we give examples of the CFAs we use for our
     experiments in Figure~\ref{Fig:STCFA} and~\ref{Fig:manyCFA}.
They have been formalized as discussed in \cite{JKSVW12b}.
In the body of the current paper, we use  slightly different CFA
     definitions, namely without increments, assignments, and
     non-deterministic choice from a range of values.

Using the algorithm from~\cite{Cytron1991}, the CFA in
     Figure~\ref{Fig:CFA_SSA} is obtained from the CFA in
     Figure~\ref{Fig:STCFA}: %.
In contrast to the original CFA from Figure~\ref{Fig:STCFA}, for the
     CFA in Figure~\ref{Fig:CFA_SSA} in every path from $q_I$ to
     $q_F$, each variable appears at most once in the left-hand side
     for every assignment in the original CFA.
Every variable~$x$ has several copies: $x$ for the initial value, $x'$
     for the final one, and $x^1, x^2, \dots$ for intermediate ones.

\section{Details of the counter abstraction}\label{sec:4Igor}

\subsubsection{Initial states.}

Let $\localstates_0$ be a set $\{\ell \mid \ell \in \localstates \wedge
    \exists s_0 \in S_0.\ \ell \gleich{\{sv\} \cup \locset} s_0 \}$; it
    captures initial local states.
Then $w_0 \in S^0_{\CntASys}$ if and only the following conditions are met:
\begin{multline*}
    \exists \param \in \AdmP\ \exists k_1 \cdots 
\exists k_{|\localstates|} . 
    \sum_{1 \le i \le |\localstates|} k_i = \syssize(\param) \, \wedge \,\\
        \forall i\colon 1 \le i \le {|\localstates|}.\;
            \abst_{\param}(k_i)  = w_0.\absCnt{\ell_i} %\label{equ:initII}
   \end{multline*}
$$    
\forall i: 1 \le i \le {|\localstates|} .\  
(\ell_i \not \in \localstates_0)
                \rightarrow (w_0.\absCnt{\ell_i} = \absZero )
%\label{equ:initIII}
$$$$
\exists s_0 \in {S}_0 .\ w_0 \gleich{\globset} s_0
%\label{equ:initI}
$$

Less formally: Concrete counter values are mapped to $w_0.\absCnt{\ell_i}$
using $\abst_\param$; We consider only combinations of counters that give a system
size $\syssize(\param)$; Every counter $\absCnt{\ell_i}$ is initialized to
zero, if the local state $\ell_i$ is met in no initial state $s_0 \in
S_0$; a shared variable $g$ of $w_0$ may be initialized to a value $v$ only if
there is some initial state $s_0 \in S_0$ with $s_0.g = v$.

\subsubsection{Transition relation.}

We now formalize the transition relation $R_{\CntASys}$ of
     $\CntASys(\Sk)$.
The formal definition of when for two states $w$ and $w'$ of the
counter abstraction it holds
     that  $(w, w') \in R_{\CntASys}$ is given  below in
     (\ref{eq:ctr-abs-proc-moves}) to (\ref{eq:ctr-abs-keep}).
We will discuss each of these formulas separately.
We start from the transition relation $R$ of the process
     skeleton~$\Sk$ from which we abstract.
Recall that $(s,s') \in R$ means that a process can go from $s$
     to~$s'$.
From (\ref{abs:existsfrom}) and (\ref{abs:existsto}) we get that,
     $\fromstate$ is the local state of $s$, and $\tostate$ is the
     local state of $s'$.

In the abstraction, if $\fromstate \ne \tostate$, a step from $s$ to
     $s'$ is represented by increasing the counter at index $\tostate$
     by $1$ and decreasing the one at $\fromstate$ by $1$.
Otherwise, that is, if $\fromstate = \tostate$, the counter of
     $\fromstate$ should not change.
Here ``increase'' and ``decrease'' is performed using the
     corresponding functions over the abstract domain $\absdomain$,
     and the mentioned updates of the counters are enforced in
     (\ref{abs:inc}),  (\ref{abs:dec}), and~(\ref{abs:nochange}).
Further, the counters of all local states different from $\fromstate$
     and $\tostate$ should not change, which we achieved by
     (\ref{eq:ctr-abs-keep}).
Performing such a transition should only be possible if there is
     actually a process in state $s$, which means in the abstraction
     that the corresponding counter is greater than $\absZero$.
We enforce this restriction by~(\ref{abs:nonzero}).

By the above, we abstract the transition with respect to local states.
However, $s$ and $s'$ also contain the shared variables.
We have to make sure that the shared variables are updated in the
     abstraction in the same way they are updated in the concrete
     system, which is achieved in  (\ref{abs:globalfrom}) and
     (\ref{abs:globalto}).

%% \hv{The above paragraph is unreadable. Shouldn't it come first to define $w,w'$ ? }

We thus arrive at the formal definition of the abstract transition
relation: $R_{\CntASys}$ consists
of all pairs $(w, w')$ for which there exist ${s}$
     and ${s}'$ in ${S}$, and $\fromstate$ and $\tostate$ in
     $\localstates$ such that
     equations~(\ref{eq:ctr-abs-proc-moves})--(\ref{eq:ctr-abs-keep}) hold:

     \begin{minipage}{0.45\linewidth}
\begin{align}
&   ({s}, {s}')  \in {R}
    \label{eq:ctr-abs-proc-moves}\\
\;&\fromstate \gleich{\{ \pc \} \cup \locset} {s}
\label{abs:existsfrom}\\
\; &w \gleich{\globset} {s} \label{abs:globalfrom}
\end{align}
     \end{minipage}
     \begin{minipage}{0.45\linewidth}
\begin{align}
& w.\absCnt{\fromstate} \ne \absZero \label{abs:nonzero}\\
\; &\tostate \gleich{\{ \pc \} \cup \locset}
   {s}' \label{abs:existsto}\\
\; &w' \gleich{\globset} {s}' \label{abs:globalto}
\end{align}
  \end{minipage}

\begin{equation}
  (\tostate = \fromstate) \; \rightarrow \;
              w'.\absCnt{\fromstate}  = w.\absCnt{\fromstate}
  \label{abs:nochange} 
\end{equation}
\begin{multline}
 (\tostate \ne \fromstate) \; \rightarrow \; \\
              (x=w.\absCnt{\tostate},x'= w'.\absCnt{\tostate} \models 
              \absEx(\{x'=x+1\}))
              %w'.\absCnt{\tostate}  \in \absPlus{1}(u.\absCnt{\tostate})
   \label{abs:inc} 
\end{multline}
\begin{multline}
  (\tostate \ne \fromstate) \; \rightarrow \;\\
              (x=w.\absCnt{\fromstate},x'= w'.\absCnt{\fromstate}
\; \models\\ 
              \absEx(\{x'=x-1\}))
              %w'.\absCnt{\fromstate}  \in \absMinus{1}(u.\absCnt{\fromstate})
   \label{abs:dec} 
\end{multline}
\begin{multline}
 \;   \forall i : 1 \le i \le {|\localstates|} . 
     (\ell_i \ne \fromstate \wedge \ell_i \ne \tostate)
         \; \rightarrow \\  w'.\absCnt{\ell_i} = w.\absCnt{\ell_i}
        \label{eq:ctr-abs-keep}
\end{multline}

\section{Detailed Proofs}\label{sec:proofs}

\mypara{Simulation}
In order to compare the behavior of system instances we use the notion
     of simulation.
Given two Kripke structures $M_1 = (S_1, S^0_1, R_1, \Prop,
     \labelfun_1)$ and $M_2 = (S_2, S^0_2, R_2, \Prop, \labelfun_2)$,
     a relation $H\subseteq S_1 \times S_2$ is a simulation relation
     with respect to a set of atomic propositions $\Prop'\subseteq \Prop$
     iff for every pair of states $(s_1, s_2) \in H$ the following
     conditions hold:

\begin{itemize}

\item $\labelfun_1(s_1) \cap \Prop' = \labelfun_2(s_2) \cap \Prop'$ 

\item for every state $t_1$, with $(s_1,t_1) \in R_1$, there is
  a state $t_2$ with the property $(s_2,t_2) \in R_2$ and
  $(t_1,t_2) \in H$. 

\end{itemize}

If there is a simulation relation $H$ on $M_1$ and $M_2$ such that, for
     every initial state $s^0_1 \in  S^0_1$ there is an initial state
     $s^0_2 \in  S^0_2$ with the property $(s^0_1,s^0_2 )\in  H$, then
     we write $M_1 \preceq M_2$.
In this case we say $M_1$ is simulated by $M_2$.

\subsection{The Proofs.}

%% \recallproposition{prop:handshake}{\propHandshake}
%% \propHandshakeProof

\recallproposition{prop:precise-abs}{\propPreciseAbs}
\propPreciseAbsProof

\recallproposition{prop:genabs}{\propGenAbs}
\propGenAbsProof

\recallthm{thm:comm}{\thmComm}
\thmCommProof

\recallthm{thm:justice}{\thmJustice}
\thmJusticeProof

\recallthm{thm:step2-general}{\thmStepTwoGeneral}
\thmStepTwoGeneralProof

\recallthm{thm:sim2}{\thmSimTwo}
\thmSimTwoProof

\recallthm{thm:justice-for-all}{\thmJusticeForAll}
\thmJusticeForAllProof

\recallthm{thm:remove-spur}{\thmRemoveSpur}
\thmRemoveSpurProof

\recallthm{thm:remove-unjustice}{\thmRemoveUnjustice}
\thmRemoveUnjusticeProof

\section{Sound refinement techniques}

\subsection{Detecting Spurious Transitions and Unjust States}\label{sec:cegar-detect}

In this section we show symbolic techniques to detect spurious
     transitions and unfair states for our specific PIA abstractions.
We are concerned with symbolic representations that can be encoded as
     a formula of an SMT solver.
While there are systems where one can encode the monster system
     $\WrapSys$ (Section~\ref{sec:Refine}) in an SMT
     solver~\cite{KP2000,PXZ02}, it is not obvious how to do this for
     threshold-based distributed algorithms, which have a
     parameterized local state space.

Our method consists of using a model for refinement that abstracts
     only local state space, but is finer than $\{\AbstSys\}_{\param
     \in \AdmP}$.
Thus, we introduce a family $\{\SemiAbstSys\}_{\param \in \AdmP}$,
     where $\SkSAAut$ is a skeleton obtained by applying a data
     abstraction similar to Section~\ref{sec:FirstAbs}, but shared
     variables $\globset$ preserve their concrete values.
Because guards operate on variables both in the abstract and concrete
     domain, we have to define a finer abstraction of guards.

\newcommand\gexpr[1]{\varepsilon_{#1}}

We need some additional notation. 
In Section~\ref{sec:AbsDomain} we introduced set $\gset$ of linear combinations
    met in threshold guards of CFA $\CFA$. 
Let $\gexpr{i}$ be such an expression that induces the threshold function
    $\thresh_j$ for $0 \le j \le \absMax$.
Note, that $\gexpr{0}$ stands for $0$ and $\gexpr{1}$ stands for $1$. 
We construct a formula $in(y, \absSym_a)$ expressing that a variable $y$
    lies within the interval captured by~$\absSym_a$. 
(Note that parameter variables are free in the formula.)
$$
    in(y, \absSym_a) \equiv (\absSym_a = \absSym_\absMax \wedge
    \gexpr{a} \le y) \;  \vee \;
    (\absSym_a \ne \absSym_\absMax \wedge \gexpr{a} \le y < \gexpr{a+1})
$$

The abstraction of CFA guards is defined in Figure~\ref{fig:abstLambda}.

\begin{figure*}[t]
$$
abst_\locset(g) =
\begin{cases}
    {\absEx(g)} & \mbox{if } g \mbox{ is a threshold guard} \\
    {\bigvee_{(\absSym_a, \absSym_b) \in ||g||_E}
    \IntAbs{x} = \absSym_a \wedge in(y, \absSym_b)}
        & \mbox{if } g \mbox{ is a comparison guard over }
        x \in \locset, y \in \globset \\
    {g} & \mbox{if } g \mbox{ is either a comparison guard over }
            x, y \in \globset \mbox { or a status guard} \\
    {abst_\locset(g_1) \wedge abst_\locset(g_2)}
        & \mbox{if } g \mbox{ is } g_1 \wedge g_2 
\end{cases}
$$

\caption{The abstraction of local variables in CFA guards.}
\label{fig:abstLambda}
\end{figure*}

Similarly to Section~\ref{sec:FirstAbs} we construct a CFA $abst_\locset(\CFA)$
    and then use a process skeleton $\SkSAAut$ induced by $abst_\locset(\CFA)$.
For every parameter values $\param \in \AdmP$ one can construct an instance
    $\SemiAbstSys$ using $\SkSAAut$. 
We are going to show that this abstraction is coarser than $\ConcSys$ and finer
    than $\AbstSys$ due to:

\begin{proposition}  %% [abstractions hierarchy]
$\SkAut \preceq \SkSAAut \preceq \SkAAut$.
\end{proposition}

Now we encode the whole family $\{\SemiAbstSys\}_{\param \in \AdmP}$ using the
    VASS representation introduced in Section~\ref{sec:CountAbs}.
A global state of the system $\LVassSys$ is represented by a vector of
    parameter values, a vector of shared variable values, and a vector of process
    counters:
$(\param, \vec{g}, \vec{K})$,
where $\param \in \AdmP$, $\vec{g} \in D^{|\globset|}$,
$\vec{K} \in \NatZero^{|\localstates|}$. Moreover,
$\syssize(\param) = \sum_{1 \le i \le |\localstates|} \vec{K}_i$.

One can define a formula $Init(\param, \vec{g}, \vec{K})$ that captures the initial
    states~$(\param, \vec{g}, \vec{K})$ similarly to the initial states of~$\CntSkAut$.

$\LVassSys$ makes a step from a global state $(\param, \vec{g}, \vec{K})$ to a
global state $(\param', \vec{g'}, \vec{K'})$ when:
\begin{itemize}
    \item $\param'=\param$;
    \item there is
        a step $(s, s') \in R$ of the skeleton $\SkSAAut$, where a process moves from
        the local state $\fromstate \gleich{\locset} s$ to the local state
        $\tostate \gleich{\locset} s'$;
    \item at least one process stays in $\fromstate$, i.e.
        $\vec{K}_{\fromstate} > 0$;
    \item the counters are updated as
        $\vec{K}'_{\fromstate} = \vec{K}'_{\fromstate} - 1$
        and  $\vec{K}'_{\tostate} = \vec{K}'_{\tostate} + 1$;
    \item 
        other counters do not change values, i.e.
        $\forall i: 1 \le i \le |\localstates|.\ 
        (i \ne \fromstate \wedge i \ne \tostate) \rightarrow \vec{K}'_i = \vec{K}_i$.
\end{itemize}

We can encode these constraints by a symbolic formula
$Step(\param, \vec{g}, \vec{K}, \param', \vec{g'}, \vec{K'})$.

The function $\labelfun_{\LVassSys}$ labels states with justice constraints
    similar to the equations that define $\labelfun_\CntASys$ in
    Section~\ref{sec:CountAbs}. We omit the formal definition here.

\begin{proposition}\label{thm:wrap-preceq-lvass}
    System $\LVassSys$ simulates system $\WrapSys$.
\end{proposition}

This allows us to use the following strategy.
We take a transition $\tau$ of $\CntSkAut$ and try to replay it
    in $\LVassSys$.
If $\tau$ is not reproducible in $\LVassSys$, due to
    Proposition~\ref{thm:remove-spur}, $\tau$ is a spurious transition in $\WrapSys$
    and it can be removed.
The following proposition provides us with a condition to check if $\tau$ can be
replayed in $\LVassSys$:

\begin{proposition}\label{thm:spurious-crit}
Let $(w, w') \in R_{\CntASys}$ be a transition of $\CntSkAut$.
If there exists a transition $(\gst, \gst') \in R_\LetterWrapSys$ such
that $w = \wrapMap(\gst)$ and $w' = \wrapMap(\gst')$, then
there exists a transition $Step(\param, \vec{g}, \vec{K}, \param, \vec{g}',
\vec{K}')$ of $\LVassSys$ satisfying the following condition:
\begin{multline*}
    \Wedge{1 \le i \le |\localstates|}{}
        \big(
        in(\vec{K}_i, w.\absCnt{i}) \wedge in(\vec{K}'_i, w'.\absCnt{i})
        \big)
    \; \wedge \\
     \Wedge{1 \le j \le |\globset|}{}
        \big(
        in(\vec{g}_i, w.g_j) \wedge in(\vec{g}'_j, w'.g_j)
        \big)
\end{multline*}
\end{proposition}

In other words, if the formula from Proposition~\ref{thm:unjust-crit} is
    unsatisfiable, the transition $(w, w')$ can be removed safely.

Further, we check whether an abstract state $w \in S_\CntASys$ is an
    unjust one.
If it is, then Proposition~\ref{thm:spurious-crit} allows us to refine justice
    constraints.
The following proposition provides us with a condition that a state is unjust
    in $\LVassSys$:

\begin{proposition}\label{thm:unjust-crit}
Let $w \in S_{\CntASys}$ be a state of $\CntSkAut$ and $q \in \PropVAR$ be a
proposition expressing a justice constraint.  If there exists a state $\gst \in
S_\LetterWrapSys$ such that $w = \wrapMap(\gst)$ and $q \in
\labelfun_\LetterWrapSys(\gst)$, then there exists a state $(\param, \vec{g},
\vec{K})$ of $\LVassSys$ satisfying the following condition:
\begin{multline*}
    q  \in \labelfun_{\LVassSys}((\param, \vec{g}, \vec{K}))
    \; \wedge \\  \Wedge{1 \le i \le |\localstates|}{}
        in(\vec{K}_i, w.\absCnt{i}) 
    \wedge \Wedge{1 \le j \le |\globset|}{}
        in(\vec{g}_i, w.g_j) 
\end{multline*}
\end{proposition}

In other words, if the formula from Proposition~\ref{thm:unjust-crit} is
    unsatisfiable, there is no state $\gst \in S_\LetterWrapSys$ with $q \in
    \labelfun_\LetterWrapSys(\sigma)$ abstracted to $w$. 
Thus, $w$ is unjust.

\begin{remark}
    The system $\LVassSys$ and the constraints of
    Propositions~\ref{thm:spurious-crit}
    and~\ref{thm:unjust-crit} can be encoded in an SMT solver. By checking
    satisfiability we detect spurious transitions and unjust states. Moreover,
    unsatisfiable cores allow us to prune several spurious transitions and
    unjust states at once.
\end{remark}

\subsection{Invariant Candidates Provided by the User}\label{sec:cegar-inv}

In the transition-based approach of the previous section we cannot detect
    paths of being spurious in the case they do not contain uniformly spurious
    transitions (cf.\ beginning of Section~\ref{sec:Refine}).
In this case a human guidance might help: An expert gives an invariant candidate.
Assuming the invariant candidate is expressed as a formula $Inv$ over a global
    state $(\param, \vec{g}, \vec{K})$ of $\LVassSys$, the invariant candidate
    can be automatically proven to indeed being an invariant by verifying
    satisfiability of the formulas:
\begin{equation}
Init(\param, \vec{g}, \vec{K}) \rightarrow Inv(\param, \vec{g}, \vec{K})
            \label{eq:inv-init}
\end{equation}
\begin{multline}
    Inv(\param, \vec{g}, \vec{K})
        \wedge Step(\param, \vec{g}, \vec{K}, \param, \vec{g}', \vec{K}')
        \\ \rightarrow Inv(\param, \vec{g}', \vec{K}')
            \label{eq:inv-trans}
\end{multline}

Then a transition $(w, w') \in R_{\CntASys}$ is spurious if the following formula
is not satisfiable:
\begin{multline*}
 Inv(\param, \vec{g}, \vec{K})
        \;  \wedge  \\
Step(\param, \vec{g}, \vec{K}, \param, \vec{g}', \vec{K}')
         \wedge
  Inv(\param, \vec{g}', \vec{K}')\ \wedge \nonumber\\
     \Wedge{1 \le i \le |\localstates|}{}
        \big(
        in(\vec{K}_i, w.\absCnt{i}) \wedge in(\vec{K}'_i, w'.\absCnt{i})
        \big)
    \; \wedge \\
     \Wedge{1 \le j \le |\globset|}{}
        \big(
        in(\vec{g}_i, w.g_j) \wedge in(\vec{g}'_j, w'.g_j)
        \big)
\end{multline*}

If we receive a counterexample $\IntAbs{T}$ that cannot be refined with the techniques
    from the previous section, we test each transition of $\IntAbs{T}$ against
    the above formula.
If the formula is unsatisfiable for a transition $(w, w') \in \IntAbs{T}$,
    it is sound to remove it from $\CntSkAut$ due to Theorem~\ref{thm:remove-spur},
    Equations~(\ref{eq:inv-init}), (\ref{eq:inv-trans}), and the formula above.

\begin{example}
To give an impression, how simple an invariant can be, for our case study
    (cf.\ Section~\ref{sec:DA}) the relay specification required us to introduce
    the following invariant candidate:
    If $L_s = \{ \ell \in \localstates \mid \ell.\pc=\SE \vee \ell.\pc=\AC \}$, then
    the following formula is an invariant $\sent = \sum_{\ell \in L_s} \vec{K}_\ell$.
Intuitively, it captures the obvious property that the number of messages sent is
    equal to the number of processes that have sent a message.
This property was, however, lost in the course of abstraction.
\end{example}

\end{document}